\documentclass[traditabstract]{aa}
\usepackage{graphicx}
\usepackage{txfonts}
\usepackage{amssymb}
\usepackage[below]{placeins}
\usepackage{natbib}
\usepackage{longtable,lscape}
\bibpunct{(}{)}{;}{a}{}{,}
\usepackage{grffile}
\usepackage{epstopdf}
\graphicspath{{}}
\usepackage[table]{xcolor}
\definecolor{tableShade}{HTML}{E4E4E4}
\usepackage{float}
\usepackage{perpage}
\MakeSorted{figure}
\MakeSorted{table}

\begin{document}


\title{Mining the gap: evolution of the magnitude gap  
in X-ray galaxy groups from the 3 square degree XMM 
coverage of CFHTLS}
\titlerunning{Evolution of the luminosity gap in X-ray galaxy groups}
\authorrunning{Gozaliasl et al. }

\author{G. Gozaliasl\inst{1,2,3}
\and 
A. Finoguenov\inst{1} 
\and 
H. G. Khosroshahi\inst{4}
\and
M. Mirkazemi\inst{5} 
\and 
M. Salvato\inst{5} 
\and 
D. M. Z. Jassur\inst{2} 
\and 
G. Erfanianfar\inst{5} 
\and 
P. Popesso\inst{11}
\and 
M. Tanaka\inst{6} 
\and 
M. Lerchster\inst{5}     
\and 
J. P. Kneib\inst{7} 
\and 
H. J. McCracken\inst{8}  
\and 
Y. Mellier\inst{8}
\and
E. Egami\inst{9}
\and 
M. J. Pereira\inst{9} 
\and
F. Brimioulle\inst{5}
\and
T. Erben\inst{10}
\and
 S. Seitz\inst{5}
}
\institute{
Department of Physics, University of Helsinki, P. O. Box 64, FI-00014, Helsinki, Finland\\
\email{ghassem.gozaliasl@helsinki.fi}
\and
Department of Theoretical Physics and Astrophysics, University of Tabriz, P.O. Box 51664, Tabriz, Iran
\and
Max-Planck-Institut f\"{u}r Plasmaphysik, D-85748 Garching, Germany
\and
School of Astronomy, Institute for Research in Fundamental Sciences (IPM), Tehran, Iran
\and
Max Planck-Institute for Extraterrestrial Physics, P.O. Box 1312, Giessenbachstr. 1., D-85741 Garching, Germany 
\and
National Astronomical Observatory of Japan 2-21-1 Osawa, Mitaka, Tokyo 181-8588, Japan
\and
Laboratoire d'Astrophysique de Marseille, UMR7326, F-13388 Marseille, France
\and
Institut d'Astrophysique de Paris, 98 bis boulevard Arago, F-75014 Paris, France
\and
Steward Observatory, University of Arizona, 933 North Cherry Avenue, Tucson, AZ 85721, USA
\and
Argelander Institute for Astronomy, University of Bonn, Auf dem H\"ugel  71, D-53121 Bonn, Germany
\and
Excellence Cluster Universe, Boltzmannstr. 2, D-85748 Garching, Germany
}
\date{Received/accepted}

\abstract {We present a catalog of 129 X-ray galaxy groups, covering a
  redshift range 0.04$<$z$<$1.23, selected in the $\sim$3 degree$
  ^{2} $ part of the CFHTLS W1 field overlapping XMM observations
  performed under the XMM-LSS project. We carry out a statistical
  study of the redshift evolution out to redshift one of the magnitude
  gap between the first and the second brightest cluster galaxies of a
  well defined mass-selected group sample. We find that the slope of
  the relation between the fraction of groups and the magnitude gap
  steepens with redshift, indicating a larger fraction of fossil
  groups at lower redshifts. We find that 22.2$\pm$6\% of our groups at
  z$\leq$0.6 are fossil groups. We compare our results with the
  predictions of three semi-analytic models based on the
  Millennium simulation. The intercept of the relation between the
  magnitude of the brightest galaxy and the value of magnitude gap
  becomes brighter with increasing redshift. This trend is steeper
  than the model predictions which we attribute to the younger stellar
  age of the observed brightest cluster galaxies. This trend argues in
  favor of stronger evolution of the feedback from active galactic nuclei
   at z$<$1 compared to the models.
  The slope of the relation between the magnitude of the brightest
  cluster galaxy and the value of the gap does not evolve with
  redshift and is well reproduced by the models, indicating that the
  tidal galaxy stripping, put forward as an explanation of
  the occurrence of the magnitude gap, is both a dominant mechanism
  and is sufficiently well modeled.}

   
   \maketitle 
   \section{Introduction}

   Groups and clusters are important environments for the formation
   and evolution of galaxies, particularly for the very bright
   galaxies. In the last few decades, several observational and
   theoretical studies have focused on the group and cluster
   environments in order to advance our understanding of their galaxy
   properties
   \citep[e.g.,][]{white78,Yang05,Yang08,Yang09,Zandivarez06,Zibetti09,Percival10,vandenBosch13,Hearin13,Scoville07}.

   While effects of gravity on galaxy formation have been modeled in
   great detail
   \citep[e.g.,][]{Zel70,Weinberg72,press74,Huntley80,Barnes89,Bekki02,Libeskind06,Junqueira13},
   processes of the galaxy evolution are also governed by the gas
   content available for star-formation. For galaxies in groups and
   clusters, the cold gas content depends on both external (cooling of
   the intracluster medium (ICM), gas stripping by galactic motion
   through ICM, observed even for the central galaxy due to sloshing),
   and internal processes galactic outflows, driven by the feedback
   from active galactic nuclei (AGN) and supernovae \citep{Croton06}. In addition, the central
   galaxy can grow by merging and tidal stripping of satellite
   galaxies.

   Dynamical friction causes massive galaxies in halos with
   sufficiently low velocity dispersion to merge in a few Gyrs with
   the central group galaxy, forming a giant elliptical galaxy
   \citep{Ponman94}. Within this galaxy-galaxy merger paradigm,
   luminosity gap may be sensitive to merger rates in groups and
   clusters of galaxies. To get the insights on the processes associated
   with the formation of central galaxies, we embark on a study of the
   occurrence of the magnitude gap.

   The magnitude gap is often used to study the formation history and
   dynamical age of the hosting halo, particularly in fossil groups
   \citep{Milosavljevic06,van den Bosch07,Dariush07}.  Simulations of
   \citet{DOnghia05} show that early formed groups ( , fossils) have
   larger magnitude gaps. Tracing fossil groups forward in time,
   \citet{VonBendaBeckmann08} find that their large magnitude gaps can
   be filled by a recent infall of galaxies. Also, some merger events
   may not be caught by the gap statistics, as indicated by a study of
   \cite{LaBarbera12}. Thus, to derive conclusions from the gap
   statistics, a comparison to numerical simulations is required. To
   date, such a comparison has only been performed on the nearby
   galaxy groups at z$\sim$0 and massive clusters at z$\sim$0.2
   \citep{Dariush10, Smith10}. This paper expands the studies of
   the cosmic evolution of the magnitude gap to a redshift of 1.1.

   To the extremes of the magnitude gap characterization belong the
   fossil groups, which have L$_{X}\gtrsim$2.13$\times$10$^{42}$
   h$^{-2}$erg s$^{-1}$ and a magnitude gap above $\Delta M_{1,2}$=2
   mag \citep{Jones03}. Origin and nature of fossil groups are still
   not completely understood. Identification of these systems is
   challenging, and further studies and observations (i.e., X-ray and
   spectroscopic) of fossil groups are of interest. According to
   \cite{Jones03}, 8 to 20 per cent of all galaxy groups are
   fossils. Simulations suggest that fossils tend to reside in less
   dense environments compared to an average group of similar total
   mass and formation history \cite{DOnghia05,Dariush07,Diaz11}. This
   is also observed in the environmental study of a fossil group, ESO
   3060170 \citep{Su12}. Fossil galaxy groups have been shown to
   exhibit interesting properties such as concentrated dark matter
   halos \citep{Khosroshahi07, Humphrey12}, disky isophotes of the
   central galaxies \citep{Khosroshahi06, Smith10}. From the age and
   metallicity gradients of fossils, in a study based on six galaxies
   \cite{Eigenthaler13} concluded that these systems form by multiple
   mergers of massive galaxies.

   Identification and detection of galaxy groups is also challenging,
   since the overdensity of galaxies per unit area in a redshift slice
   is more sensitive to the projection effects by foreground and
   background galaxies compared to clusters. In addition, to study the
   galaxy properties it is advantageous to have a group selection that
   is independent of galaxy properties. Extended X-ray emission from
   groups and clusters of galaxies offers such a selection
   \citep{Borgani01, Rosati02}. Deep XMM and Chandra surveys such as
   Chandra Deep Field South \citep[CDFS;][]{Giacconi02}, Chandra Deep
   Field North \citep[CDFN;][]{Bauer02}, Lockman Hole
   \citep[]{Finoguenov05}, the Cosmic Evolution Survey
   \citep[COSMOS;][]{Finoguenov07}, XMM-Large Scale Structure
   \citep[LSS;][]{Pacaud07}, the Canadian Network for Observational
   Cosmology \citep[CNOC2;][]{Finoguenov09, Connelly12}, Subaru-XMM
   Deep Field \citep[SXDF;][]{Finoguenov10}, the XMM-Newton-Blanco
   Cosmology Survey project \citep[XMM-BCS;][]{suhada12} reveal new
   capabilities of X-ray selection of groups. Catalogs available from
   these surveys have already made an important contribution to
   studies of galaxy formation and evolution \citep[e.g.,][]{Tanaka08,
     Tanaka12, Giodini09,Silverman09,Allevato12}.

   The main goal of this paper is to identify X-ray galaxy groups in
   the part of the CFHTLS W1 field covered by XMM observations. We
   carry out a statistical study of the redshift evolution, up to a
   redshift of 1.1, of the magnitude gap and the volume abundance of
   the detected X-ray galaxy groups. We combine the observational
   measurements of the distribution of the luminosity gap of groups
   and the absolute magnitude of the brightest galaxies, and compare
   this to predictions of semi-analytic models (SAMs) from \citet[][hereafter B06]{Bower06},
   \citet[][hereafter DLB07]{De Lucia07} and \citet[][hereafter
   G11]{Guo11}.  The study of magnitude gap statistics and X-ray
   properties of galaxy groups enable us to identify fossil group
   candidates in our sample. We also report on the redshift evolution
   of the extent of X-ray detection from galaxy groups.
 
   The structure of this paper is as follows: we present the
     XMM-Newton and CFHTLS data in \S 2. \S 3 gives a brief
     description of the SAMs we use in this paper, highlighting their
     differences and similarities.  In \S 4, we describe the group
     identification technique and present a catalog of galaxy groups.
     \S 5 discusses group membership contamination. We present
     magnitude gap statistics and relation between the magnitude of
     the first and second brightest group galaxies (BGGs) and the magnitude gap in \S 6. We summarize our results in \S 7.  
     
     Unless stated otherwise, we adopt a \textquoteleft concordance\textquoteright cosmological
   model, with ($\Omega_{\Lambda}, \Omega_{M}$, h)=(0.75, 0.25, 0.71),
   where the Hubble constant is characterized as 100 h km s$^{-1}$
   Mpc$^{-1}$ and quote uncertainties on 68\% confidence level.
  
\section{Data}
\subsection{XMM-Newton data }
A description of the XMM--Newton observatory is given by
\cite{Jansen01}. We use the data collected by the European photon
imaging cameras (EPIC): the pn-CCD camera \citep{Struder01} and the
MOS-CCD cameras \citep{Turner01}.

In this paper we analyzed the XMM-Newton observations of the CFHTLS
wide (W1) field as a part of the XMM--LSS survey \cite{Pierre07}. The
details of observations and data reduction are presented in
\cite{Bielby10}. We concentrate on the low-z counterparts of the X-ray
sources and use all XMM observations performed till 2009, 
  covering an area of 2.276$^{\circ}\times $2.276$^{\circ}$.
  Altogether, $ \sim$3 degrees$^{2}$ of this area is covered by the
  optical data of the CFHTLS survey. The primary cluster catalog of
the XMM-LSS project is published in \cite{Adami11}. The group catalog
presented in this paper extends the \cite{Adami11} work to low-mass
systems, as well as presents X-ray flux estimates with contribution
from point sources removed. This allows us to self-consistently infer
cluster masses, using the calibrations achieved in the COSMOS survey
using a similar flux extraction technique \citep{Leauthaud10}. In our
modeling of the survey sensitivity, we account for both the variation
in the exposure due to differences in the flare removal, vignetting,
and the aperture flux loss for nearby groups.
  
The modeling of the flux limit is defined by the spatial scales used
for the detection and the sensitivities achieved on these scales. The
change in the flux limit with redshift occurs at redshifts below 0.2
and is due to the correction for the missing flux, while the aperture
flux limit is considered to be redshift independent. The correction
for the missing flux is done following the tabulations of
\cite{Finoguenov07}.

\begin{figure}[ht!]
 \begin{center}  
 \leavevmode
 \includegraphics[width=9.5cm]{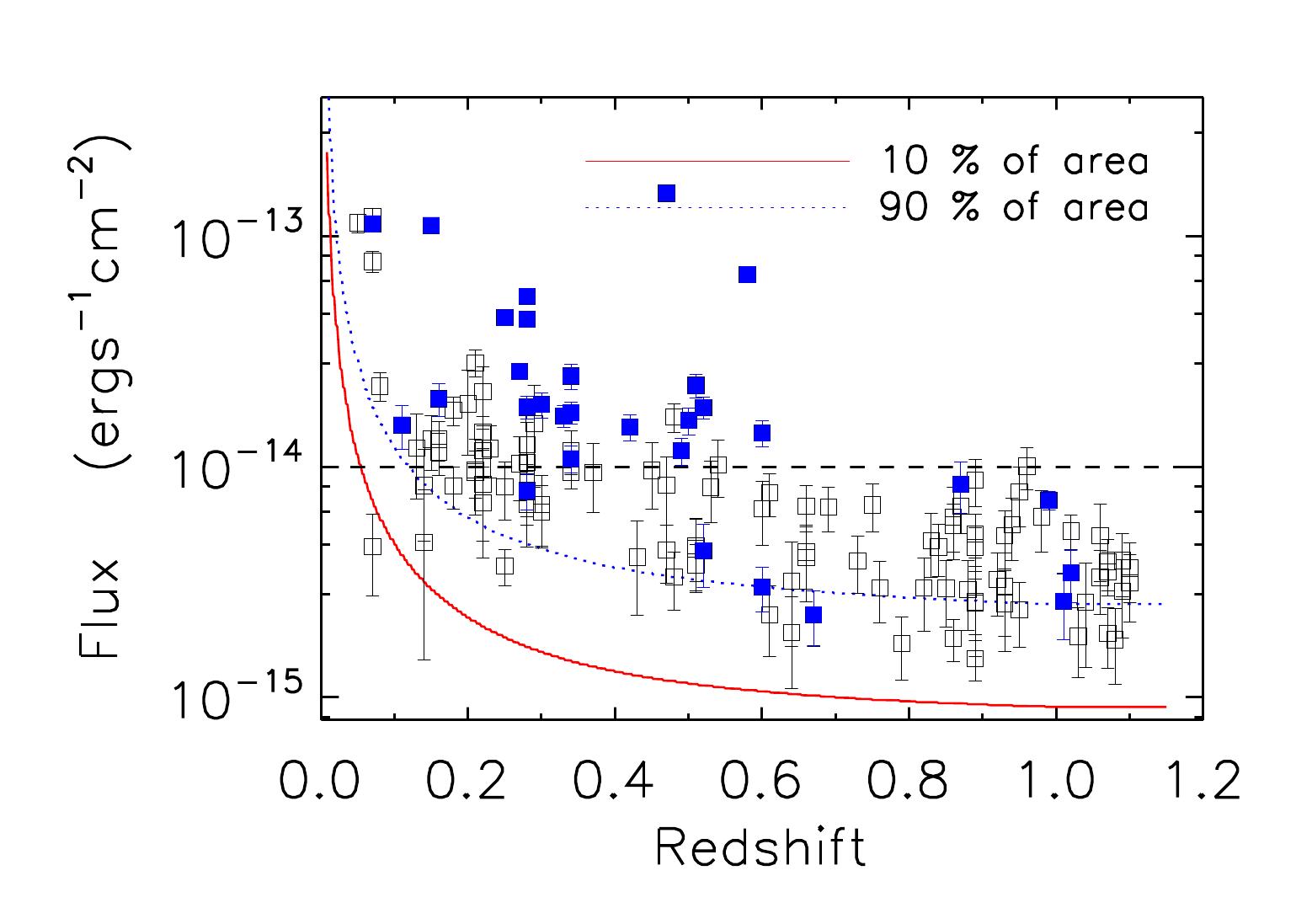}
 
  \end{center} \caption[flag]{Detected flux (0.5-2 keV band) as
  a function of redshift for the sample and its modeling. The open
  black squares with error bars present our identified galaxy
  groups. The filled blue squares mark the galaxy groups in
  common with \cite{Adami11} catalog. Solid red (dotted blue) curve
  shows the sensitivity achieved at the deepest 10 (90)\% area of the
  survey. The dashed black horizontal line shows the $10^{-14}$ ergs
  s$^{-1} $ cm $^{-2}$ flux threshold adopted for the C1 XMM--LSS
  sample.} \label{flux} \end{figure}

In Fig. \ref{flux} we present the flux limit corresponding to the best
10\% (solid red curve) and 90\% (dotted blue curve) of the area.  In
addition to an increase in the flux, very low-z systems are hard to
identify in the photometric data due to the projection effects. The
volume covered at low-z (e.g., z $< 0.05 $) is very small, 0.02\% of
the total survey volume, while formal galaxy group luminosities are
below $10^{41}$ ergs s$^{-1}$ level, beyond the explored level for
galaxy groups. The dashed black horizontal line in Fig.\ref{flux}
illustrates the $10^{-14}$ ergs s$^{-1} $ cm $^{-2}$ flux threshold
adopted for the C1 XMM-LSS sample \citep{Pacaud07}. The open black
squares with error bars show our flux (ergs s$^{-1} $ cm $^{-2}$,
0.5--2 keV band) estimates versus redshift. We show 30 galaxy groups
in common with \cite{Adami11} with filled blue squares. These 30
galaxy groups include one system from C0, all 17 systems from C1 which
are within the CFHTLS coverage, five systems from C2 and seven systems
from C3 in the XMM-LSS classification. Using simulations,
\cite{pacaud06} classified extended sources as C1 and C2. The C1 class
is a purely X-ray selected cluster (i.e., less than 1\% of clusters can
be found as miss-classified point-like source). The C2 class allows
for the 50\% contribution from the miss-classified point-liked
sources. The C3 clusters are classified as faint objects with
less-well characterized X-ray properties. These objects have been
selected mostly by visual inspection of the optical and X-ray
data. The C0 class considered as the low mass cluster category without
clear X-ray emission. The detailed information is presented in
\cite{Adami11,Pacaud07,pacaud06}.

\begin{figure}[ht!]
  \begin{center}  
  
 \includegraphics[width=9.5cm]{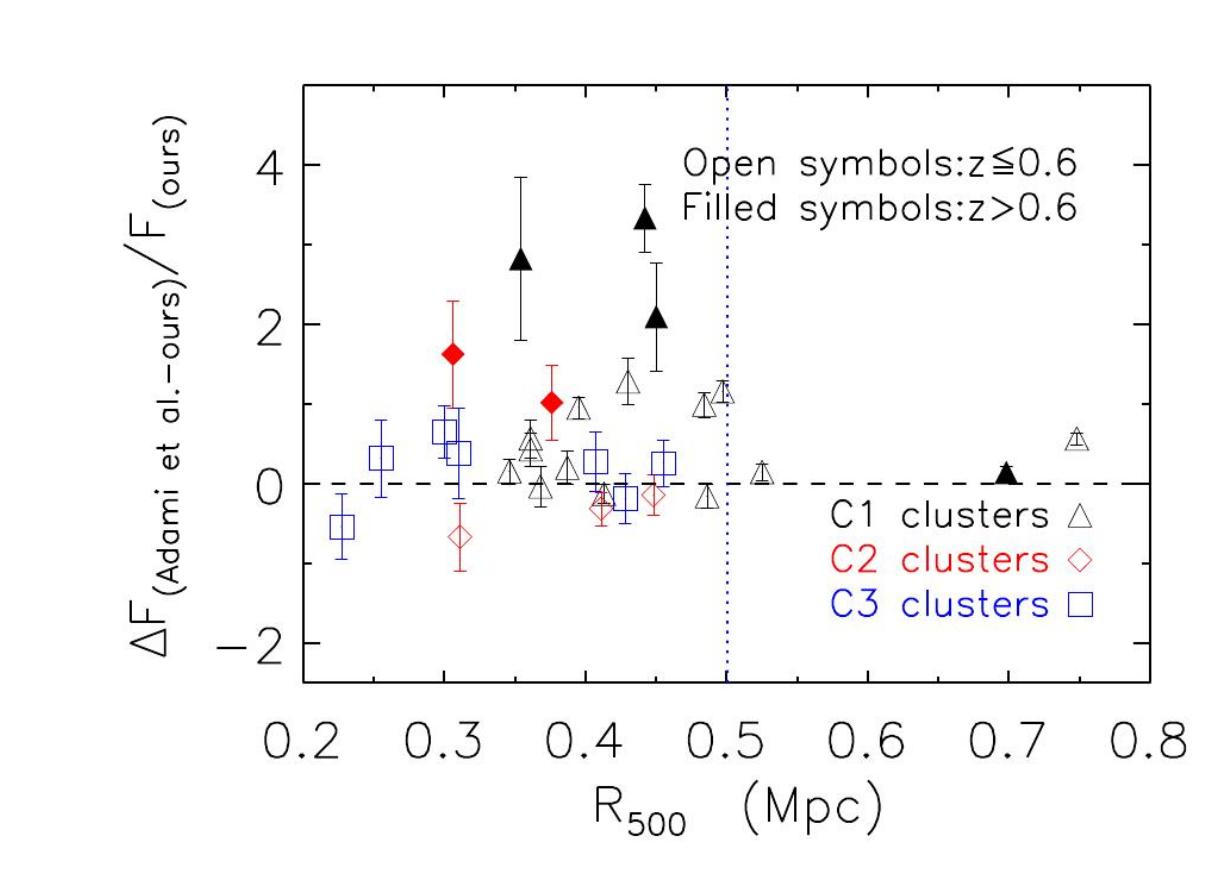}
  
\end{center} \caption[flag]{Comparison of the reported flux in the
  0.5--2 keV band for galaxy groups and clusters in common between
  Adami et al. (2011) and our catalog, ($Flux_{(Adami}/Flux_{Ours}-1$)
  versus $ R_{500} $ (Mpc). We illustrate C1 clusters with black
  triangles, C2 clusters with red diamonds and C3 clusters with blue
  squares. The filled and open symbols show galaxy groups and clusters
  at redshift ranges z$>$0.6 and z$\leq$0.6, respectively. The dashed
  vertical blue line at 0.5 Mpc, marks the fixed radius used for the flux
  estimates in Adami et al.(2011).}  \label{r500}
   \end{figure}

In Fig. \ref{r500} we compare our flux estimates in the 0.5--2 keV
band for the galaxy groups and clusters in common with
\cite{Adami11}. As can be seen, our estimates of flux for these groups
are generally consistent, but some groups, notoriously all the groups
at z$>$0.6 (filled symbols in Fig.\ref{r500}), have lower flux in
our analysis. In Fig. \ref{r500}, we also study the possibility of
these differences to stem from the extrapolation of the flux, and find
it unlikely, as the values for radii of $R_{500}$ we use are
comparable to the 0.5 Mpc radius, used in \cite{Adami11}. All these
high-z groups have strong contribution from point sources to the
emission, which we detect and remove, which would explain the
differences in the flux estimates. Most of these point sources are not
aligned with cluster center and can hardly be associated with
unresolved cool cores. A previous case of such disagreement in the
flux, related to the z=1.6 cluster (compare
\cite{Tanaka10,Papovich10}), has been settled by Chandra observation
in favor of our flux \citep{Pierre12}.  Our own correction for the
removal of cool core flux of high-z groups, based on the Chandra data
on COSMOS, included in the analysis of \cite{Leauthaud10}, is 10\% on
average. This correction is applied when using the scaling relations.

Our modeling of the survey volume, used for refined selection of the
mock galaxy catalogs, takes into account the relation between the
luminosities we extract from our flux estimates and a total mass,
inferred by the weak lensing analysis \citep{Leauthaud10} on systems
of similar mass and redshift found in the COSMOS field. The
two-dimensional information on the sensitivity toward the flux
detection and the resulting flux limits as a function of redshift
discussed above.

   \begin{figure}[ht!]
\begin{center}
\includegraphics[width=9.5cm]{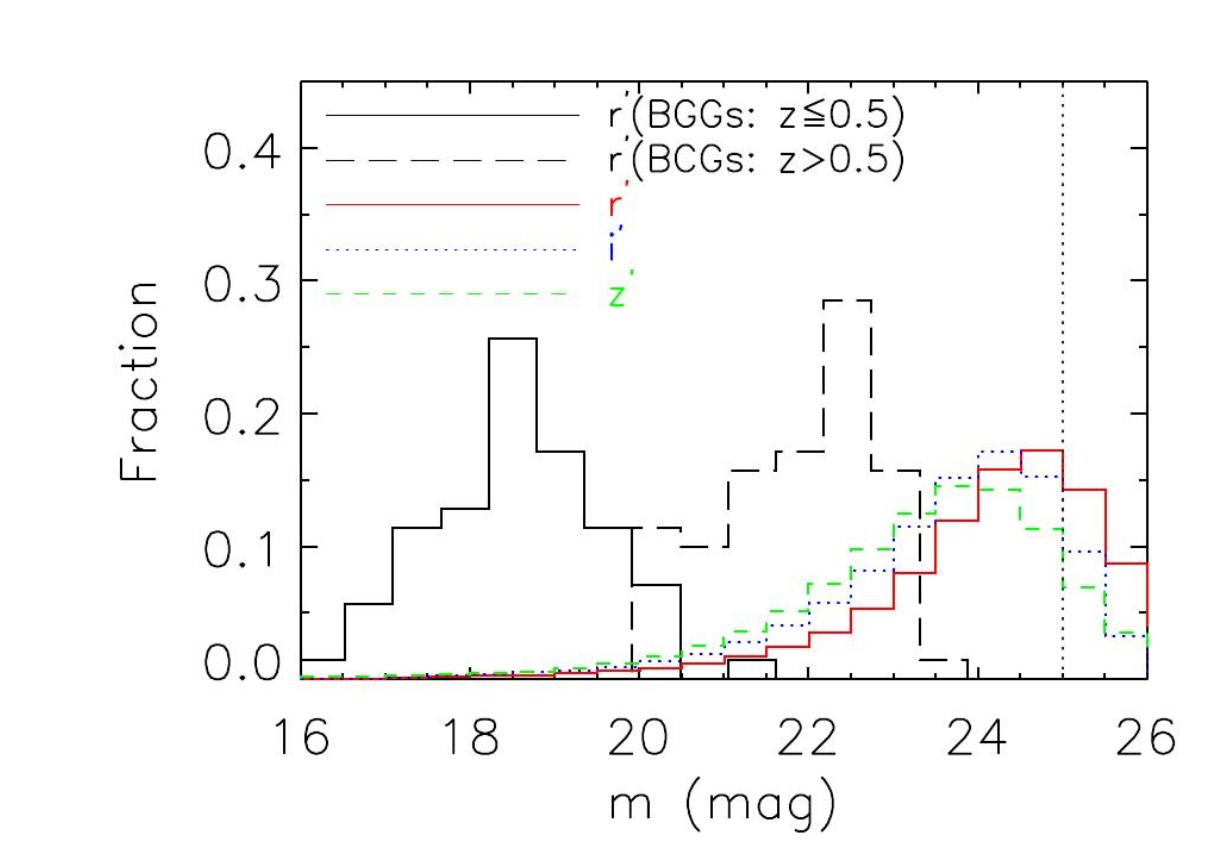}
\end{center}
\caption[completeness]{Magnitude distribution of the CFHTLS-W1 
photometric data in r$^{\prime}$-, i$^{\prime}$- and z$^{\prime} $ band. 
The solid and dashed black histograms indicate the r$^{\prime}$-band
magnitude distribution of BGGs and BCGs in our catalog in the redshift
ranges, z$\leqslant 0.5$ and z$>$0.5, respectively. The dotted
black vertical line represents the magnitude completeness of the CFHTLS
data in r$^{\prime}$-band.}
\label{comp}
\end{figure}

\subsection{Photometric and spectroscopic data}

The CFHTLS wide observations have been carried out in the period
between 2003 and 2008, covering an effective survey area of $\sim$154
square degrees. The W1 field is the widest field of the CFHTLS survey
and covers an area of $ \sim$64 square degrees around RA = $ 02^{h}
18^{m}00^{s} $, Dec = $ 07^{h}00^{m}00^{s}$. The optical images and
data of the CFHTLS were obtained with the MegaPrime instrument mounted
on the CFHT in the five filters u$^{*}$, g$^{\prime}$, r$^{\prime}$,
i$^{\prime}$ and z$^{\prime} $ \citep[e.g.,][]{Erben13}.

The completeness limits and image quality of the photometric survey
data are important for the detection of the optical counterparts of
the X-ray sources and for the accuracy of the photometric redshift of
galaxies \citep[e.g.,][]{Mirkazemi14,Brimioulle08,Bordoloi12}.

For the purposes of our study the completeness toward the
detection of the second brightest galaxies is important for the large
magnitude gap systems at high-z.

In Fig. \ref{comp} we show the completeness limits, r$^{\prime}\sim$25
mag (vertical dotted black line), i$^{\prime}\sim$24.5 mag and
z$^{\prime}\sim$24 mag for the optical data of CFHTLS-W1 field
overlapping XMM observations. We also compare the r$^{\prime}$-band
magnitude distribution of BGGs in our
catalog in two redshift ranges, z$\leq$0.5 (solid black histogram) and
z$>$0.5 (dashed black histogram), to the r$^{\prime}$-band
completeness limit of the CFHTLS.
 
This comparison indicates that completeness limit has no effect on the
magnitude gap calculation of groups with $ \Delta $M$
_{1,2}<$2. However for the z$>0.5$ groups with $ \Delta $M$ _{1,2}>2$,
the determination of the gap is affected by the completeness. In
  order to account for this effect in the plots, we assigned all $
\Delta $M$ _{1,2}>2$ systems to a single bin (see Fig. \ref{gap}).

In this paper, we use the photometric redshift catalogs of the CFHTLS
survey by \cite{Brimioulle08,Brimioulle13}. The XMM-CFHTLS W1
field, has a good spectroscopic coverage of $ \sim $ 0.64
degrees$ ^{2} $ with the VIMOS-VLT Deep Survey (VVDS)
\citep{LeFevre04,LeFevre05} and the targeted cluster follow-up of
\cite{Adami11}.
\section{Semi-analytic galaxy catalogs}

In order to decipher the observational trends it is instructive to
compare them to predictions of theory. To this end we have embarked on
the semi-analytic galaxy catalogs of B06, DLB07, and G11.  All three
models implemented on the dark matter halo merging trees obtained from
the Millennium simulation  which was described in detail in \cite{Springel05}.

The B06 model improves Durham SAMs \citep[GALFORM;  e.g.,][]{Cole00,Benson03} notably the black hole growth, AGN
feedback \citep{Kauffmann00} and disk instability
\citep{Cole00,Mo98}. The DLB07 model is a
developed SAM of those presented in \cite{Springel05},
\cite{DeLucia06} and \cite{Croton06} models. G11 is an updated version
of these models.
   
To estimate the photometric properties of galaxies, DLB07 and
  G11 apply the stellar population synthesis model of
  \cite{Bruzual03} and the initial mass function (IMF) of
  \cite{Chabrier03}. B06 uses a \cite{Kennicutt83}
  IMF with no correction for brown dwarf stars and outputs magnitudes
  in the Vega system, which we convert to the AB system to match two
  other models.
     
G11 refines the definition of a satellite galaxy in halos and
  treats a galaxy as a satellite, applying the tidal and ram-pressure
  stripping, only at times when it is located within $R_{vir}$. In two
  other models a galaxy acts as a satellite when it is ascribed to a
  larger friends-of-friends halo and its hot gas atmosphere is immediately stripped,
  leading a reddening in color and a rapid decline in the star
  formation.  Thus, modifications of G11 lead galaxies to retain more gas
  for star formation.
 
  The merger trees \citep{Harker06} used in the B06 model
  differ from that implemented in DLB07 and G11
  \citep{Springel05}. All three models include star-bursts triggered
  by major merger, associated with the mass ratio of the merging
  progenitors in excess of 0.3. When a major merger occurs, all the
  stellar objects in progenitors are transferred into the bulge of the
  remnant galaxy. In G11 and DLB07 a fraction of cold gas turns into
  the bulge stars \citep{Somerville01}, while in B06 all the
  cold gas is converted into stars. Among these three models only G11
  considers the satellite-satellite mergers, which can reduce the
  number of satellite galaxies in massive halos
  \citep[e.g.,][]{Kim09}.
 
  Another relevant physical process for galaxy evolution which has
  only been taken into account in G11 model is a tidal disruption of
  the stellar component from merging satellites
  \citep[e.g.,][]{Font08,Henriques10}.
       
  Following \cite[e.g.,][]{white91}, all three SAMs apply a model for
  the cooling from hot gas halo. They all define a cooling radius to
  distinguish between the rapid cooling and static hot halo regimes
  and compare it with the virial radius (DLB07 and G11) or the
  free-fall radius (B06). B06 assumes that cooling flows in halos with
  virial velocities below 50 km s$ ^{-1} $ are suppressed at z$\leq$6
  (see B06 and G11).  In addition, all the models include the AGN
  feedback based on the model introduced by \cite{Kauffmann00}
  which suppresses cooling flows in massive halos. The feedback used
  in B06 is similar to the radio mode feedback considered in
  \cite{Croton06}, but the details of implementation are different. In
  B06 AGN feedback is more effective at low-z during the
  quasi-hydrostatic phase.  G11 and DLB07 apply both the radio mode
  and quasar mode following \cite{Croton06}. In G11 the radio mode
  feedback and gas cooling are also applied on massive satellite
  galaxies.
             
  Finally, all three SAMs implement SNe feedback from massive stars,
  which can heat gas and eject it from the cold disc into the hot
  halo. G11 uses a model of the SNe feedback which depends on the
  galaxy circular velocity (V$ _{max} $) and gives stronger feedback
  at low-V$ _{max} $ and dwarf galaxies.
           
  In summary, treatments of both minor and major mergers as well as
  dry and wet mergers, cooling flows, AGN and stellar feedback can
  affect galaxy evolution and its properties (e.g., stellar mass,
  luminosity and stellar age). In this paper, we perform a statistical
  study of the redshift evolution of the magnitude gap out to z=1.1 to
  compare the galaxy properties in massive halos predicted by
  different SAMs and examine consistency with observations.

\begin{figure}[ht!]
\begin{center}  
\leavevmode
\includegraphics[width=9cm]{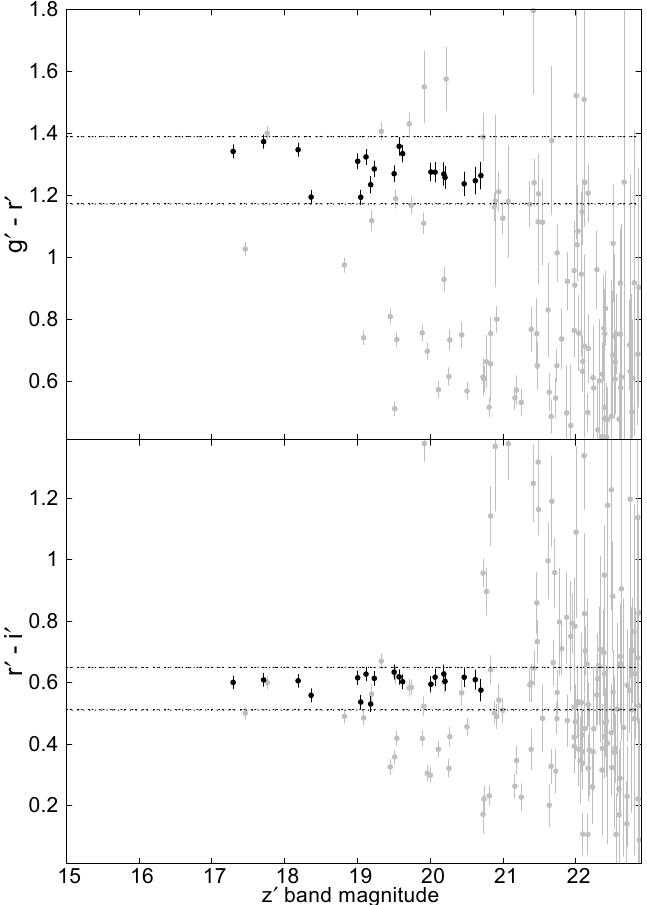}
\end{center}
\caption[flag]{Example of using two color magnitude diagrams to
  identify the red sequence of a galaxy cluster at z=0.28 (cluster
  100320). Top and bottom panels show g$^\prime$-r$^\prime$ and
  r$^\prime$-i$^\prime$ versus z$^\prime$-band magnitude,
  respectively. Gray points are all galaxies within the radius of 0.5
  Mpc from the X-ray source center. Black dots are red galaxy group
  members selected according to the method described in
    \S4.1. The upper and the lower ranges of colors are shown with
  the horizontal dashed lines.}
\label{rs}
\end{figure}
\section{Group Identification}
\subsection{Red-sequence Method}

The full details of our red sequence finder are presented in
\cite{Mirkazemi14}, but here we provide an outline of the method. We
measure an excess of red sequence galaxies over the background
within the 0.5 Mpc radius from the X-ray center. The background is
computed using 200 random areas in the CFHTLS survey and repeating the
selection of galaxies used in the red sequence. We define the red
sequence significance as a ratio of the excess number density of red
sequence galaxies to a dispersion in the number density of background
galaxies. The redshift at which the red sequence significance reaches
its maximum value is taken as the redshift of a cluster. Our red
sequence filter is defined as sliding (as a function of redshift)
upper and lower limits for galaxy colors and a selection of bright, L$\ge0.4$
L$_{\ast}$, galaxies.
\begin{figure}[ht!]
 \begin{center}  
 \leavevmode
 \includegraphics[width=9.5cm]{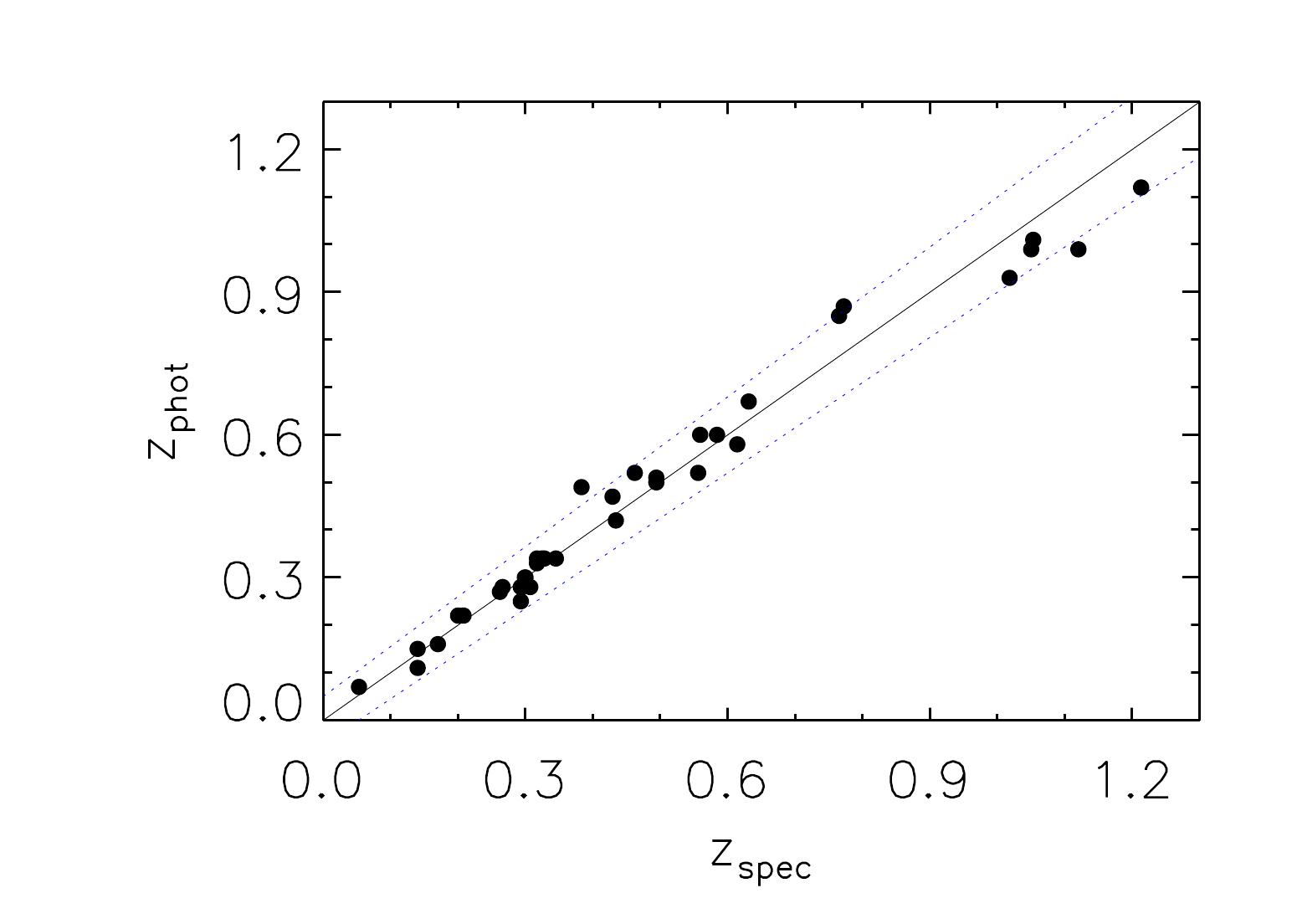}
 \end{center}
 \caption[flag]{Photometric redshift versus spectroscopic redshift for our 
 sample of galaxy groups. Black line gives the 1:1 relation and the
 dotted blue lines show the adopted $\pm$0.05(1 + z) uncertainty of
 our red sequence redshift.}
 \label{zs_zp}
 \end{figure}
Using \cite{Maraston09} stellar population models for the red galaxies
we derived the characteristic luminosity $L^{\ast}$ for the MegaCam
g$^\prime$, r$^\prime$, i$^\prime$ and z$^\prime$ filters at each
redshift. A model for the colors of red sequence galaxies is derived by
a sample of red galaxies with spectroscopic data from the SDSS III
\citep{Aihara11} and Hectospec \citep{Mirkazemi14}. This sample of red
galaxies gives a model for colors of red galaxies up to a redshift of
0.75. Beyond this redshift, \cite{Maraston09} stellar population model
was used. The upper and lower ranges on colors of red galaxies were
defined as $\pm$2$\Delta_{rs} $ around the color model at each
redshift. The dispersion $\Delta_{rs}$ has two terms: an intrinsic
color dispersion and the measurement uncertainty. The first term is
modeled by varying the metallicity in the PEGASE.2 stellar
population models \citep{Fioc99}. To compute the second term, we use
the error in measuring the magnitudes in the CFHTLS data, as detailed
in \cite{Bielby10}.

  \begin{figure*}[ht!]
  \begin{center}  
  \leavevmode
  \includegraphics[width=0.48\textwidth]{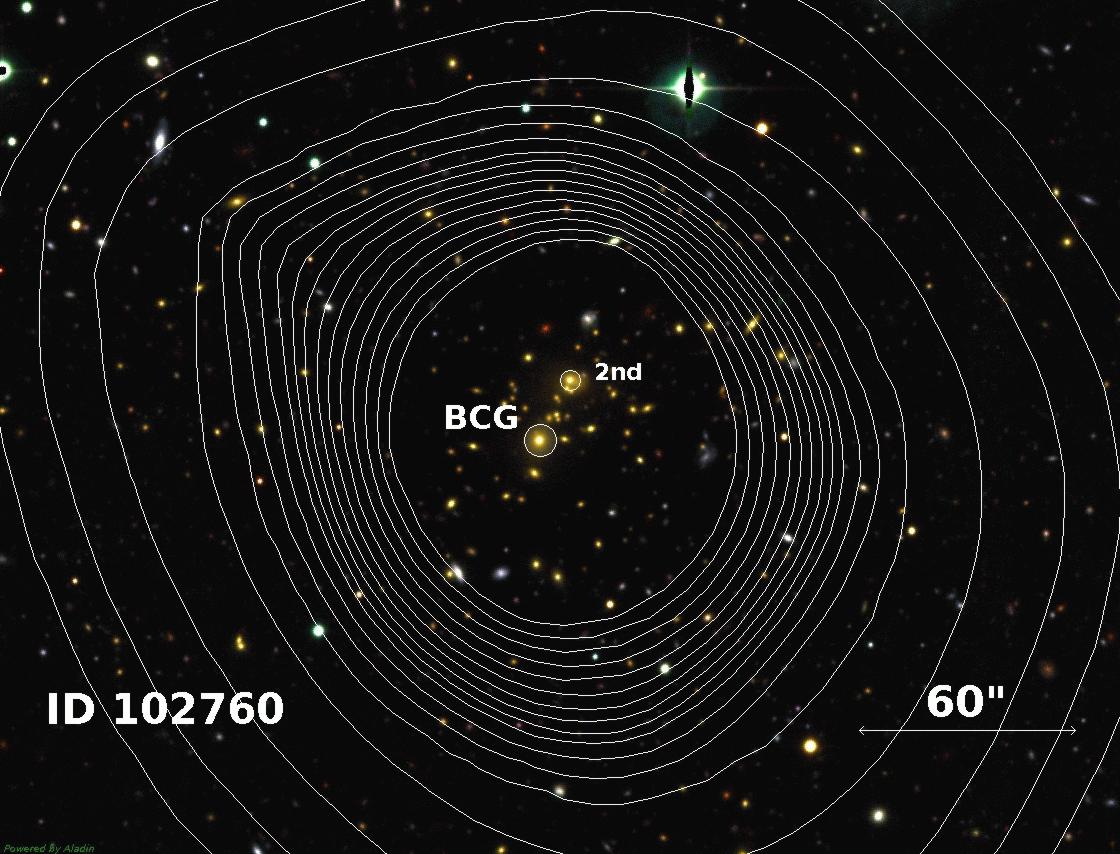}
  \includegraphics[width=0.48\textwidth]{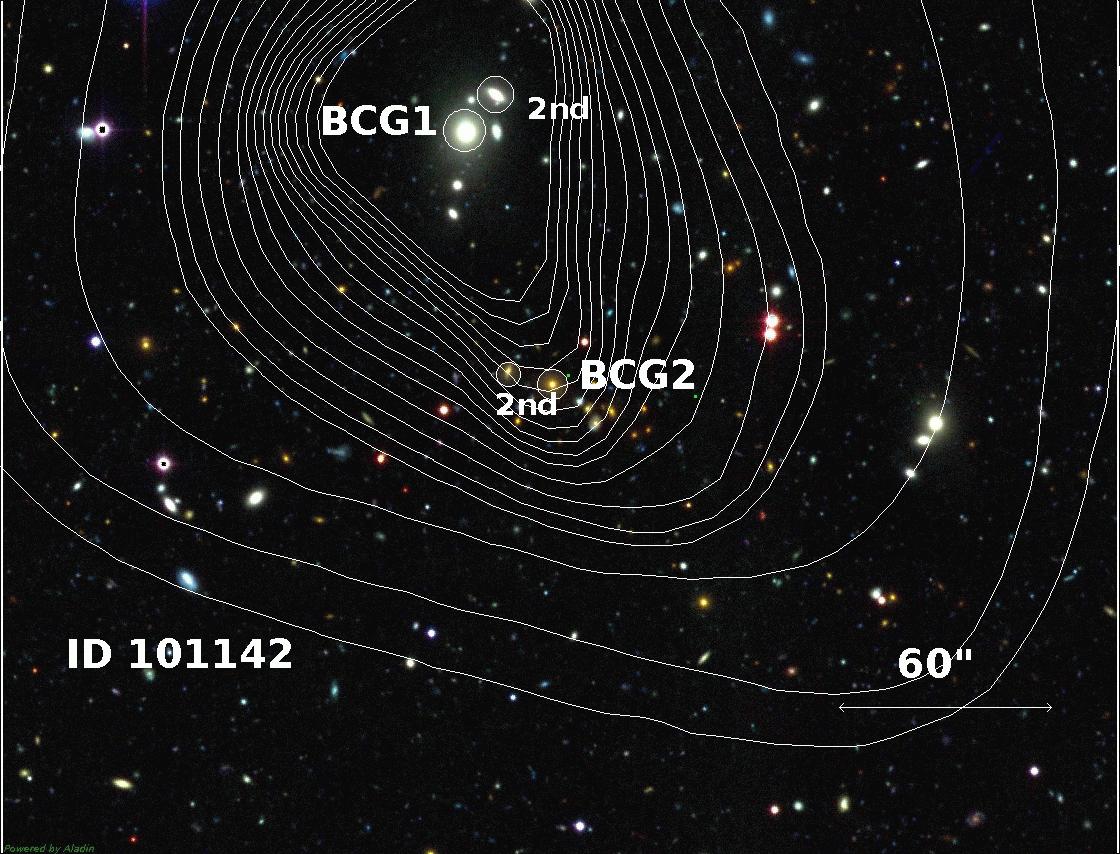}
  \includegraphics[width=0.48\textwidth]{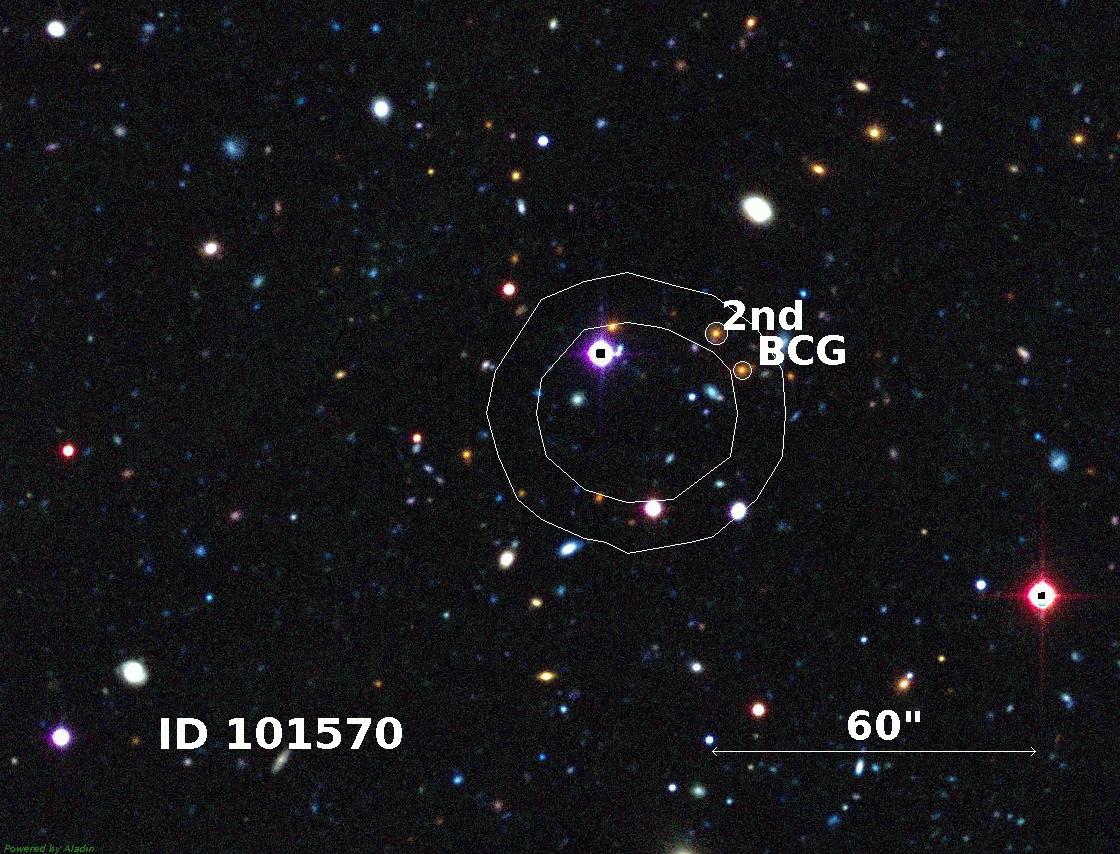}
  \includegraphics[width=0.48\textwidth]{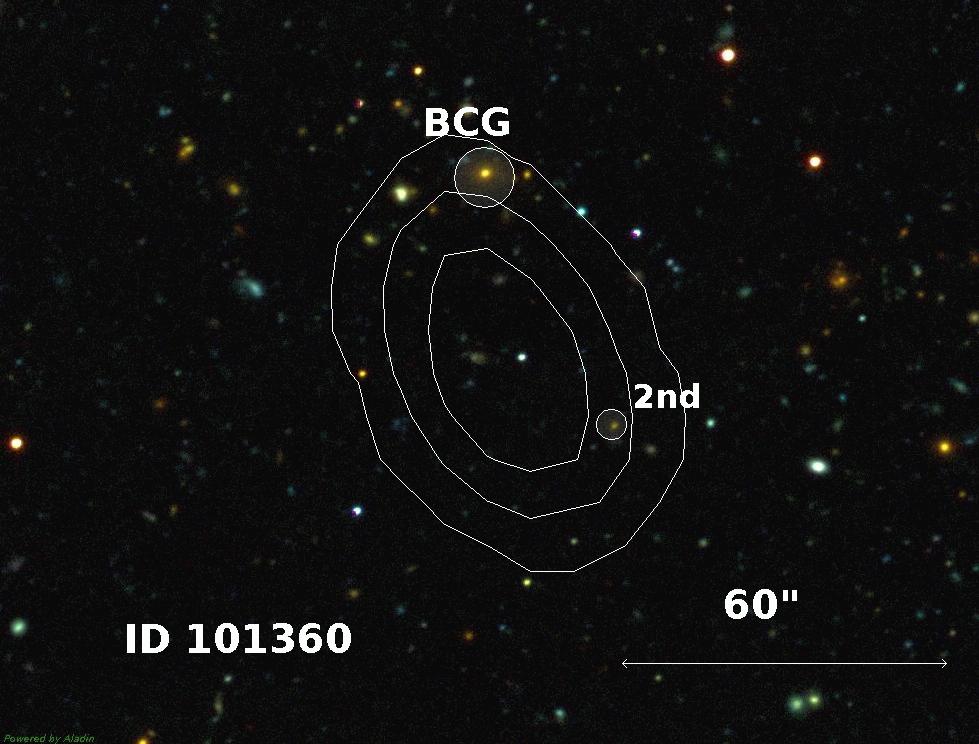}
   
  \end{center}
  \caption[flag]{Contours of the extended X-ray emission overlaid on
    the CFHTLS RGB image in i$^{\prime}$, r$^{\prime}$, g$^{\prime}$
    filters. Top left panel shows a spectroscopically confirmed
    cluster at z=0.47 within the center of X-ray emission and having a
    single optical counterpart, with a high significance of X-ray flux
    estimate (flag=1 and flag=3 if without the spectroscopic
    confirmation). Top right panel, assigned with a flag=2, indicates
    a presence of two overlapping X-ray sources at z=0.52 and z=0.15.
    The bottom left image shows the multiple optical counterparts
    within the X-ray emission with no possibility of separating out
    their contributions (flag=4). The bottom right panel shows a
    galaxy group at z=0.60 located at the edge of of X-ray emission
    (flag=5). The scale bar presented in each image.}
  \label{flag}
  
  \end{figure*}
     
  We chose the following combination of filters as a function of
  redshift for the red sequence algorithm:\\ 0.05$\le$z$\le$0.66 :
  g$^\prime$, r$^\prime$, i$^\prime$ \\ 0.66$<$z$\le$1.10 :
  r$^\prime$, i$^\prime$, z$^\prime$.\\
  Fig. \ref{rs} shows the g$^\prime$-r$^\prime$ and
  r$^\prime$-i$^\prime$ colors versus z$^\prime$ band magnitude for a
  cluster at a redshift of 0.28.

  For computing the absolute magnitude of red sequence galaxies, we
  applied a $\chi^2$ template-fitting method using the {\it Le Phare}
  package \citep{Arnouts02} and templates of \cite{Ilbert06}. We
  applied the \cite{Calzetti00} extinction law to the star-forming,
  irregular and star burst templates. Absolute magnitudes of member
  galaxies were computed assuming the redshift of their host galaxy
  group. Similar to \cite{2009ApJ...690.1236I} and
  \cite{2009ApJ...690.1250S}, we applied the AUTOADAPT mode of the
  {\it Le Phare} code on a sample of galaxies with spectroscopic
  redshifts from the VVDS survey to reduce the systematic
  offset between photometric and spectroscopic redshifts by adjusting
  photometric zero points. Further details of the procedure can be
  found in \cite{Ilbert06}.

  For galaxy groups, located within the VVDS survey, we compute the
  mean spectroscopic redshift of red sequence galaxies and identify as
  member the galaxies located within $R_{200}$ from the center of the
  group and obeying
   
 \begin{equation}
 |z_{member}-z_{group}|<3\frac{\sigma_{x}(v)}{c}
 \end{equation}
 We estimate the galaxy velocity dispersion $ \sigma_{x}(v)$ using the
 virial mass of groups (estimated from the X-ray luminosity), as
 introduced in \cite{Erfanianfar13}. In Tab.\ref{cat} we present the
 values of $R_{200}$ and $\sigma_{x}(v)$ for each galaxy group.

 We were able to assign a spectroscopic redshift to 15 galaxy groups
 using the spectroscopic data of the VVDS survey. For further 30
 groups and clusters in common with \cite{Adami11} catalog,
 spectroscopic redshifts are adopted from that study.  For the sample
 of spectroscopically identified systems, in Fig. \ref{zs_zp} we
 compare the red sequence redshift to the spectroscopic one.  The
 standard deviation between the red sequence and spectroscopic
 redshifts is 0.04. Black line gives the 1:1 relation and blue dotted
 lines correspond to the adopted uncertainty in our red sequence
 redshift determination of $\pm0.05(1 + z)$.

 \subsection{Visual inspection and identification of group candidates}

 For identification of galaxy groups, assignment of a photometric
 redshift and determination of the contamination from several optical
 groups, we used the red-sequence technique as described in \S 4.1.
  
 We visually inspected optical counterparts of each X-ray source using
 the photometric redshift catalog of the CFHTLS
 \citep{Brimioulle08,Brimioulle13} and the CFHTLS RGB image in
 i$^{\prime}$, r$^{\prime}$ and g$^{\prime}$ filters. The visual
 inspection helps us to assign a redshift to each X-ray source, and to
 evaluate the correspondence between the galaxy distribution and the
 X-ray emission.  For some sources where we found more than one
 counterpart and where the shape of X-ray emission allows us to
 separate the contribution from several counterparts, we specify a new
 ID in our X-ray source catalog.
  
 We define a flag for each extended X-ray source based on the visual
 inspection and the position of the over-density of galaxies inside
 each X-ray source. We use flag values 1 and 3 to mark the best
 identifications. This requires having both a unique X-ray source with
 a well-defined center, and a unique optical counterpart with the
 spectroscopic confirmation for flag=1 or without one -- for flag=3.  A
 flag=2 is assigned when a single X-ray source has been split into
 several sources (e.g., top right panel in Fig.\ref{flag}). A flag=4
 indicates a presence of multiple optical counterparts, whose
 contribution to the observed X-ray emission is not possible to
 separate or rule out (e.g., bottom left panel in Fig.\ref{flag}). 
 
 In cases where the X-ray emission covers a part of the group area and a
  concentration of galaxies is located at the edge of X-ray emission,
  we assign a flag=5 (e.g., bottom right panel in Fig
  \ref{flag}). Finally, systems with a potentially wrong assignment of
  optical counterpart are also flagged as 5.
  
\subsection{Catalog of identified groups}
We present a catalog of 129 X-ray selected galaxy groups in
Tab. \ref{cat}. Columns 1 and 2 indicate the internal X-ray source ID
and our defined group ID. Columns 3 and 4 are the RA (J2000) and Dec
(J2000) of the X-ray center. Column 5 shows the group photometric
redshift estimated according to the method outlined in \S 4.1.  

      \begin{figure}[ht!]
      \begin{center}  
      \leavevmode
      \includegraphics[width=9.5cm]{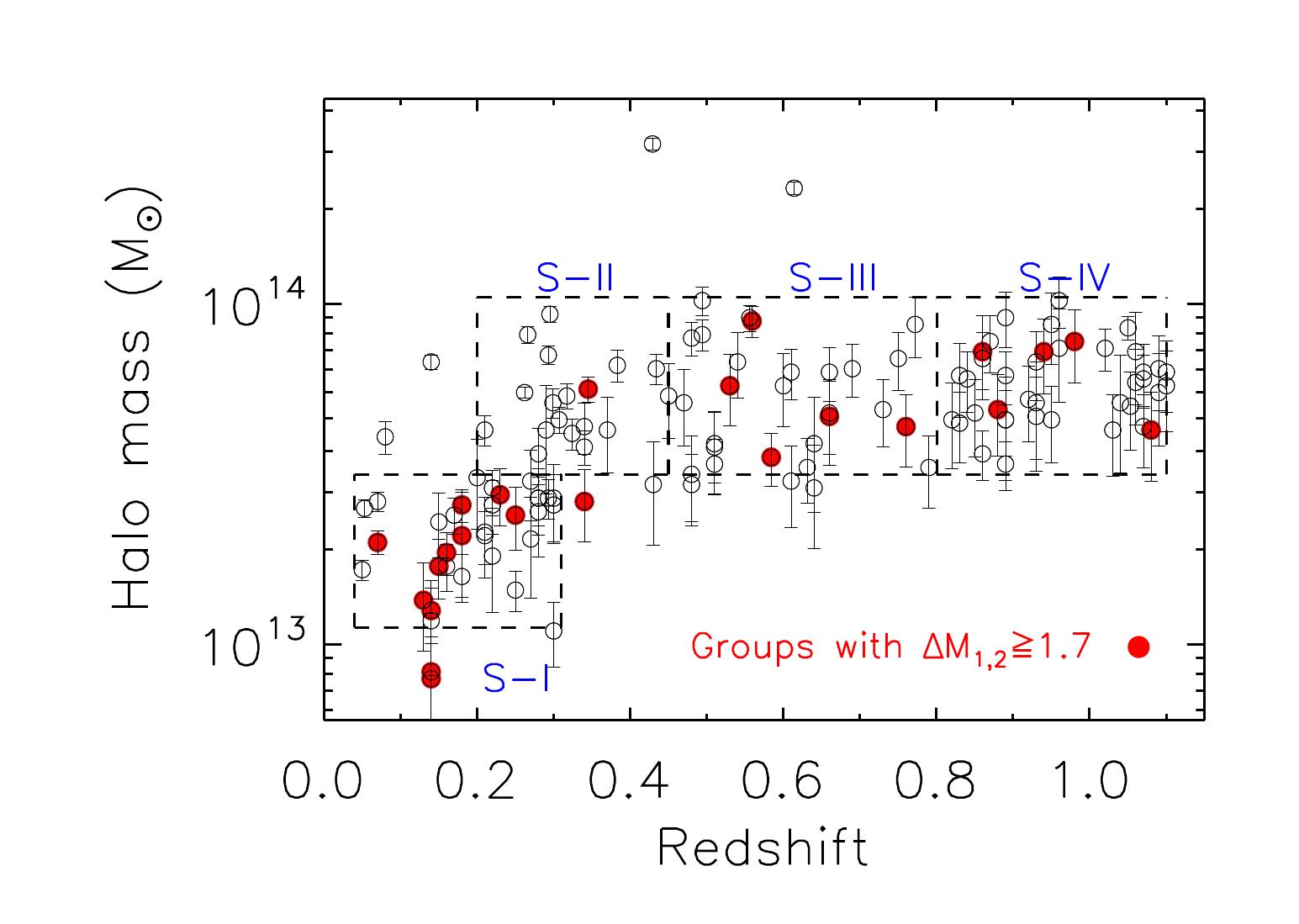}
      \end{center}
      \caption[flag]{Halo mass versus redshift for the detected X-ray
        groups.  Groups with $\Delta M_{1,2}$$\geq$1.7 have been
        marked with red filled circles. The dashed boxes show the four
        subsamples, defined in \S 4.3.}
      \label{mass_z}
      \end{figure}
      
      \begin{figure}[]
      \begin{center}  
      \leavevmode
      \includegraphics[width=9.5cm]{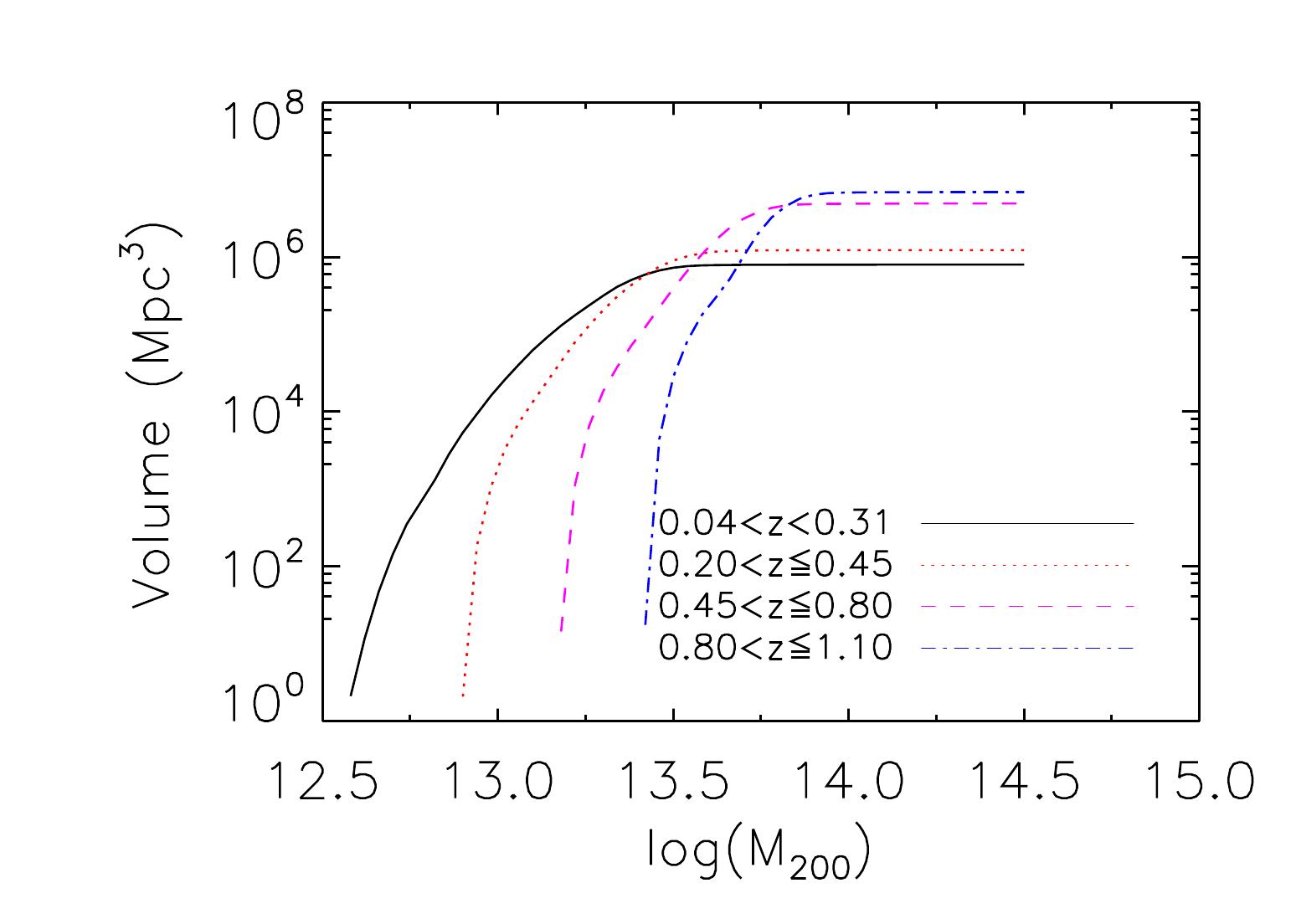}
      \end{center}
      \caption[flag]{Survey volume as a function of halo mass.  Solid
        black,dotted red, dashed magenta and dashed-dotted blue curves
        present trends for the four redshift intervals, S-I to S-IV,
        which are described in \S 4.3.  }
      \label{volume}
      \end{figure}

      Column 6 is the total group mass, $M_{200c}$, in $M_{\odot}$ and
      a corresponding statistical error, obtained using the $ L_{x} -
      M$ relation of \cite{Leauthaud10}. The systematic uncertainty on
      inferring $M_{200c}$ using $L_{x}$ is 20\% \citep{Allevato12}.
      Column 7 lists L$_{x,0.1-2.4}$ in the rest-frame 0.1--2.4 keV
      energy range. Column 8 presents the galaxy velocity dispersion
      in km s$ ^{-1} $, estimated from $L_X$ using the scaling
      relations \citep{Erfanianfar13}. Column 9 shows group's
      $R_{200c}$ in degrees. Column 10 lists the mean group
      temperature in keV estimated using L--T relation as in
      \cite{Finoguenov07}.
    
      The group X-ray flux in the observed 0.5-2 keV band (in units of
      $10^{-14}$ ergs cm$^{-2}$ s$^{-1})$ and a corresponding
      $1\sigma$ error is listed in column 11. Column 12 provides the
      significance of the flux estimate. Column 13 reports the visual
      flag as defined in \S 3.1. Column 14 reports the spectroscopic
      redshift, available for 45 groups. We mark the spectroscopic
      redshift of groups adopted from with \cite{Adami11} with $ '*' $.

To test the evolution of properties of galaxy groups
in observations and SAMs, we define four subsamples using the following
redshift and halo mass ranges:

(S--I)   0.0$4<$z$<$0.31  \& 13.00$<log(\frac{M_{200}}{M_{\odot}})\le$13.45 
 
(S--II) 0.20$<$z$\leq$0.45 \& 13.45$<log(\frac{M_{200}}{M_{\odot}})\le$14.02
 
(S--III) 0.45$<$z$\leq$0.80 \& 13.45$<log(\frac{M_{200}}{M_{\odot}})\le$14.02
  
(S--IV) 0.80$<$z$\leq$1.10 \& 13.45$<log(\frac{M_{200}}{M_{\odot}})\le$14.02
             
S--I includes 30 galaxy groups, and other three subsamples, S--II,
S--III and S--IV, contain 23, 29 and 41 massive groups, respectively.
In Fig. \ref{mass_z}, we illustrate subsamples with dashed boxes.

  \subsection{Abundance of galaxy groups }
     
  In order to study the abundance of galaxy groups in our survey, we
  compute the survey volume using the sensitivity toward the group
  detection as a function of coordinate in four selected redshift
  intervals. The components of the calculation consist of the
  countrate limits as a function of coordinate and a predicted
  countrate from a galaxy group given its redshift and mass in the
  $16^{\prime\prime}-128^{\prime\prime}$ aperture defined by the
  detection scales. The X-ray luminosity of the group is computed for
  each mass and redshift using the \cite{Leauthaud10} scaling
  relation, and the surface brightness profile of the group is
  computed using the parametrization of \cite{Finoguenov07}. The
  countrate limits correspond to a four sigma deviation in the
  background RMS predicted on the scale of the detection, which
  include the local residual background estimate by the wavelets and
  the systematic errors on AGN removal.  The analysis of the noise of
  the image on the detection scales has been presented in
  \cite{Connelly12} and its negative side is well described by the
  Gaussian. Fig.\ref{volume} shows the computed survey volume of the
  X-ray galaxy groups as a function of the halo mass, M$_{200}$, for
  different subsamples.
  
\begin{figure*}[t]
                    \begin{center}  
                    \leavevmode
                    \includegraphics[width=0.48\textwidth]{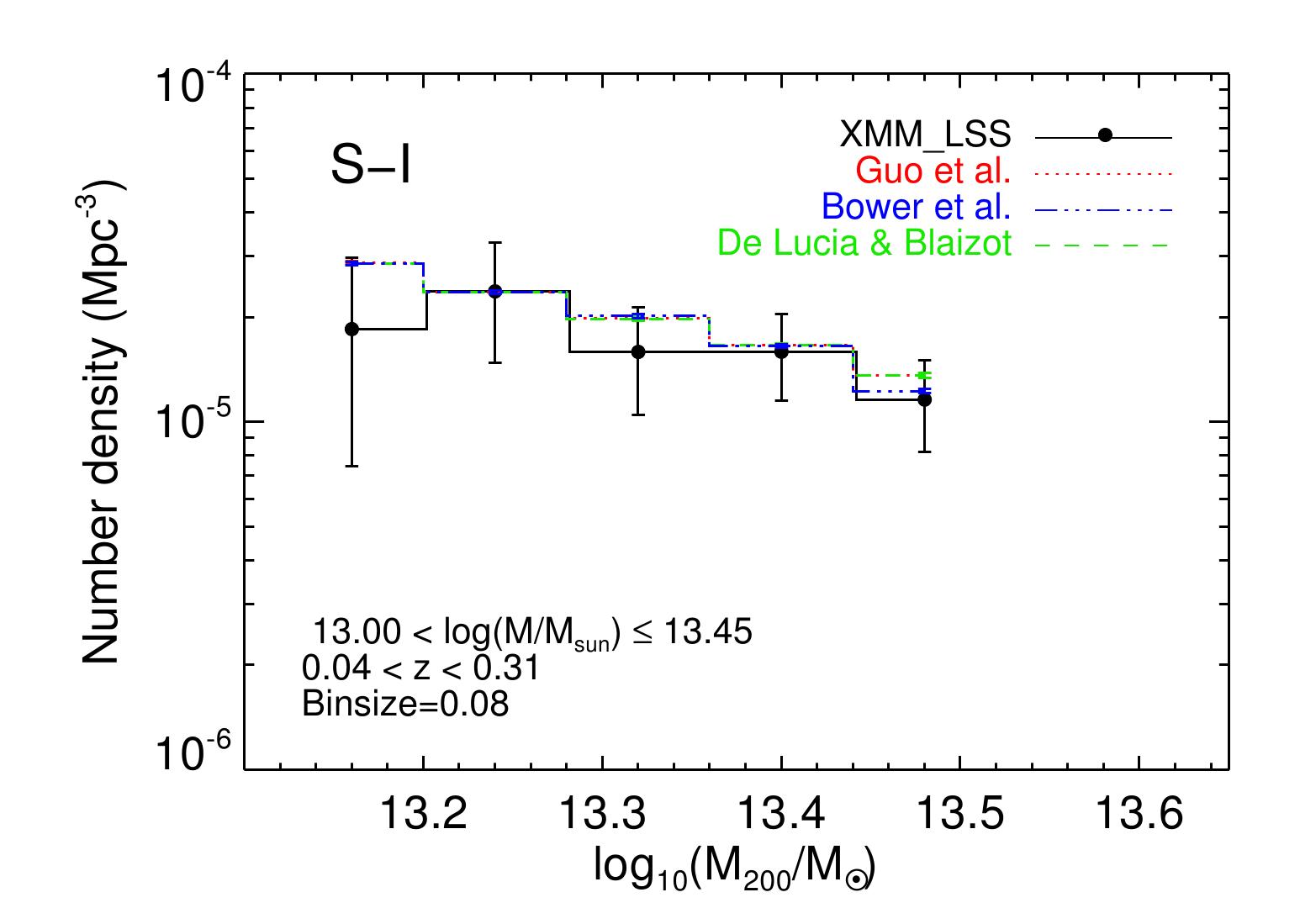}                    \includegraphics[width=0.48\textwidth]{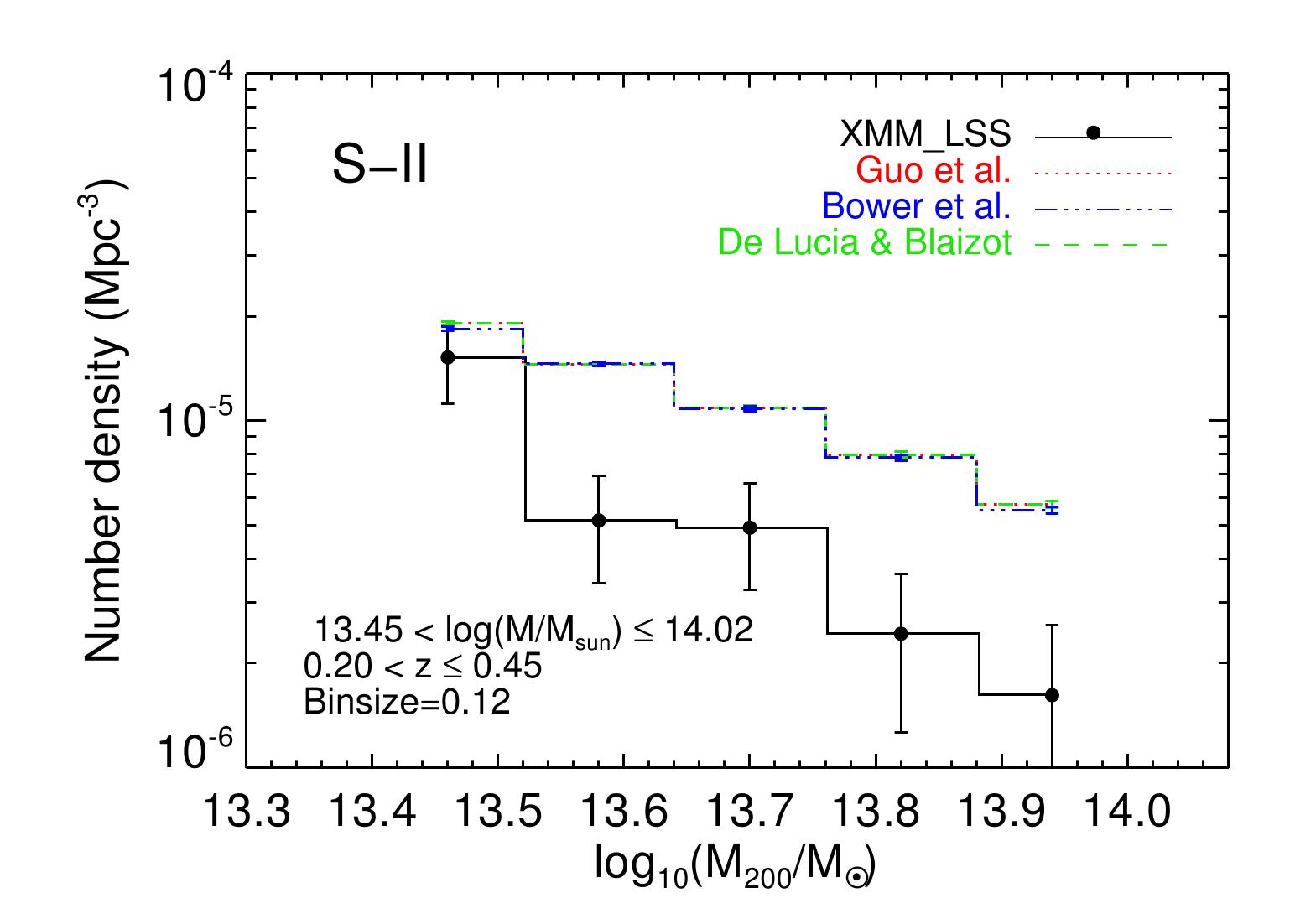}                    \includegraphics[width=0.48\textwidth]{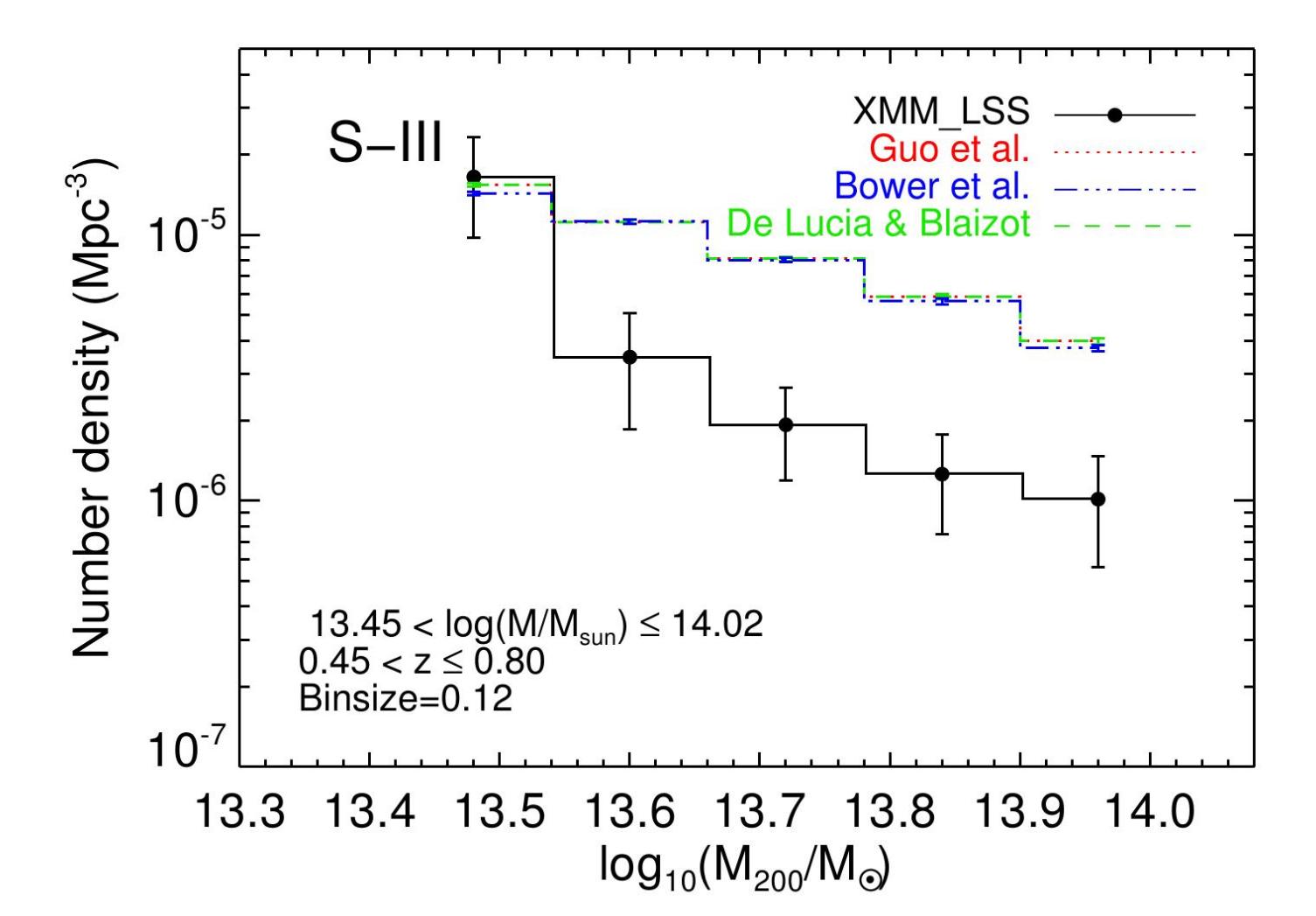}
                    \includegraphics[width=0.48\textwidth]{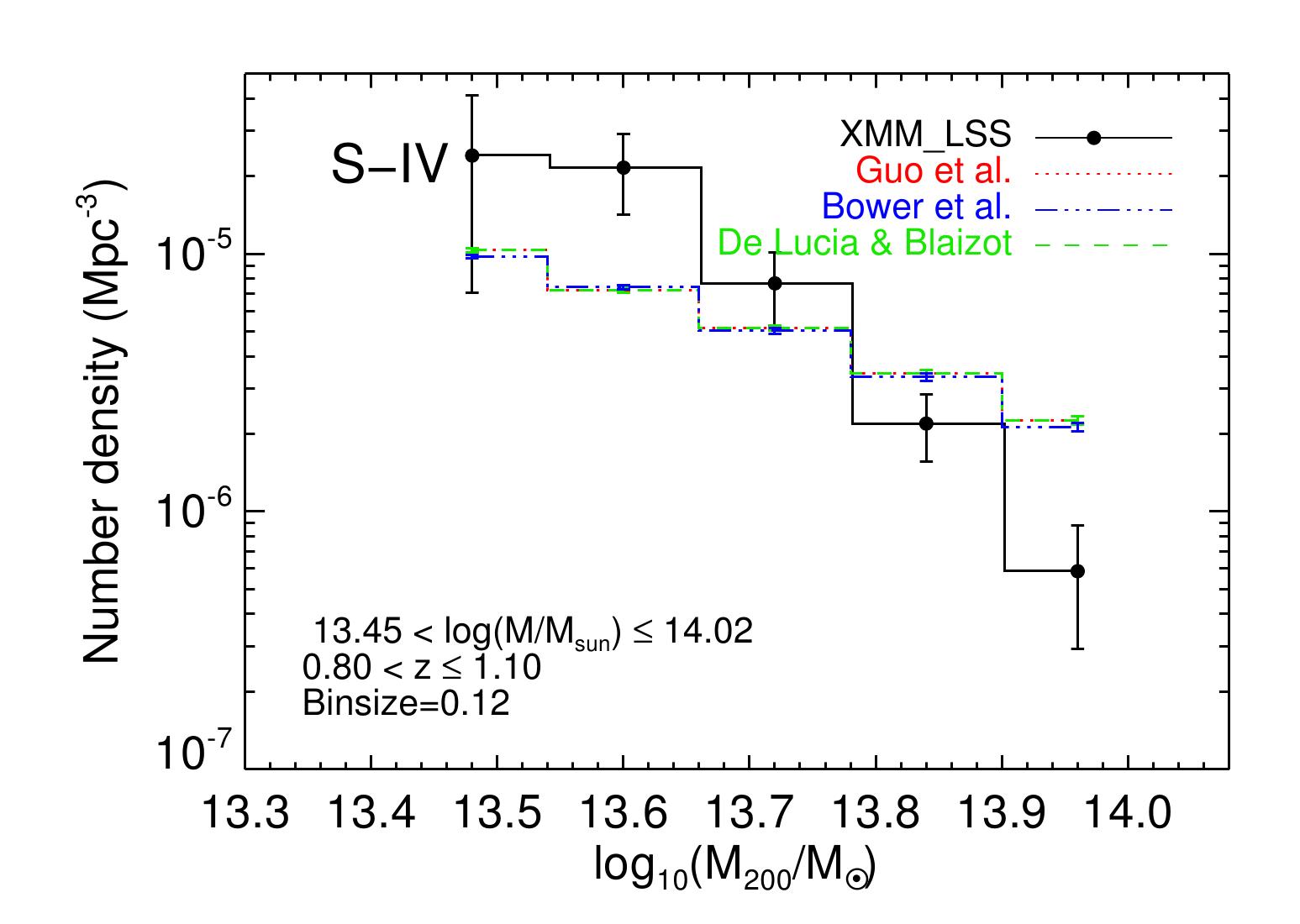}
                    \end{center}
                    \caption[flag]{The observed number density of
                      galaxy groups as a function of halo mass (black
                      histogram with error bars) compared to the halo
                      catalogs of B06, DLB07 and G11 for S--I (top
                      left panel), S--II (top right panel), S--III
                      (bottom left panel) and S--IV (bottom right
                      panel). Number density of halos in B06, G11 and
                      DLB07 have been plotted with blue dashed-dotted,
                      red dotted and green dashed histograms and are
                      identical. Halo abundance in simulations is
                      consistent with the observed number density of
                      galaxy groups in S--I, but this consistency is
                      reduced at high masses for S--II and
                      S--III. Subsamples are defined in the legend of
                      each plot.}
                    \label{hab}

\end{figure*}

  In Fig.\ref{hab} we compare the number density of galaxy groups as a
  function of halo mass (solid black histogram with error bar) with
  that from the SAMs of G11 (red dotted histogram), B06 (dash-dotted
  blue histogram) and DLB07 (dashed green histogram). Number of
  observed galaxy groups in each halo mass bin is normalized to the
  corresponding volume as shown in Fig.\ref{volume}. For the
  normalization of the number of galaxy groups in the models we use
  the constant volume of the Millennium simulation, a box of size 500
  h$^{-1}$Mpc on a side. Halo abundance in models is consistent with
  the observed trend for S--I (left top panel). For S--II (right top
  panel ) and S--III (left bottom panel), this consistency becomes
  worse at high masses, but the agreement reappears at higher redshift
  traced by the subsample S--IV (right bottom panel).

A detailed assessment of the cosmological importance of the observed
halo abundances is outside the scopes of this paper and would need to
include the effect of the sample variance at high halo masses. The
main conclusion we draw is that for further comparison we should
rather use fractions determined separately for the group catalogs in
observations and simulations.   

\section{Membership contamination and  magnitude gap calculation }
         
\subsection{Contamination of group membership}
         
The purity of the galaxy selection using the red-sequence method is
important for calculating the magnitude gap of galaxy groups.  In the
absence of spectroscopic group membership, the calculation using red
sequence galaxies may be contaminated by dusty star forming galaxies
(DSFGs).
  
We examine the impact of contamination of group members by DSFGs using
the X-ray galaxy group catalog in the redshift range 0.066$ < $z$ <
$1.544 presented in \cite{Erfanianfar13} and the data of galaxies in
the AEGIS Herschel survey located within a CFHTLS field covering the
redshift range 0.05$ < $z$ < $1.10. We define as DSFGs those galaxies
which have star formation rate above $\approx$100 M $ _{\odot} $ yr$
^{-1} $ as inferred from FIR observations by Herschel space telescope
\citep{Erfanianfar14}.
             
\begin{figure}[ht!]
    \begin{center}  
   \leavevmode
    \includegraphics[width=9.5cm]{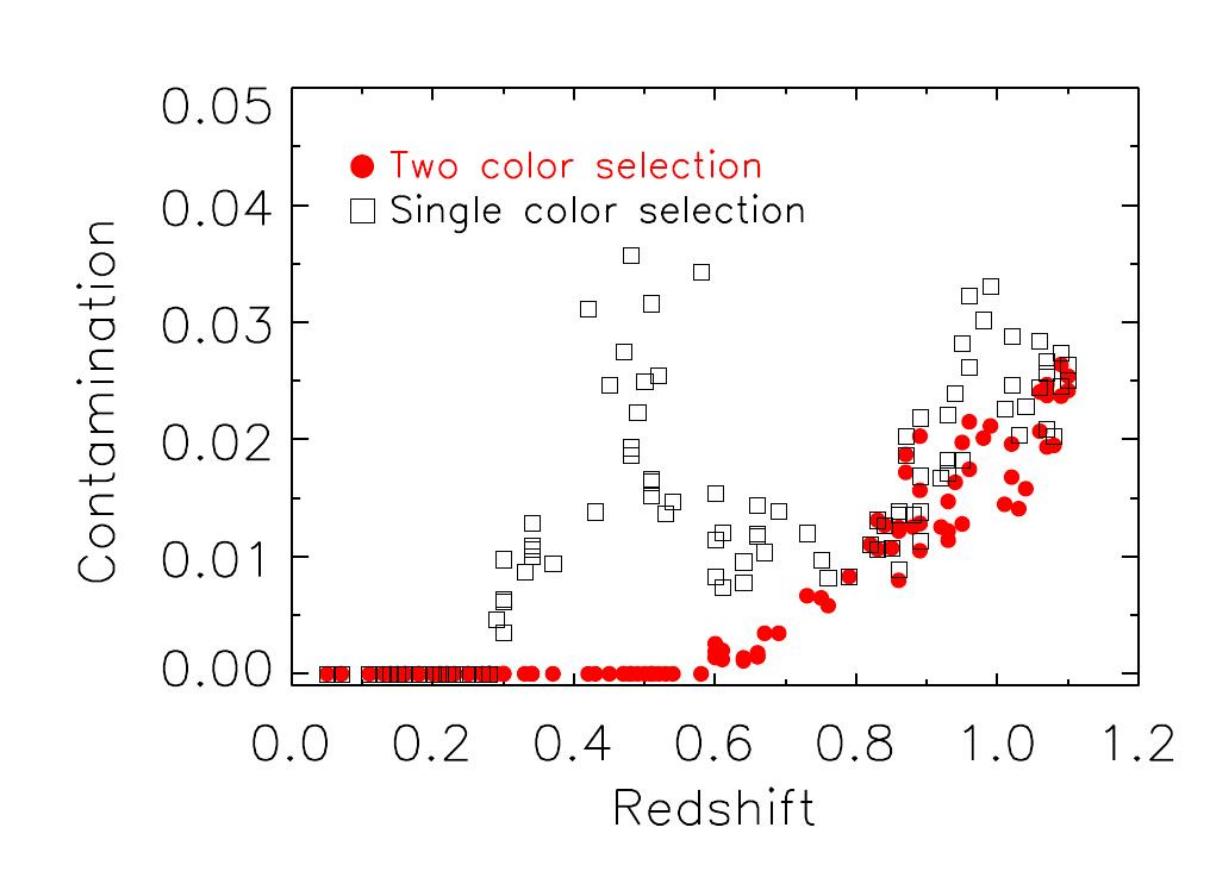}
    \end{center}
    \caption[flag]{ Model prediction for red sequence contamination by
      dusty star-burst galaxies versus redshift of group. The filled
      red circles and open black open squares show the contamination
      for single color and two color selection of group
      members.}  \label{sfr}
                                   
 \end{figure}

For each group in our CFHTLS catalog we test whether any of DSFGs
would be selected as a member. We apply a luminosity cut of
L$>$0.4L$^{*} $, and use two red-sequence finders: a single
r$^{\prime}$-i$^{\prime}$ color, and our two color selection (\S
4.1). The contamination for each group is estimated by multiplying the
number of DSFGs selected by the red sequence method to the ratio of
the area of group which was used to define the magnitude gap (0.5R$
_{200}$) and the area of the Herschel data coverage of the
AEGIS field, $ \sim$0.35 degrees$^{2}$. Fig. \ref{sfr} shows the
expected contamination for each group versus its redshift for the single
color and the two color selection of the group members. Two color
selection completely removes the contamination at z$<$0.6. The
expected contamination by DSFGs at 0.6$<$z$<$1.1 is low,
$\lesssim$3\%, to significantly affect the magnitude gap measurement.             
         
To examine the effect of the contamination by DSFGs on the magnitude
gap calculation, we use data of nine galaxy groups with spectroscopic
membership from the AEGIS X-ray groups catalog within the redshift
range of 0.65$<$z$<$1.1 \citep{Erfanianfar13,Erfanianfar14}. The
redshift range of these groups of galaxies covers the range where the
contamination by DSFGs is the highest.
       
Although, the redshifts of these groups do not span the full redshift
range of our catalog, they can still give us a good estimate of
contamination of the star forming foreground or background galaxies to
the red-sequence. For these nine groups, we only find two galaxies as
DSFGs which are group members, but they are not selected as a brightest cluster galaxy (BCG) or
the second brightest satellite.
\newpage
       \setcounter{table}{0} 
        
         \begin{landscape} 
         \begin{table}[ht!]
         \caption{\footnotesize
         Catalog of XMM X-ray galaxy groups (for more detail see text in \S 3.3).
         \label{cat}}
         \tiny
         \centering
         \renewcommand{\arraystretch}{0.9}\renewcommand{\tabcolsep}{0.12cm}
          
         \begin{tabular}[tc]{ccccc ccccc cccc cc}
          \hline
          \hline\\
          XMM ID & Group ID & RA (J2000) & Dec (J2000) & z$ _{phot} $& $M_{200c}$ & $L_{x,(0.1-2.4 keV)}$& $ \sigma_{v} $&$R_{200}$ & kT& Flux & Sig.& Flag & z$ _{spec} $   \\
             &  & degrees &degrees& & $ 10^{13}M_{\odot} $ & $10^{42}erg s^{-1}$&kms$ ^{-1} $ &degree & KeV& $10^{-14}erg s^{-1}cm^{-2}$\\
        
                        (1) & (2)  & (3) & (4) & (5) & (6) & (7) & (8) & (9) & (10) & (11) & (12) & (13)&(14)  \\
           \hline \\
         XMMU J022538.8-050133 &100010  &36.4119 &-5.0261 &0.18  &3.133  $\pm $0.284 &2.41    $\pm $0.35 &272.3  &0.056& 0.8$  \pm $0.04  &1.757  $\pm $0.255 &6.89  &3 & 0.192 \\
          XMMU J022330.4-050415 &100020 &35.877  &-5.0708 &0.48  &3.811  $\pm $0.658 &4.875   $\pm $1.38 &306.2  &0.027& 0.98$ \pm $0.1   &0.334  $\pm $0.094 &3.54  &5& --  \\
          XMMU J022617.1-050447 &100030 &36.5712 &-5.0797 &0.07  &3.162  $\pm $0.171 &2.148   $\pm $0.185&268.6  &0.134& 0.78$ \pm $0.02  &12.204 $\pm $1.051 &11.61 &3& --   \\
          XMMU J022402.2-050528 &100040 &36.0092 &-5.0913 &0.34  &5.248  $\pm $0.439 &6.653   $\pm $0.887 &332.4 &0.04&  1.13$ \pm $0.06  &1.084  $\pm $0.145 &7.49  &1& 0.324*  \\
          XMMU J022448.4-050510 &100050 &36.2019 &-5.0863 &0.64  &3.75   $\pm $0.966 &5.998   $\pm $2.588 &313.8  & 0.022& 1.02$ \pm $0.15  &0.191  $\pm $0.082 &2.32  &5& --  \\
          XMMU J022054.0-050610 &100060 &35.225  &-5.1028 &0.66  &6.237  $\pm $1.161 &13.677  $\pm $4.178 & 373.3  & 0.026& 1.41$ \pm $0.16  &0.427  $\pm $0.131 &3.27  &3& --  \\
          XMMU J022416.3-050321 &100090 &36.0682 &-5.0559 &0.16  &2.061  $\pm $0.267 &1.222   $\pm $0.256&236.1  &0.054& 0.64$ \pm $0.04  &1.158  $\pm $0.242 &4.78  &3& -- \\
          XMMU J022428.9-050545 &100100 &36.1204 &-5.0959 &0.51  &4.385  $\pm $0.619 &6.368   $\pm $1.455&322.8  &0.027& 1.07$ \pm $0.09  &0.375  $\pm $0.086 &4.37  &5& --  \\
          XMMU J022846.9-050310 &100110 &37.1956 &-5.0528 &0.29  &5.37   $\pm $1.503 &6.442   $\pm $3.027&331.9  &0.045& 1.13$ \pm $0.19  &1.537  $\pm $0.723 &2.13  &3 & -- \\
          XMMU J022127.9-050340 &100120 &35.3663 &-5.0613 &0.88  &6.457  $\pm $1.315 &19.907  $\pm $6.683& 393.7 &0.022& 1.56$ \pm $0.2   &0.294  $\pm $0.099 &2.98  &5 & -- \\
          XMMU J022728.1-050319 &100130 &36.8673 &-5.0554 & 0.13  &1.629  $\pm $0.392 &0.817   $\pm $0.327&217.2  &0.061& 0.56$ \pm $0.06  &1.208  $\pm $0.484 &2.49  &4 & 0.143\\
          XMMU J022632.6-050019 &100140 &36.6359 &-5.0055 &0.5   &10.399 $\pm $0.863 &24.155  $\pm $3.206&429.7  &0.037& 1.86$ \pm $0.1   &1.59   $\pm $0.211 &7.54  &1& 0.494*  \\
          XMMU J022105.9-050012 &100150 &35.2746 &-5.0035 &0.16  &2.239  $\pm $0.268 &1.39    $\pm $0.269&242.6  &0.056& 0.66$ \pm $0.04  &1.318  $\pm $0.255 &5.17  &3 & -- \\
          XMMU J022628.0-045959 &100180 &36.6169 &-4.9999 &0.38  &8.492  $\pm $0.718 &17.338  $\pm $2.344&400.7  &0.035& 1.62$ \pm $0.09  &1.175  $\pm $0.159 &7.4   &1 & 0.383*  \\
          XMMU J022307.7-045950 &100190 &35.7821 &-4.9974 &0.34  &4.808  $\pm $0.451 &5.794   $\pm $0.871&322.8  &0.039& 1.07$ \pm $0.06  &0.945  $\pm $0.142 &6.65  &3 & -- \\
          XMMU J022342.5-050203 &100200 &35.9273 &-5.0343 &0.89  &6.039  $\pm $1.002 &18.197  $\pm $4.932&385.8  &0.021& 1.5$  \pm $0.16  &0.257  $\pm $0.07  &3.69  &3 & -- \\
          XMMU J022609.8-045812 &100210 &36.5409 &-4.97   &0.07  &3.013  $\pm $0.152 &1.991   $\pm $0.159& 264.3 &0.132& 0.76$ \pm $0.02  &11.337 $\pm $0.904 &12.54 &1 & 0.053*   \\
          XMMU J022153.1-050132 &100220 &35.4715 &-5.0258 &0.27  &3.784  $\pm $0.71  &3.622   $\pm $1.117&294.3  &0.043& 0.91$ \pm $0.1   &1.031  $\pm $0.318 &3.24  &3 & -- \\
          XMMU J022740.4-045738 &100240 &36.9186 &-4.9606 &0.93  &9.057  $\pm $1.556 &36.308  $\pm $10.186&444.8 &0.023& 1.99$ \pm $0.22  &0.503  $\pm $0.141 &3.56  &4 & -- \\
          XMMU J022740.3-045132 &100270 &36.9181 &-4.8591 &0.25  &8.79   $\pm $0.434 &13.183  $\pm $1.03&388.5   &0.06&  1.52$ \pm $0.05  &4.444  $\pm $0.347 &12.8  &1 & 0.293*  \\
          XMMU J022427.8-045014 &100300 &36.1161 &-4.8375 &0.48  &10.186 $\pm $0.904 &22.699  $\pm $3.228& 424.9 &0.038& 1.82$ \pm $0.11  &1.644  $\pm $0.234 &7.03  &3 & -- \\
          XMMU J022803.7-045102 &100320 &37.0154 &-4.8506 &0.28  &11.535 $\pm $0.514 &20.989  $\pm $1.479& 427.6 &0.06&  1.84$ \pm $0.05  &5.492  $\pm $0.386 &14.21 &1& 0.295*  \\
          XMMU J022805.3-045355 &100340 &37.0222 &-4.8988 &0.87  &10.186 $\pm $1.472 &39.995  $\pm $9.397& 457.5 &0.025& 2.1$  \pm $0.2   &0.675  $\pm $0.159 &4.26  &3 & -- \\
          XMMU J022222.7-045314 &100370 &35.5946 &-4.8874 &0.37  &5.346  $\pm $1.072 &7.129   $\pm $2.355& 336.1 &0.037& 1.16$ \pm $0.14  &0.946  $\pm $0.313 &3.03  &3& --  \\
          XMMU J022500.4-044838 &100410 &36.2519 &-4.8106 &0.22  &2.259  $\pm $0.575 &1.524   $\pm $0.647&245.9  &0.043& 0.68$ \pm $0.09  &0.699  $\pm $0.297 &2.35  &3 & -- \\
          XMMU J022648.4-044940 &100420 &36.7019 &-4.8279 &1.07  &5.902  $\pm $1.052 &22.699  $\pm $6.637& 395.8 &0.019& 1.58$ \pm $0.18  &0.19   $\pm $0.056 &3.42  &5 & -- \\
          XMMU J022431.6-044815 &100430 &36.1319 &-4.8042 &0.98  &10.399 $\pm $1.879 &48.417  $\pm $14.355& 470.2&0.024& 2.22$ \pm $0.27  &0.608  $\pm $0.18  &3.37  &2 & -- \\
          XMMU J022236.8-044857 &100450 &35.6537 &-4.816  &0.84  &7.889  $\pm $1.216 &25.704  $\pm $6.457&417.8  &0.024& 1.76$ \pm $0.18  &0.449  $\pm $0.113 &3.98  &5 & -- \\
          XMMU J022110.4-044206 &100480 &35.2936 &-4.7019 &0.75  &9.162  $\pm $1.334 &28.445  $\pm $6.73& 431.7  &0.027& 1.87$ \pm $0.18  &0.683  $\pm $0.162 &4.23  &5 & -- \\
          XMMU J022525.1-044044 &100510 &36.3547 &-4.679  &0.28  &10.0   $\pm $0.471 &16.788  $\pm $1.25 &407.6  &0.058& 1.67$ \pm $0.05  &4.364  $\pm $0.325 &13.42 &1 & 0.261*  \\
          XMMU J022746.4-044523 &100520 &36.9435 &-4.7565 &1.04  &6.792  $\pm $1.928 &27.164  $\pm $12.972&412.7 &0.02&  1.71$ \pm $0.31  &0.259  $\pm $0.124 &2.09  &5 & -- \\
          XMMU J022422.6-044311 &100540 &36.0944 &-4.7199 &0.47  &6.561  $\pm $1.303 &11.272  $\pm $3.681&366.4  &0.033& 1.36$ \pm $0.17  &0.832  $\pm $0.272 &3.06  &3 & -- \\
          XMMU J022643.3-044142 &100550 &36.6805 &-4.6952 &0.92  &7.047  $\pm $1.318 &24.21   $\pm $7.43 & 408.5 &0.022& 1.68$ \pm $0.2   &0.325  $\pm $0.1   &3.26  &1 & 0.958\\
          XMMU J022222.3-044039 &100560 &35.593  &-4.6777 &0.83  &8.091  $\pm $1.517 &26.363  $\pm $8.110 &420.6  &0.024& 1.78$ \pm $0.22  &0.479  $\pm $0.148 &3.25  &4& --  \\
          XMMU J022357.3-043526 &100570 &35.9891 &-4.5906 &0.51  &13.092 $\pm $0.989 &35.075  $\pm $4.227&464.6  &0.04&  2.17$ \pm $0.11  &2.255  $\pm $0.272 &8.3   &1& 0.494*   \\
          XMMU J022438.4-043759 &100580 &36.1602 &-4.6333 &0.61  &8.241  $\pm $1.072 &19.634  $\pm $4.140& 405.7  &0.03&  1.66$ \pm $0.14  &0.771  $\pm $0.162 &4.75  &3 & -- \\
          XMMU J022135.3-044004 &100590 &35.3973 &-4.6679 &0.28  &3.999  $\pm $0.714 &4.009   $\pm $1.175&300.4  &0.042& 0.94$ \pm $0.1   &1.044  $\pm $0.306 &3.42  &5  & --\\
          XMMU J022120.6-043729 &100600 &35.3361 &-4.6248 &0.14  &1.422  $\pm $0.364 &0.668   $\pm $0.286&208.0  &0.054& 0.53$ \pm $0.06  &0.831  $\pm $0.356 &2.34  &5 & -- \\
          XMMU J022830.7-043942 &100620 &37.1283 &-4.6618 &0.66  &6.039  $\pm $1.312 &13.002  $\pm $4.677& 369.2 &0.026& 1.38$ \pm $0.19  &0.405  $\pm $0.146 &2.78  &3 & -- \\
          XMMU J022200.2-043808& 100640& 35.50081 &-4.63559 & 0.34& 3.336 $ \pm $ 0.637& 3.271$ \pm $ 0.102& 285.7 & 0.0341&  0.87$ \pm $  0.09& 0.534 $ \pm $ 0.167&   3.18& 1 & 0.317 \\          
          XMMU J022204.3-043243 &100660 &35.5179 &-4.5453 &0.33  &6.653  $\pm $0.454 &9.484   $\pm $1.033& 359.0 &0.044& 1.31$ \pm $0.06  &1.665  $\pm $0.181 &9.2   &1 & 0.317*   \\
          XMMU J022225.5-043741 &100690 &35.6066 &-4.6281 &1.1   &7.78   $\pm $1.545 &36.475  $\pm $11.94 &436.5 &0.02&  1.91$ \pm $0.25  &0.315  $\pm $0.103 &3.06  &5 & -- \\
          
            \hline
            Continued on next page
            \end{tabular}
            
            \end{table}
           
            \end{landscape}

            \setcounter{table}{0} 
            
        \begin{landscape}
         
        \begin{table}[ht!]
         \caption{\footnotesize
         Catalog of XMM-CFHTLS X-ray selected galaxy groups, continued from previous page
            \label{cat1}}
            \tiny
            \centering
            \renewcommand{\arraystretch}{0.9}\renewcommand{\tabcolsep}{0.12cm}
            \begin{tabular}[tc]{ccccc ccccc cccc cc}
           \hline 
           \hline \\
         XMM ID & Group ID & RA(J2000)  & Dec (J2000) & z$ _{phot} $& $M_{200c}$ & $L_{x,(0.1-2.4 keV)}$& $ \sigma_{v} $&$R_{200}$ & kT& Flux & Sig.& Flag & z$ _{spec} $   \\
             &  & (degrees &(degrees)& & $ 10^{13}M_{\odot} $ & $10^{42}erg s^{-1}$&kms$ ^{-1} $ &degree & KeV& $10^{-14}erg s^{-1}cm^{-2}$\\
        
                        (1) & (2)  & (3) & (4) & (5) & (6) & (7) & (8) & (9) & (10) & (11) & (12) & (13)&(14)  \\
           \hline \\
       
             XMMU J022700.2-043422 &100700 &36.751  &-4.573  &0.05  &1.963  $\pm $0.113 &0.998   $\pm $0.091& 228.4 &0.157& 0.6$  \pm $0.02  &11.442 $\pm $1.047 &10.93 &3 & --  \\
             XMMU J022303.2-043621 &100740 &35.7634 &-4.6059 &1.02  &7.834  $\pm $1.23  &32.961  $\pm $8.433& 431.1 &0.021& 1.87$ \pm $0.19  &0.347  $\pm $0.089 &3.91  &1 & 1.213*   \\
             XMMU J022620.3-043633 &100750 &36.5849 &-4.6092 &0.51  &4.864  $\pm $1.045 &7.465   $\pm $2.655& 334.0 &0.029& 1.14$ \pm $0.15  &0.442  $\pm $0.157 &2.81  &5 & -- \\
             XMMU J022727.0-043223 &100760 &36.8629 &-4.54   &0.28  &5.715  $\pm $0.406 &7.015   $\pm $0.791& 338.4 &0.048& 1.17$ \pm $0.05  &1.815  $\pm $0.205 &8.86  &1 & 0.307*   \\
             XMMU J022306.0-043403 &100780 &35.7751 &-4.5677 &0.82  &6.109  $\pm $1.294 &16.711  $\pm $5.861&382.2  &0.022& 1.47$ \pm $0.2   &0.298  $\pm $0.104 &2.85  &5 & -- \\
             XMMU J022712.9-043536 &100792 &36.804  &-4.5936 &0.3   &3.334  $\pm $0.687 &3.105   $\pm $1.054& 283.7 &0.038& 0.86$ \pm $0.1   &0.687  $\pm $0.233 &2.94  &3 & 0.30  \\
             XMMU J022331.5-043439 &100810 &35.8815 &-4.5776 &1.07  &8.531  $\pm $1.321 &40.457  $\pm $10.186&447.7 &0.021& 2.01$ \pm $0.2   &0.387  $\pm $0.097 &3.97 &5 & -- \\
             XMMU J022148.6-043110 &100820 &35.4526 &-4.5196 &0.21  &2.582  $\pm $0.368 &1.854   $\pm $0.429& 256.6 &0.047& 0.72$ \pm $0.05  &0.95   $\pm $0.22  &4.32  &3 & --   \\
          XMMU J022254.1-043308 &100830 &35.7255 &-4.5524 &1.03  &5.662  $\pm $1.148 &20.137  $\pm $6.730 & 387.7 &0.019& 1.52$ \pm $0.19  &0.184  $\pm $0.062 & 2.99  &5 & -- \\
          XMMU J022540.0-042841 &100860 &36.4168 &-4.4781 &0.22  &3.556  $\pm $0.362 &3.09    $\pm $0.505& 285.9 &0.05&  0.87$ \pm $0.05  &1.418  $\pm $0.232 &6.11  &1 & 0.203 \\
          XMMU J022556.6-042834 &100870&36.486  &-4.4764 &0.89  &8.204  $\pm $1.091 &29.376  $\pm $6.324& 427.4 &0.023& 1.84$ \pm $0.16  &0.447  $\pm $0.096 &4.65  &1 & 0.979 \\
          XMMU J022859.9-042936 &100880 &37.2499 &-4.4935 &0.54  &8.65   $\pm $1.469 &19.143  $\pm $5.321 &406.9 &0.033& 1.67$ \pm $0.18  &1.022  $\pm $0.284 &3.6   &5 & -- \\
          XMMU J022235.9-042854 &100910 &35.6496 &-4.4818 &1.08  &5.754  $\pm $1.245 &22.182  $\pm $7.943& 393.4 &0.018& 1.56$ \pm $0.21  &0.178  $\pm $0.064 &2.79  &3 & -- \\
          XMMU J022132.7-042655 &100920 &35.3865 &-4.4489 &0.3   &3.184  $\pm $0.59  &2.884   $\pm $0.877&279.3  &0.037& 0.83$ \pm $0.09  &0.638  $\pm $0.194 &3.29  &3 & -- \\
          XMMU J022827.9-042600 &100940 &37.1163 &-4.4335 &0.42  &8.279  $\pm $0.675 &15.066  $\pm $1.963& 392.2 &0.039& 1.55$ \pm $0.08  &1.489  $\pm $0.194 &7.68  &1 & 0.434*  \\
          XMMU J022142.0-042739 &100950 &35.425  &-4.4608 &0.28  &3.062  $\pm $0.65  &2.642   $\pm $0.927& 274.8 &0.039& 0.81$ \pm $0.09  &0.69   $\pm $0.242 &2.85  &3 & -- \\
          XMMU J022523.5-042649 &100960 &36.3482 &-4.447  &0.52  &4.909  $\pm $0.908 &7.691   $\pm $2.333& 335.7 &0.028& 1.15$ \pm $0.13  &0.433  $\pm $0.132 &3.29  &1& 0.462*   \\
          XMMU J022610.7-042433 &100970 &36.545  &-4.4094 &0.66  &8.185  $\pm $1.138 &20.941  $\pm $4.721& 408.8 &0.028& 1.68$ \pm $0.15  &0.676  $\pm $0.152 &4.44  &3 & -- \\
          XMMU J022325.6-042721 &100980 &35.8568 &-4.4559 &0.61  &3.963  $\pm $0.817 &6.266   $\pm $2.128& 317.9 &0.023& 1.04$ \pm $0.13  &0.228  $\pm $0.078 &2.94  &3 & -- \\
          XMMU J022332.5-042550 &101000 &35.8854 &-4.4308 &0.76  &5.728  $\pm $1.04  &13.868  $\pm $4.121& 369.8 &0.023& 1.38$ \pm $0.16  &0.299  $\pm $0.089 &3.36  &5 & -- \\
          XMMU J022711.8-425457 &101020 &36.79945& -4.42938 &0.21 &1.935$\pm$0.423 & 1.178 $\pm$ 0.426& 233.0&  0.042&  0.62$ \pm $0.07 &  0.599$\pm$ 0.216& 2.76 & 3& -- \\     
          
          XMMU J022433.8-041427 &101030 &36.1409 &-4.2411 &0.27  &6.871  $\pm $0.195 &9.205   $\pm $0.412& 359.1 &0.052& 1.31$ \pm $0.02  &2.595  $\pm $0.116 &22.37 &1 & 0.262*   \\
          XMMU J022139.2-042423 &101040 &35.4136 &-4.4066 &0.28  &3.342  $\pm $0.6   &3.027   $\pm $0.891&282.2  &0.04&  0.85$ \pm $0.09  &0.79   $\pm $0.233 &3.4   &4 & -- \\
          XMMU J022539.8-042255 &101060 &36.4159 &-4.382  &0.89  &11.995 $\pm $1.687 &53.088  $\pm $12.134&484.8 &0.026& 2.37$ \pm $0.22  &0.875  $\pm $0.2   &4.37  &3 & -- \\
         XMMU J022127.8-042342 &101080 &35.366  &-4.395  &1.09  &7.396  $\pm $1.675 &33.266  $\pm $12.503&428.4 &0.02&  1.85$ \pm $0.27  &0.289  $\pm $0.108 &2.66  &5 & -- \\
          XMMU J022536.3-042152 &101090 &36.4013 &-4.3645 &0.22  &2.78   $\pm $0.489 &1.928   $\pm $0.555& 260.5 &0.064& 0.74$ \pm $0.07  &2.117  $\pm $0.609 &3.48  &1 & 0.142 \\
          XMMU J022208.1-042113 &101120 &35.534  &-4.3536 &1.1   &8.395  $\pm $1.489 &41.115  $\pm $11.967&447.6 &0.021& 2.01$ \pm $0.23  &0.364  $\pm $0.106 &3.44  &3 & -- \\
          XMMU J022328.3-042132 &101130 &35.868  &-4.3591 &0.79  &4.315  $\pm $0.794 &9.311   $\pm $2.812& 338.5 &0.02&  1.17$ \pm $0.13  &0.171  $\pm $0.052 &3.31  &5 & -- \\
          XMMU J022531.8-041351 &101141 &36.3825 &-4.2311 &0.15  &8.222  $\pm $0.33  &10.495  $\pm $0.667& 373.9 &0.091& 1.41$ \pm $0.04  &11.144 $\pm $0.707 &15.75 &1 & 0.140*   \\
          XMMU J022529.0-041540 &101142 &36.3708 &-4.2613 &0.52  &11.695 $\pm $0.798 &29.923  $\pm $3.243& 448.5 &0.038& 2.02$ \pm $0.09  &1.81   $\pm $0.196 &9.21  &1 & 0.556*   \\
          XMMU J022506.7-041851 &101150 &36.2781 &-4.3144 &0.89  &8.892  $\pm $1.186 &33.266  $\pm $7.178&438.8  &0.024& 1.93$ \pm $0.17  &0.513  $\pm $0.111 &4.63  &4 & -- \\
          XMMU J022403.8-041319 &101181 &36.016  &-4.2221 &0.99  &11.429 $\pm $0.659 &56.885  $\pm $5.212& 486.2 &0.024& 2.38$ \pm $0.09  &0.715  $\pm $0.065 &10.92 &1 & 1.050*  \\
          XMMU J022406.0-041443 &101182 &36.0254 &-4.2453 &0.89  &4.487  $\pm $0.551 &11.429  $\pm $2.270 & 349.3 &0.019& 1.24$ \pm $0.09  &0.147  $\pm $0.029 &5.04  &2 & -- \\
          XMMU J022239.6-041819 &101200 &35.6651 &-4.3055 &0.47  &4.385  $\pm $0.75  &5.998   $\pm $1.679& 320.2 &0.029& 1.06$ \pm $0.11  &0.436  $\pm $0.122 &3.57  &3 & 0.493\\
          XMMU J022519.4-041908 &101210 &36.331  &-4.3191 &0.15  &2.056  $\pm $0.34  &1.205   $\pm $0.326& 235.6 &0.058& 0.63$ \pm $0.05  &1.318  $\pm $0.356 &3.7   &3 & 0.143 \\
          XMMU J022220.2-041641 &101230 &35.5843 &-4.2782 &0.28  &4.487  $\pm $0.689 &4.797   $\pm $1.199& 312.1 &0.044& 1.01$ \pm $0.09  &1.248  $\pm $0.312 &4.0   &3 & -- \\
          XMMU J022309.5-041717 &101240 &35.7899 &-4.2883 &0.95  &6.109  $\pm $1.156 &20.23   $\pm $6.295& 391.8 &0.02&  1.55$ \pm $0.19  &0.24   $\pm $0.075 &3.22  &3 & -- \\
          XMMU J022709.6-041810 &101250 &36.7901 &-4.303  &1.01  &6.683  $\pm $1.297 &25.351  $\pm $8.110& 408.2  &0.02&  1.68$ \pm $0.21  &0.261  $\pm $0.083 &3.13  &1 & 1.053*  \\
          XMMU J022253.1-041618 &101270 &35.7213 &-4.2718 &0.18  &1.945  $\pm $0.251 &1.146   $\pm $0.238& 232.3 &0.048& 0.62$ \pm $0.04  &0.828  $\pm $0.172 &4.8   &3 & -- \\
          XMMU J022105.7-041831 &101280 &35.2739 &-4.3087 &0.93  &6.237  $\pm $1.439 &20.277  $\pm $7.780& 392.9  &0.021& 1.56$ \pm $0.23  &0.256  $\pm $0.098 &2.61  &5 & -- \\
          XMMU J022643.0-041621 &101290 &36.6794 &-4.2727 &0.25  &2.951  $\pm $0.506 &2.393   $\pm $0.673& 270.0 &0.042& 0.79$ \pm $0.07  &0.819  $\pm $0.23  &3.56  &3 & -- \\
          XMMU J022637.1-041651 &101310 &36.6548 &-4.281  &0.93  &6.855  $\pm $1.629 &23.496  $\pm $9.290& 405.4  &0.021& 1.65$ \pm $0.25  &0.304  $\pm $0.12  &2.53  &1 & 1.018 \\
          XMMU J022630.6-041312 &101320 &36.6278 &-4.2201 &0.22  &3.155  $\pm $0.456 &2.559   $\pm $0.603& 274.7 &0.048& 0.81$ \pm $0.06  &1.179  $\pm $0.277 &4.25  &1 & 0.208 \\
        
             \hline
                        Continued on next page
           \end{tabular}
            \end{table}
           \setcounter{table}{0} 
            \end{landscape}
               
          \begin{landscape}
                   \begin{table}[ht!]
                   \caption{\footnotesize
                   Catalogue of XMM-LSS X-ray galaxy groups.
                   \label{cat2}}
                   \tiny
                  \centering
                   \renewcommand{\arraystretch}{0.9}\renewcommand{\tabcolsep}{0.12cm}
                   \begin{tabular}[tc]{ccccc ccccc cccc cc}
                   \hline
                   \hline\\
         XMM ID & Group ID & RA(J2000)  & Dec (J2000) & z$ _{phot} $& $M_{200c}$ & $L_{x,(0.1-2.4 keV)}$& $ \sigma_{v} $&$R_{200}$ & kT& Flux & Sig.& Flag & z$ _{spec} $   \\
             &  & (degrees &(degrees)& & $ 10^{13}M_{\odot} $ & $10^{42}erg s^{-1}$&kms$ ^{-1} $ &degree & KeV& $10^{-14}erg s^{-1}cm^{-2}$\\
        
                        (1) & (2)  & (3) & (4) & (5) & (6) & (7) & (8) & (9) & (10) & (11) & (12) & (13)&(14)  \\
                      \hline \\   
         XMMU J022129.4-041229 &101340 &35.3727 &-4.2082 &0.86  &9.078  $\pm $1.318 &32.961  $\pm $7.762& 439.5 &0.025& 1.94$ \pm $0.18  &0.559  $\pm $0.132 &4.24  &3 & -- \\  
          XMMU J022307.4-041309 &101350 &35.7809 &-4.2192 &0.67  &4.355  $\pm $0.708 &7.907   $\pm $2.104& 331.8 &0.023& 1.13$ \pm $0.11  &0.228  $\pm $0.061 &3.76  &1 & 0.631*  \\
          XMMU J022725.0-041130 &101360 &36.8543 &-4.1919 &0.6   &4.592  $\pm $0.627 &7.78    $\pm $1.722& 333.3 &0.025& 1.14$ \pm $0.09  &0.301  $\pm $0.067 &4.52  &5 & 0.584*   \\
          XMMU J022301.8-040904 &101370 &35.7578 &-4.1511 &1.07  &8.11   $\pm $1.072 &37.411  $\pm $7.998& 440.3 &0.021& 1.95$ \pm $0.17  &0.351  $\pm $0.075 &4.68  &2 & -- \\
          XMMU J022432.2-041001 &101380 &36.1343 &-4.1671 &0.25  &1.795  $\pm $0.196 &1.102   $\pm $0.194& 228.8 &0.036& 0.61$ \pm $0.03  &0.371  $\pm $0.065 &5.68  &3& --
           \\
          XMMU J022447.7-040911 &101391 &36.1988 &-4.1532 &0.07  &0.953  $\pm $0.21  &0.357   $\pm $0.131& 181.9 &0.047& 0.44$ \pm $0.04  &0.423  $\pm $0.155 &2.73  &3 & 0.097 \\
          XMMU J022257.6-040709 &101400 &35.7403 &-4.1193 &0.28  &3.334  $\pm $0.361 &3.013   $\pm $0.526& 282.6 &0.04&  0.85$ \pm $0.05  &0.786  $\pm $0.137 &5.73  &1 & 0.293*  \\
          XMMU J022419.0-040814 &101420 &36.0794 &-4.1374 &0.86  &4.797  $\pm $0.605 &12.162  $\pm $2.477& 355.3 &0.02&  1.28$ \pm $0.1   &0.18   $\pm $0.037 &4.91  &2 & -- \\
          XMMU J022635.1-040408 &101480 &36.6465 &-4.0691 &0.34  &7.047  $\pm $0.485 &10.52   $\pm $1.151& 366.4 &0.044& 1.36$ \pm $0.06  &1.716  $\pm $0.188 &9.12  &1 & 0.345*   \\
          XMMU J022129.7-040543 &101490 &35.3741 &-4.0954 &0.94  &9.705  $\pm $1.782 &40.926  $\pm $12.359& 456.0&0.024& 2.09$ \pm $0.25  &0.562  $\pm $0.17  &3.31  &3 & -- \\
          XMMU J022201.0-040641 &101500 &35.5045 &-4.1115 &0.73  &6.471  $\pm $1.096 &16.069  $\pm $4.446& 383.0 &0.024& 1.48$ \pm $0.16  &0.391  $\pm $0.108 &3.61  &3& --  \\
          XMMU J022552.1-040711 &101510 &36.4675 &-4.1199 &0.89  &6.067  $\pm $1.538 &18.365  $\pm $7.745& 386.5 &0.021& 1.51$ \pm $0.24  &0.26   $\pm $0.11  &2.37  &5 & -- \\
          XMMU J022510.0-040148 &101530 &36.2918 &-4.0302 &0.16  &2.891  $\pm $0.29  &2.075   $\pm $0.334& 264.3 &0.061& 0.76$ \pm $0.04  &1.972  $\pm $0.318 &6.21  &1 & 0.17*  \\
          XMMU J022529.4-040040 &101540 &36.3727 &-4.0112 &0.07  &2.366  $\pm $0.159 &1.365   $\pm $0.146& 243.9 &0.121& 0.67$ \pm $0.02  &7.779  $\pm $0.833 &9.33  &3 & -- \\
          XMMU J022158.5-040248 &101550 &35.494  &-4.0467 &1.06  &9.795  $\pm $2.193 &49.545  $\pm $18.365&468.0 &0.022& 2.2$  \pm $0.33  &0.504  $\pm $0.187 &2.69  &4 & -- \\
          XMMU J022702.1-040509 &101570 &36.7589 &-4.0859 &0.85  &6.281  $\pm $1.227 &18.239  $\pm $5.875& 387.9 &0.022& 1.52$ \pm $0.19  &0.297  $\pm $0.095 &3.11  &1 & 0.765 \\               
        XMMU J022425.6-040057 &101580 &36.1069 &-4.016  &1.02  &9.84   $\pm $1.045 &46.99   $\pm $8.035& 465.1 &0.023& 2.18$ \pm $0.15  &0.527  $\pm $0.09  &5.85  &3 & -- \\
          XMMU J022619.8-040008 &101610 &36.5827 &-4.0024 &0.21  &5.236  $\pm $0.434 &5.585   $\pm $0.738& 324.8 &0.059& 1.09$ \pm $0.05  &2.827  $\pm $0.374 &7.57  &3 & -- \\
          XMMU J022606.9-040315 &101620 &36.5291 &-4.0544 &0.2   &3.758  $\pm $0.908 &3.289   $\pm $1.321& 290.3 &0.055& 0.89$ \pm $0.12  &1.879  $\pm $0.755 &2.49  &5 & -- \\
          XMMU J022222.9-040112 &101630 &35.5955 &-4.02   &0.69  &8.472  $\pm $1.146 &23.014  $\pm $5.058& 415.7 &0.028& 1.74$ \pm $0.15  &0.67   $\pm $0.147 &4.56  &3 & -- \\
          XMMU J022631.6-035640 &101720 &36.6319 &-3.9446 &0.34  &5.534  $\pm $0.741 &7.211   $\pm $1.567& 338.2 &0.04&  1.17$ \pm $0.1   &1.177  $\pm $0.255 &4.61  &3 & -- \\
          XMMU J022406.3-035507 &101730 &36.0263 &-3.9187 &0.6   &11.535 $\pm $0.94  &32.809  $\pm $4.266& 453.1 &0.034& 2.06$ \pm $0.11  &1.4    $\pm $0.182 &7.69  &1 & 0.559*   \\
          XMMU J022532.7-035506 &101741 &36.3864 &-3.9184 &0.87  &11.508 $\pm $1.762 &48.417  $\pm $12.078&476.6 &0.026& 2.29$ \pm $0.23  &0.839  $\pm $0.209 &4.02  &1 & 0.772*  \\
          XMMU J022821.2-035223 &102090 &37.0884 &-3.8733 &0.28  &3.327  $\pm $0.462 &2.999   $\pm $0.678& 282.4 &0.04&  0.85$ \pm $0.07  &0.783  $\pm $0.177 &4.43  &5 & -- \\
          XMMU J022229.2-035251 &102130 &35.622  &-3.881  &0.23  &3.381  $\pm $0.525 &2.897   $\pm $0.731& 281.7 &0.047& 0.85$ \pm $0.07  &1.201  $\pm $0.303 &3.96  &3 & -- \\
          XMMU J022303.4-034654 &102220 &35.7643 &-3.7819 &0.86  &9.506  $\pm $2.0   &35.4    $\pm $12.303&446.2 &0.025& 2.0$  \pm $0.28  &0.606  $\pm $0.211 &2.88  &5 & -- \\
          XMMU J022441.5-034858 &102290 &36.1732 &-3.8163 &0.51  &4.954  $\pm $0.906 &7.674   $\pm $2.307& 336.0 &0.029& 1.16$ \pm $0.13  &0.455  $\pm $0.137 &3.33  &3 & -- \\
          XMMU J022357.5-034311 &102570 &35.9897 &-3.7198 &0.95  &11.668 $\pm $2.061 &55.59   $\pm $16.069&486.0 &0.025& 2.38$ \pm $0.28  &0.777  $\pm $0.225 &3.46  &3 & -- \\
          XMMU J022644.2-034100 &102650 &36.6843 &-3.6834 &0.34  &8.872  $\pm $0.693 &15.066  $\pm $1.879& 395.7 &0.047& 1.58$ \pm $0.08  &2.475  $\pm $0.309 &8.01  &1 & 0.328*   \\
          XMMU J022539.2-034147 &102700 &36.4133 &-3.6966 &1.06  &7.816  $\pm $1.466 &34.834  $\pm $10.715& 434.2&0.021& 1.89$ \pm $0.23  &0.332  $\pm $0.102 &3.25  &3 & -- \\
          XMMU J022420.5-034135 &102750 &36.0855 &-3.6933 &0.64  &5.093  $\pm $1.439 &9.661   $\pm $4.603& 347.5 &0.025& 1.23$ \pm $0.21  &0.319  $\pm $0.152 &2.1   &3 & -- \\
          XMMU J022145.4-034616 &102760 &35.4392 &-3.7712 &0.47  &38.019 $\pm $1.164 &175.388 $\pm $8.453& 658.1 &0.06&  4.49$ \pm $0.1   &15.374 $\pm $0.741 &20.75 &1 & 0.429*   \\
          XMMU J022402.4-034622 &102780 &36.0103 &-3.7728 &0.6   &7.379  $\pm $1.393 &16.331  $\pm $5.058& 390.4 &0.029& 1.54$ \pm $0.18  &0.659  $\pm $0.204 &3.23  &3 & -- \\
          XMMU J022119.1-034451 &102810 &35.3297 &-3.7476 &0.43  &3.758  $\pm $0.991 &4.457   $\pm $1.968& 302.1 &0.03&  0.95$ \pm $0.14  &0.406  $\pm $0.179 &2.27  &3 & -- \\
          XMMU J022752.2-035044 &102820 &36.9678 &-3.8458 &0.11  &1.517  $\pm $0.203 &0.714   $\pm $0.155& 211.6 &0.069& 0.54$ \pm $0.03  &1.518  $\pm $0.329 &4.61  &1 & 0.140*   \\
          XMMU J022520.6-034809 &102840 &36.3359 &-3.8026 &0.3   &6.353  $\pm $0.5   &8.472   $\pm $1.064& 351.6 &0.047& 1.26$ \pm $0.06  &1.864  $\pm $0.234 &7.96  &1 & 0.299*  \\
          XMMU J022457.1-034905 &102900 &36.2379 &-3.8181 &0.58  &28.445 $\pm $0.979 &130.317 $\pm $7.096& 609.7 &0.047& 3.82$ \pm $0.09  &6.827  $\pm $0.372 &18.38 &1 & 0.614*  \\
          XMMU J022246.8-035156 &102910 &35.6953 &-3.8657 &0.08  &5.07   $\pm $0.452 &5.445   $\pm $0.778& 322.4 &0.054& 1.07$ \pm $0.06  &2.238  $\pm $0.319 &7.01  &3 & -- \\
          XMMU J022242.9-035529 &102951 &35.6789 &-3.9249 &1.09  &8.71   $\pm $1.6   &42.954  $\pm $12.942&452.4 &0.021& 2.06$ \pm $0.25  &0.393  $\pm $0.119 &3.31  &3 & -- \\
          XMMU J022225.2-035701 &103010 &35.605  &-3.9504 &0.21  &2.618  $\pm $0.57  &1.888   $\pm $0.682& 257.7 &0.047& 0.73$ \pm $0.08  &0.969  $\pm $0.35  &2.77  &5 & -- \\
          XMMU J022623.7-034309 &103320 &36.5988 &-3.7194 &0.53  &7.413  $\pm $1.358 &14.825  $\pm $4.457& 385.8 &0.032& 1.5$  \pm $0.17  &0.817  $\pm $0.246 &3.33  &3 & 0.558 \\
          XMMU J022134.7-040350 &103330 &35.3946 &-4.0641 &0.14  &1.012  $\pm $0.404 &0.392   $\pm $0.270 & 185.6 &0.048& 0.45$ \pm $0.07  &0.47   $\pm $0.324 &1.45  &3 & -- \\
          XMMU J022358.4-045607 &103340 &35.9936 &-4.9355 &0.96  &13.583 $\pm $1.742 &71.285  $\pm $14.791& 512.0&0.026& 2.65$ \pm $0.23  &1.006  $\pm $0.209 &4.82  &5 & -- \\ 
          \hline
                      
                     \end{tabular}
                     
                     \end{table}
                
           \end{landscape} 
\newpage 
 Thus, in this test we also find that
the contamination by DSFGs has no effect on the magnitude gap
estimate. In addition, one way to inspect the effect of the red
sequence selection on the magnitude gap estimate is to compare values
of the magnitude gap which are calculated using the photometric
membership to that of the spectroscopic membership.
 
 We use the X-ray group catalogs from the AEGIS field
 \citep{Erfanianfar13}, the COSMOS field \citep{Finoguenov07,George11},
 including all spectroscopic observations, available within the
 collaboration. We also use all spectroscopic data of groups in our
 catalog provided by the VVDS data and \cite{Adami11}.

We only consider the $\sim$1 degree$^{2}$ area of the COSMOS survey,
that is in common with CFHTLS survey, where a comparison to our red
sequence selection using CFHT colors is possible. Altogether, we
obtain a catalog of 84 galaxy groups (42 groups from the COSMOS
catalog, 26 groups from our catalog and 16 galaxy groups from the
AEGIS catalog), having the spectroscopic confirmation of the magnitude
gap. These groups span a redshift range of 0.05$<$z$<$1.22, which is
similar to the redshift range of our XMM-LSS catalog.

                      \begin{figure}[ht!]
                      \begin{center}  
                      \leavevmode
                     \includegraphics[width=9.5cm]{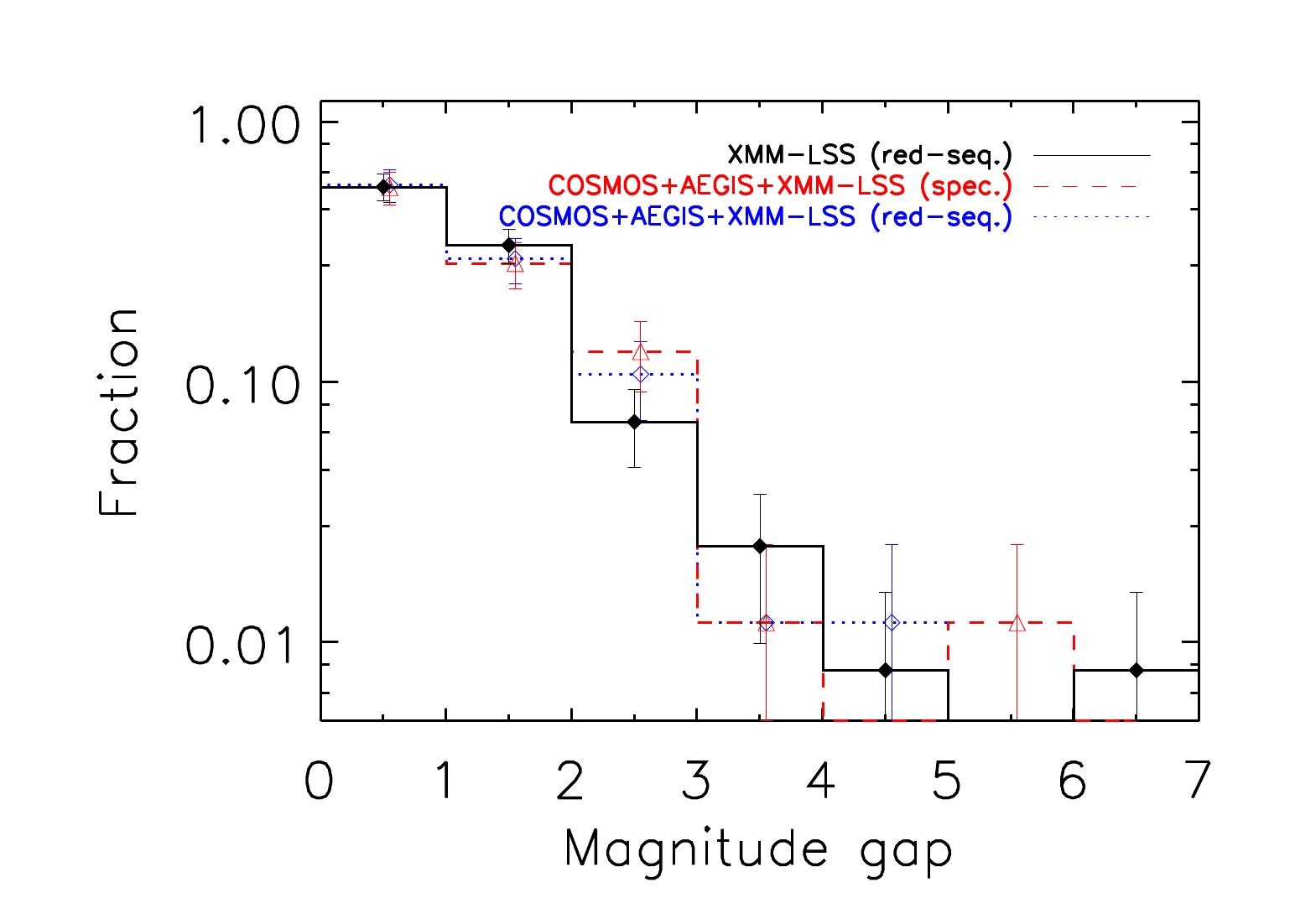}
                       \end{center}
                       \caption[flag]{Comparison of the magnitude gap
                         distributions between our X-ray group
                         catalog (XMM-LSS) with the red sequence
                         membership (solid black histogram) and 84
                         X-ray galaxy groups from the
                         COSMOS+AEGIS+XMM-LSS with the spectroscopic
                         membership (dashed red histogram) and the red
                         sequence membership (dotted blue histogram). 
                         The redshift range is 0.05$<$z$<$1.22. A good
                         agreement is seen between the two
                         distributions, implying that the results
                         obtained with the red sequence selection are
                         robust. }  \label{specgap}
\end{figure}

          \begin{figure*}[ht!]
     \begin{center} 
     \leavevmode
    \includegraphics[width=9cm]{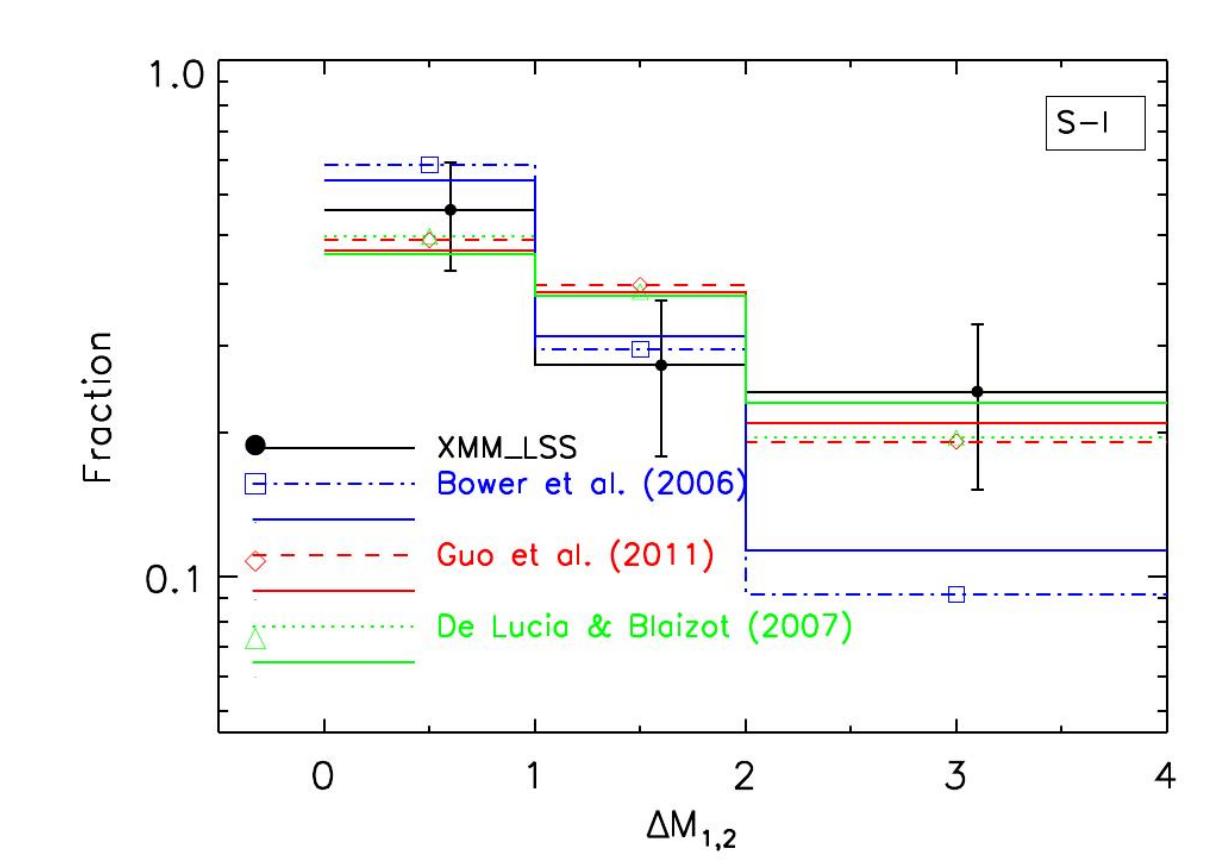}
    \includegraphics[width=9cm]{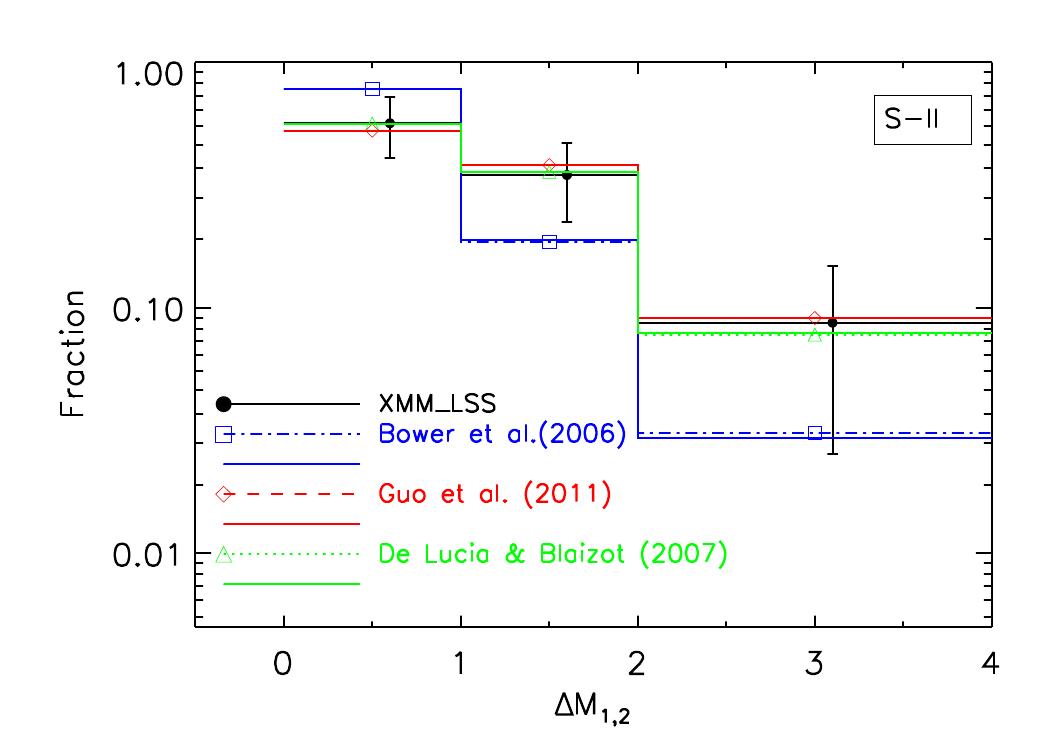}
    \includegraphics[width=9cm]{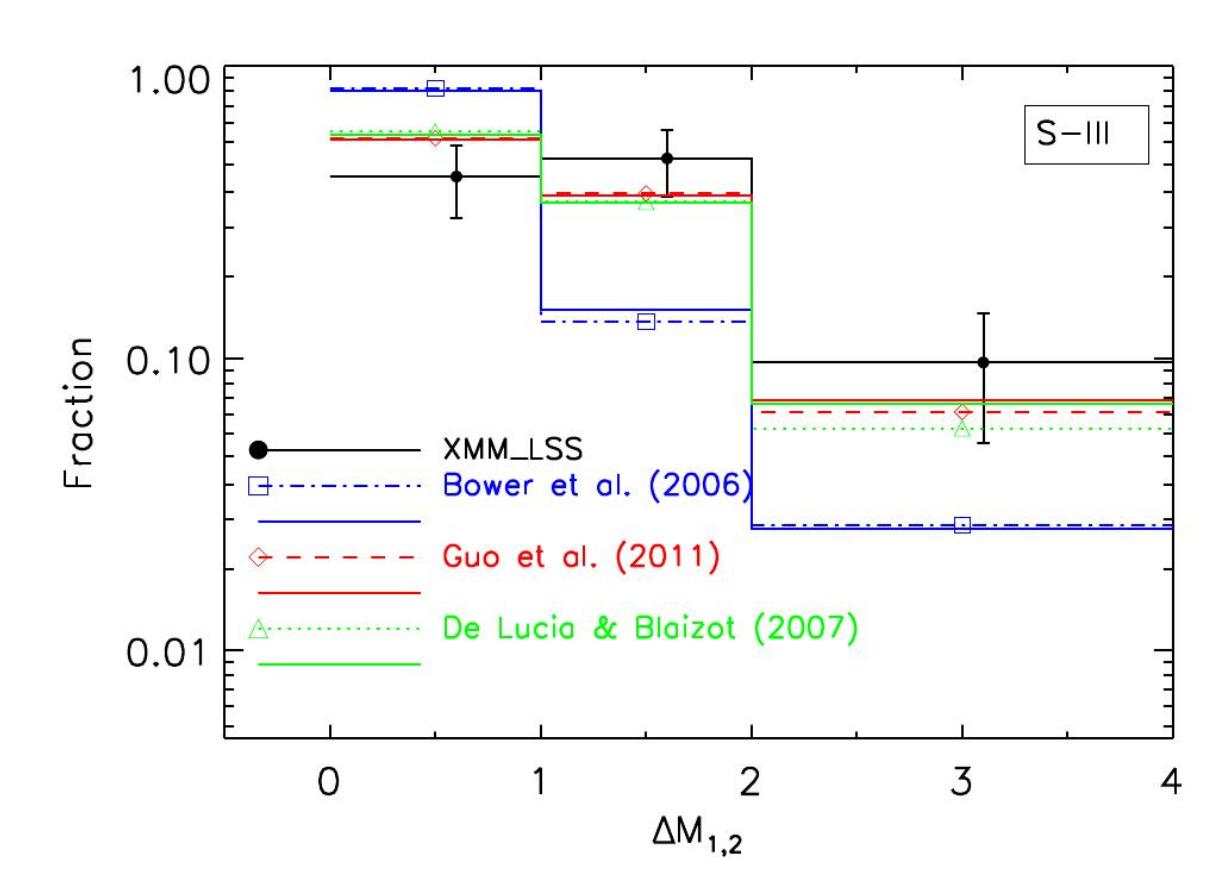}
    \includegraphics[width=9cm]{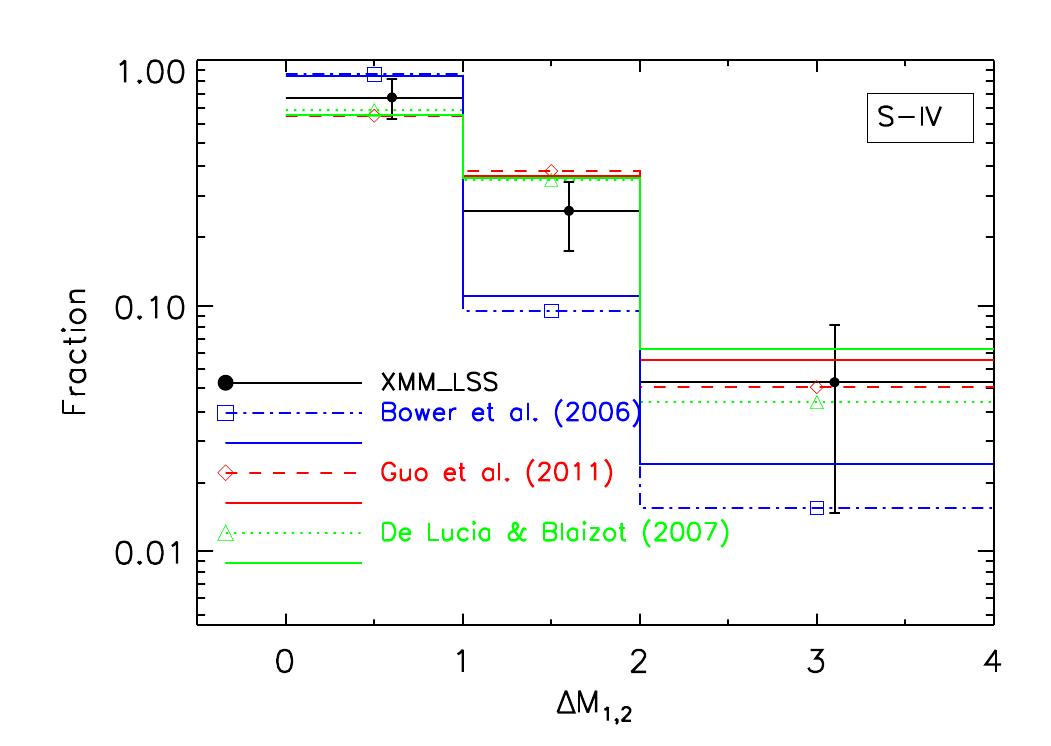}
   \end{center}
   \caption[flag]{The observed fraction of galaxy groups as a function
     of their magnitude gap (black points with error bars) compared to
     the predictions of semi-analytic galaxy formation models, B06
     (solid and dashed-dotted blue histograms), DLB07 (solid and
     dotted green histograms) and G11 (solid and dashed red
     histograms) for S--I (top left panel), S--II (top right panel),
     S--III (bottom left panel) and S--IV (bottom right panel). We
     find that the fraction of groups declines with magnitude gap at
     all redshifts. Magnitude gap bin sizes are equal to one magnitude
     for $ \Delta M_{1,2}<$2. $ \Delta M_{1,2}\geq$2 systems are
     plotted as a 2 magnitude wide bin.}
   \label{gap}
\end{figure*}   
         
For this sample of 84 groups (COSMOS+AEGIS+XMM-LSS) we select the red
sequence and spectroscopic members of groups according to the approach
described in \S 4.1.  We compute the magnitude gap between the first
and the second brightest group galaxies using both spectroscopic and
photometric group members, separately. The magnitude gaps of the
spectroscopic sample range from 0.02 to 5.75.
 
In Fig.\ref{specgap} we compare the magnitude gap distribution of the
COSMOS+AEGIS+XMM-LSS groups for two calculations of gap, one -- using
red sequence membership (dotted blue histogram) and the other -- using
the spectroscopic membership (dashed red histogram). It appears that
the two distributions do not deviate considerably.  In
Fig.\ref{specgap} we also compare these distributions to that of our
XMM-LSS galaxy groups catalog with red sequence members (solid black
histogram). 
 
By comparing the spectroscopic and photometric membership assignment
for the COSMOS+AEGIS+XMM-LSS groups, we find that redshifts of the
second ranked red sequence galaxies in six groups deviates from the
redshift of the hosting galaxy group. The magnitude gaps of these
contaminated groups, which have been computed using red sequence
galaxies range as follows: three galaxy groups with $ \Delta
M_{1,2}\geq$1.7 (classified as fossils) and three galaxy groups with
$\Delta M_{1,2}\leq$1.28 (classified as non-fossils). Differences
between the gap values in two calculations range from 0.05 to 1.2
mag. These deviations do not affect the classification by $\Delta
M_{1,2}$ in 5 out of 6 cases. From finding one system to change the
classification, we place a 95\% confidence limit on the effect of
$<6$\%, based on the Poisson PDF. We added the uncertainty associated
with this to each magnitude bin. Fractionally the effect is strongest
for the bin containing the large magnitude gap systems.
          
 \subsection{Determination of $\Delta M_{1,2}$ and definition of  fossil groups}

 In a number of studies different fossil group criteria for the
 magnitude gap and search radius have been used. For instance,
 \cite{LaBarbera12} defined as fossil groups those elliptical galaxies
 with no satellite brighter than a given magnitude gap of $\Delta
 M_{1,2}\ge$1.75, within a search radius of 0.35 Mpc and a maximum
 redshift difference of 0.001. \cite{Voevodkin10} classified an object
 as a fossil group if $\Delta$M$_{1,2} \ge 1.7$ for galaxies inside
 0.7r$_{500}$. \cite{Smith10} used a fixed projected physical radius
 of R = 640 kpc ($\sim$ 0.4r$_{200} $) for a sample of massive
 clusters at z$\sim$0.2 to study the distribution of luminosity gap.
                 \begin{figure*}[ht!]
                  \begin{center} 
                  \leavevmode
                \includegraphics[width=9cm]{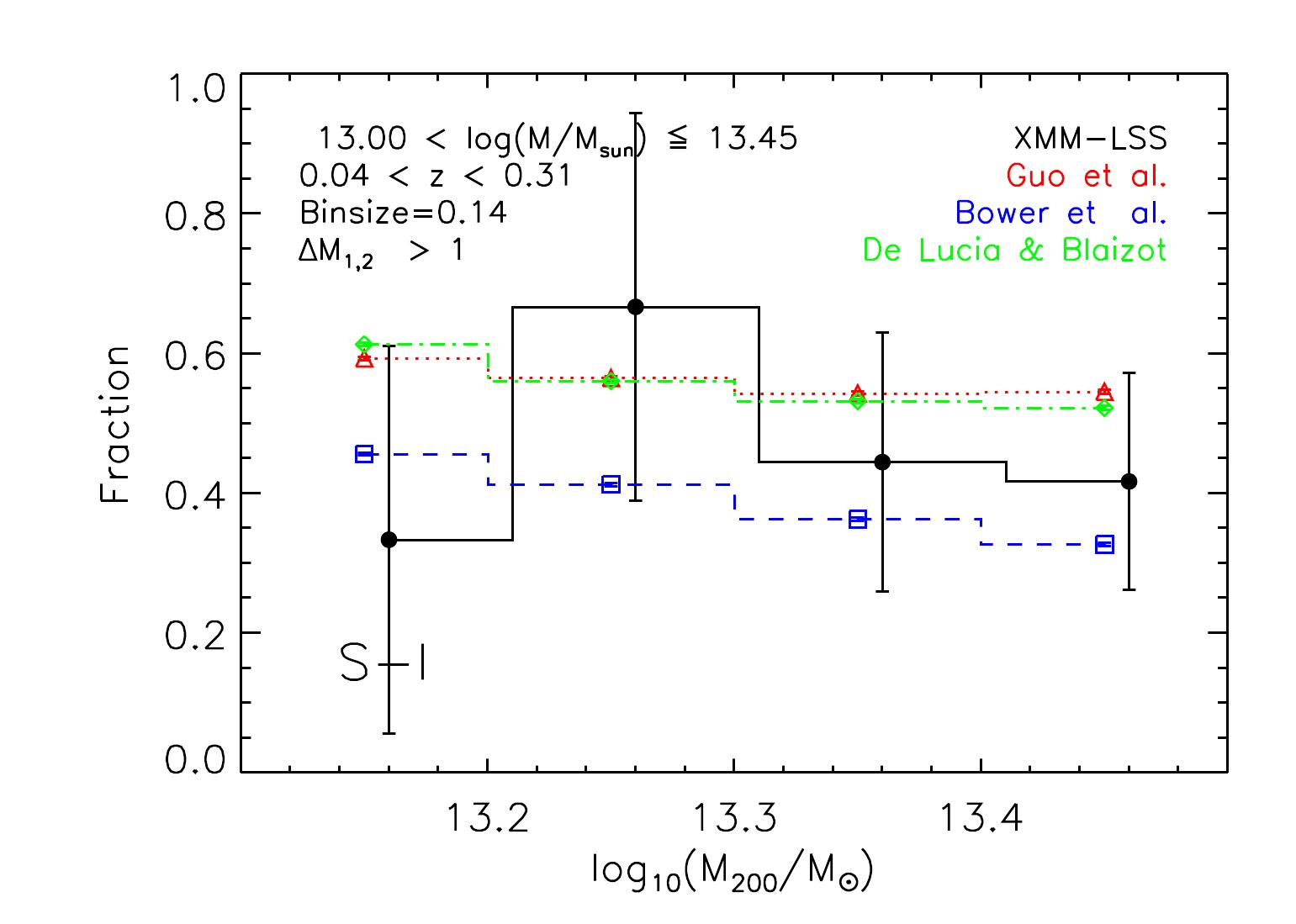}
                \includegraphics[width=9cm]{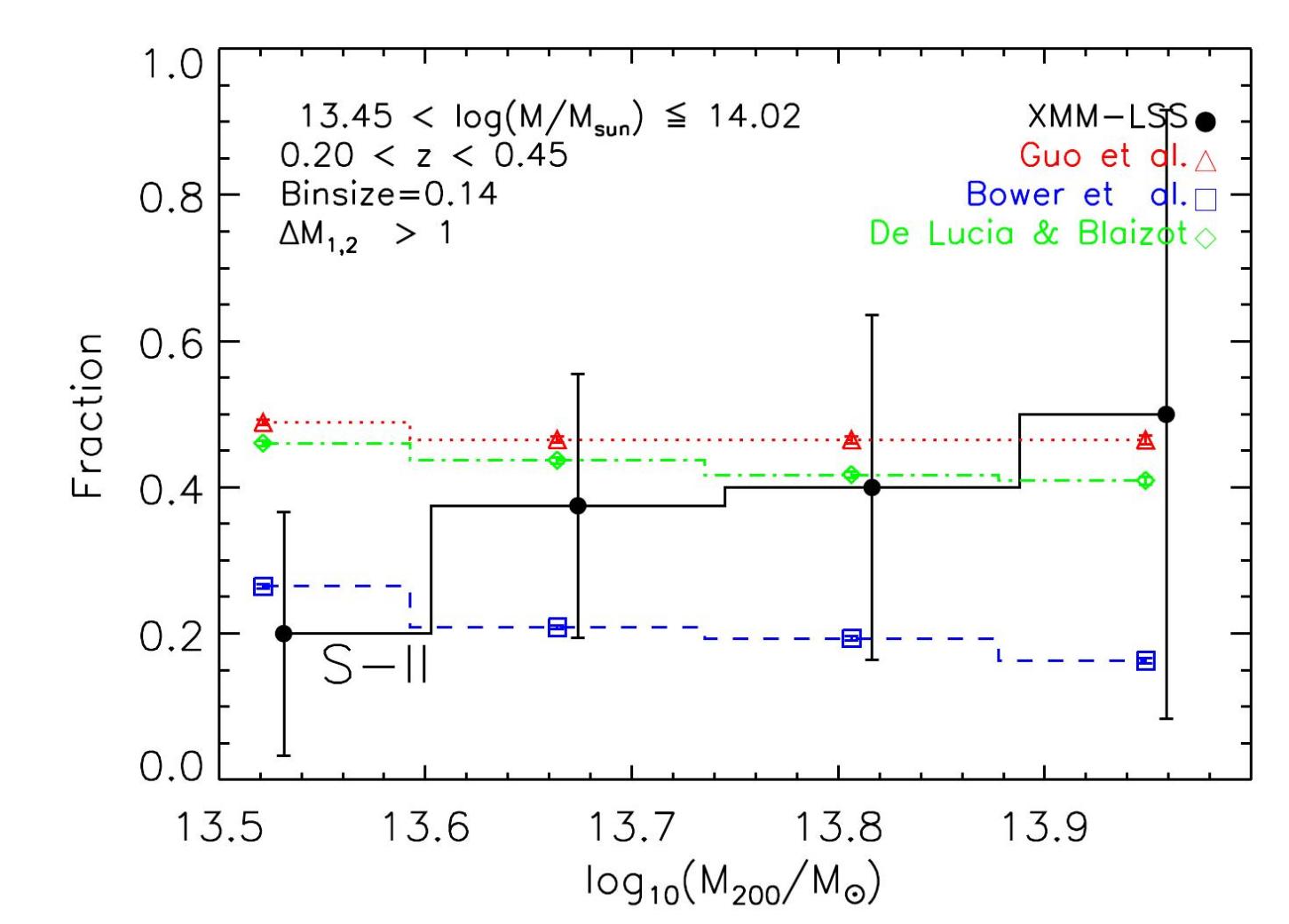}
                \includegraphics[width=9cm]{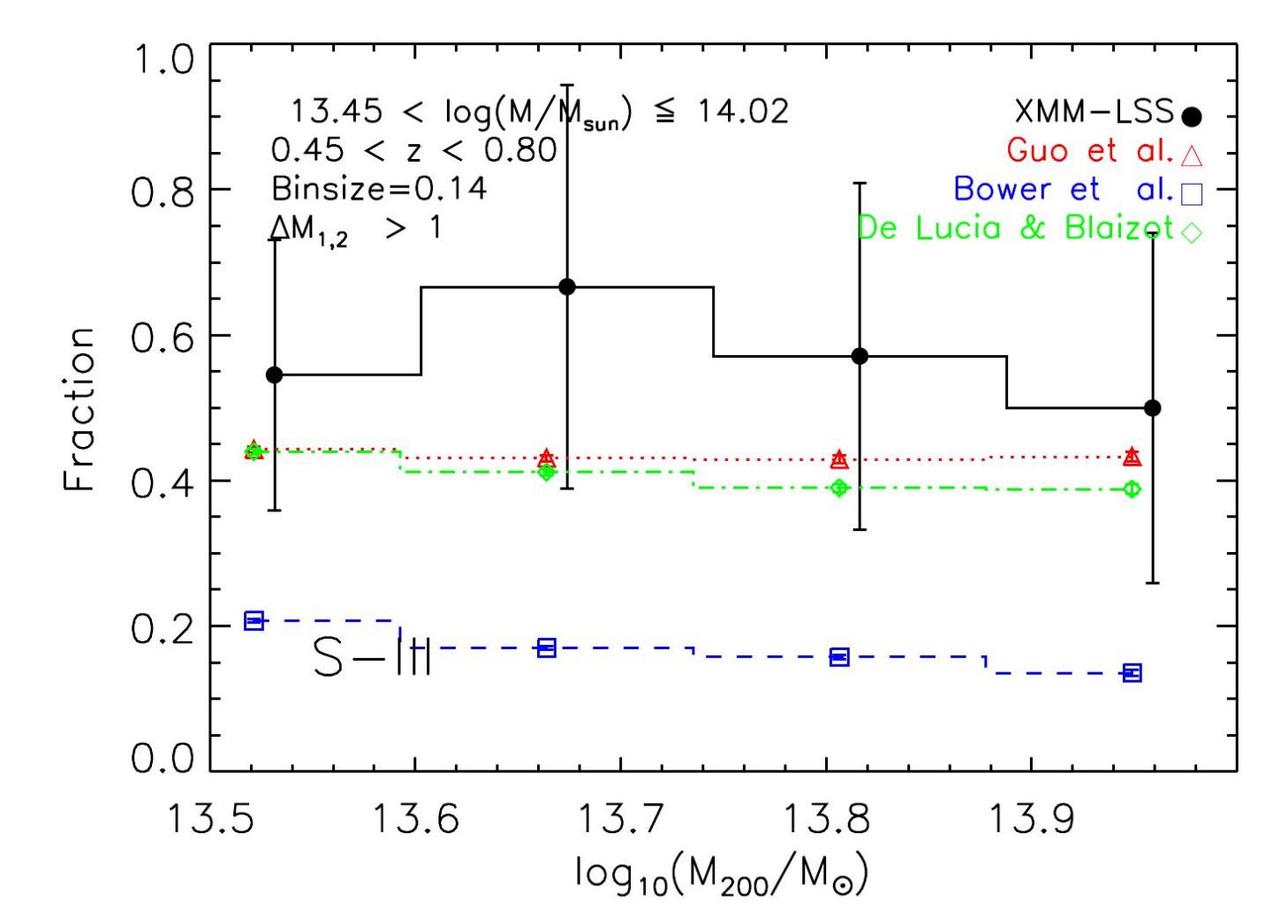}
                \includegraphics[width=9cm]{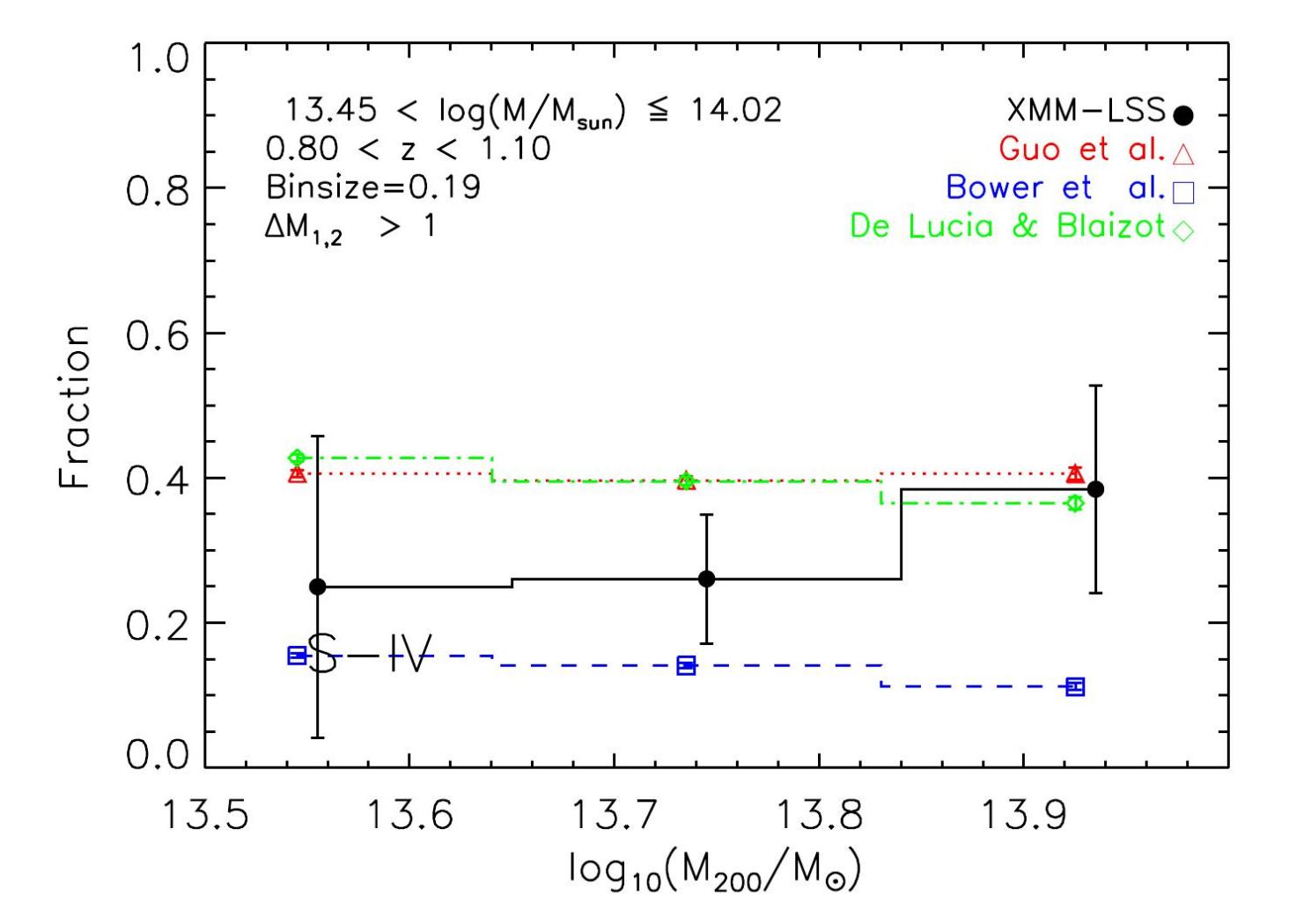} 
               \end{center}
               \caption[flag]{Halo mass dependence of the fraction of
                 groups with $\Delta M_{1,2}>$1. Black points with
                 error bars show the data and dotted red, dashed blue
                 and dash-dotted green histograms present model
                 predictions of G11, B06 and DLB07, respectively for
                 S--I (top left panel), S--II (top right panel),
                 S--III (bottom left panel) and S--IV (bottom right
                 panel). While the data are in agreement with the
                 models, the low statistics of the data prevents us
                 from choosing the best model. }
                \label{massgap}
\end{figure*}

We define the spatial selection of galaxies for the $\Delta M_{1,2}$
calculation in such a way that it allows us to select a subsample of
fossil groups candidates. For computing the magnitude gap of a galaxy
group, we adopt the coordinates of the BGG as a center of group and measure
the magnitude gap, $\Delta M_{1,2}$, using the rest frame absolute
r-band magnitude of galaxies between the BGG and the second brightest
galaxy within 0.5R$_{200}$, following \cite{Jones03}. The same
definition of the magnitude gap has been applied to the SAM catalogs.

We define as a fossil group candidate, the X-ray group of galaxies
that has a magnitude gap exceeding $\Delta M_{1,2}\approx$1.7 mag
within 0.5R$_{200}$. The low $L_x$ cut has been historically done to
avoid inclusion of X-ray emitting galaxies. However, by implying this
cut one restricts the studies only to massive groups. Since we remove
the galactic contribution to the X-ray emission by using only large
spatial scales in the flux estimates, we do not have to limit the
selection of fossil groups. Our lowest $L_x$ group has a luminosity in
the rest-frame 0.1--2.4 keV band of (3.6$\pm$1.3)$\times$10$^{41}$
ergs s$^{-1}$.

\section{Results and discussion}
\subsection{ Magnitude gap distribution}

We calculate the magnitude gap between the first and the second
brightest galaxies in groups and clusters according to the method
described in \S5.2. For 26 groups in our catalog we calculate the
magnitude gap using a spectroscopic membership assignment and for the
remaining 102 groups the gaps are calculated using the red-sequence
membership assignment.
 
In Fig.\ref{gap}, we compare the distribution of the magnitude gap
between X-ray groups and the predictions of SAMs, G11 , DLB07 and
B06. As discussed in \S 2.2, the completeness limit may affect the
magnitude gap calculation of groups with $ \Delta$ M$_{1,2}>$ 2 at
z$>$0.5. To account for this effect we define two different bin sizes,
with 1 mag for systems with $\Delta M_{1,2}<$2, while assigning all
$\Delta M_{1,2}\geq $2 groups to a single bin (see
Fig.\ref{comp}). Such a binning scheme is introduced to remove the
effect of completeness on the binned fractions.

For SAMs we use both the formal selection of the subsamples (solid
histograms) and modeling of the X-ray selection. For the later, we
introduce mass-dependent weights,
$\frac{V_{obs}(M_{200},z)}{V_{sim}}$. V$_{obs}$ is the survey volume
for detecting groups as a function of $ M_{200} $ and z, where
V$_{sim}$ is equal to the (500 Mpc h$^{-1})^{3}$ volume of the
Millennium simulation. We evaluate the total number of groups
corresponding to each $\Delta M_{1,2}$ bin by summation of the
weighted number of groups. We illustrate the so constructed model
magnitude gap distributions with dashed red (G11), dashed-dotted blue
(B06) and dotted green (DBL07) histograms in Fig. \ref{gap}. It can be
seen that the differences in the resulting magnitude gap distributions
are small compared to our observational errors.
 
In agreement with the findings of \cite{Smith10}, the fraction of
detected groups, $ f(\Delta M_{1,2}) $, in all histograms of
Fig. \ref{gap} declines with $ \Delta M_{1,2}$. A linear relation
provides a good fit to the data. We provide the fit parameters
corresponding to each subsample in Tab.\ref{gap_param}. The slope of
the relation is shallower for (S--I) compared to other subsamples,
indicating a larger contribution of fossil groups. The trends seen in
the data for groups and clusters with $\Delta M_{1,2} \leqslant$3 are
well reproduced by both DLB07 and G11 models, while the B06 model
predicts the slopes that are too steep. In addition, as previously
noticed by \cite{Dariush10}, the B06 model over-predicts the fraction
of clusters with $\Delta M_{1,2}\leqslant$1 for S-II to S-IV.
       
             \setcounter{table}{1}
              \begin{table}[ht!]
                        \caption{\footnotesize
                       Parameters of the linear fit to a distribution of groups over the value of magnitude gap, $ \Delta M_{1,2} $.
                        \label{gap_param}}
                        \tiny
                        \centering
                        \renewcommand{\arraystretch}{0.9}\renewcommand{\tabcolsep}{0.12cm}
                         
                        \begin{tabular}[tc]{ccccc ccccc cccc cc}
                         \hline
                        \hline\\
                       Subsample   &  Intercept     &   slope \\
                       \hline \\
                                   
                       S--I & $ 0.51 \pm 0.12 $ & $ - 0.11 \pm 0.06 $ \\
                       S--II &$  0.70 \pm 0.01 $ & $ - 0.22 \pm 0.01 $ \\
                        S--III & $ 0.57 \pm 0.14 $ & $ - 0.14 \pm 0.07 $ \\
                       S--IV & $  0.75 \pm 0.17 $ &  $ - 0.25 \pm 0.09 $ \\
                       \\
                       \hline
                       
                       \end{tabular}
\end{table}

In Fig. \ref{massgap}, we investigate a mass dependence of the
fraction of groups characterized by a large magnitude gap ($ \Delta
M_{1,2}\ge1 $).  The relation reveals a mild dependence on mass,
justifying our use of wide mass bins.

\begin{figure}[H]
   \begin{center} 
   \leavevmode
   \includegraphics[width=9.5cm]{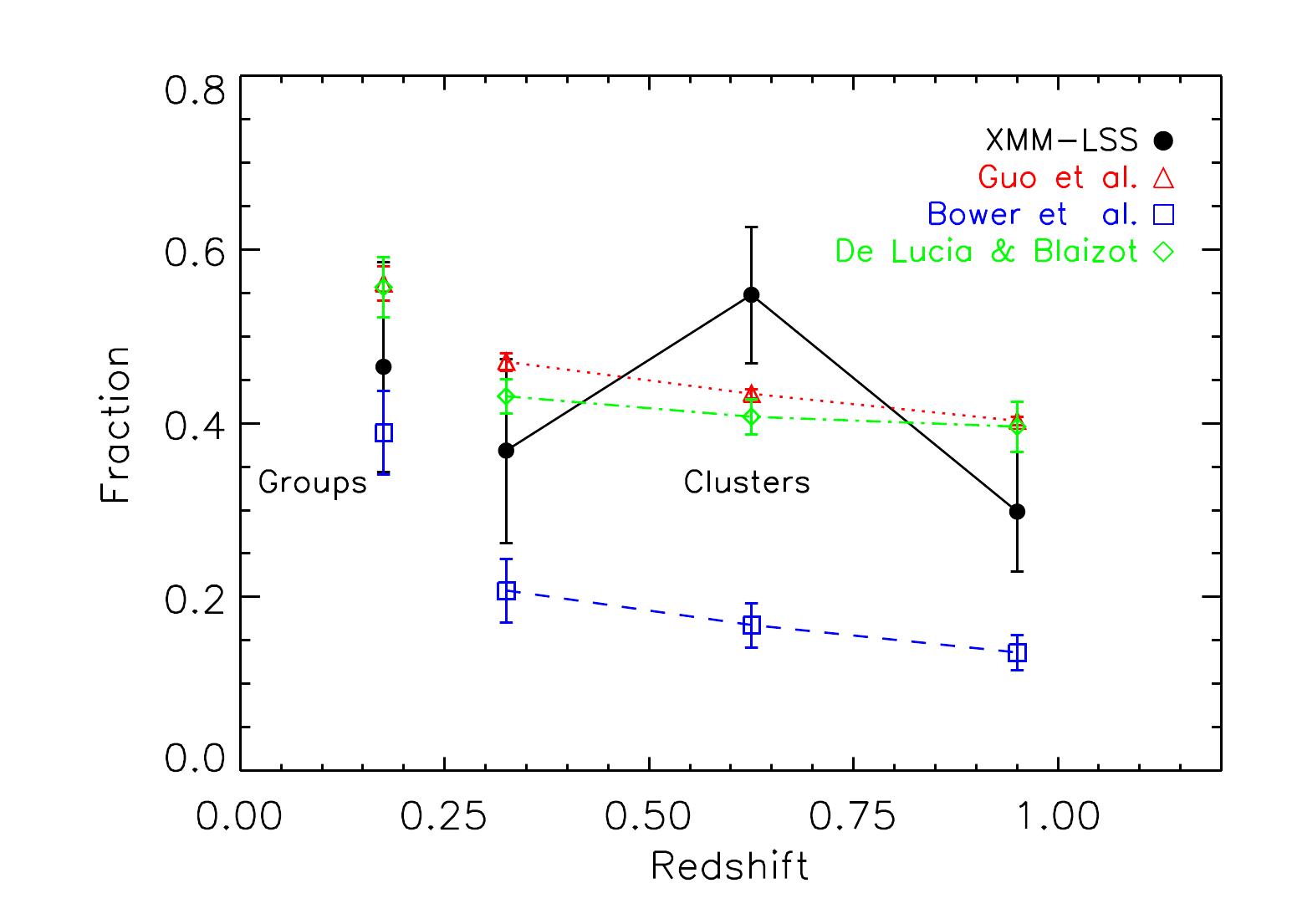}
  \end{center}
  \caption[flag]{The redshift evolution of the fraction of groups and
    clusters with $\Delta M_{1,2} > 1 $. The error bars correspond to the
    95\% confidence limits of the standard error of the mean. }
  \label{sumfr}
 \end{figure}

Fig. \ref{sumfr} summarizes the trends presented in Fig. \ref{massgap}
and shows the redshift evolution of the mean fraction of large
magnitude gap groups/clusters, n(z). We find that n(z) evolves slowly
with redshift: n(z)=0.47-0.12$\times$z, in agreement with all models.
B06 underestimates fraction of clusters having a large magnitude gap.
     
Behavior of the luminosity gap in groups and clusters can depend on
several effects, which we describe below. If a galaxy brighter than
the second brightest galaxy falls into the cluster or group, the
luminosity gap of the system changes to a lower value, as soon as the
infalling galaxy passes into $ (0.5R_{200}) $, within which the $
\Delta M_{1,2} $ is calculated. For the galaxy group members, the
dynamical friction causes satellite galaxies to merge together or with
the central galaxy. Mergers between satellites can reduce the number
of satellite galaxies in SAMs and in massive halos, especially
reducing the number of the low-luminosity satellite galaxies; the product
of merger is more luminous and massive. Finally, evolution of galaxy
properties, most notably the stellar mass and age of the stellar
population in both central and satellite galaxies also affects the
magnitude gap. This in turn depends on the gas cooling rate, tidal
stripping and disruption, AGN and SNe feedback.

The G11 model shows the best agreement with our observational results,
possibly due to its advances in modeling the satellite disruption
and gas striping due to the tidal and ram-pressure forces and the
satellite-satellite merger. These processes are not taken into account
in B06 and DLB07, but play an important role in the evolution of the
galaxy luminosity in massive halos
\citep{Conroy07a,Conroy07b,Henriques10,Liu10}.

 \subsection{Magnitude gap and  BCG luminosity relation }
 \subsubsection{$ M_{r} $ - $\Delta M_{1,2}$}

\begin{figure}[ht!]
           \begin{center}  
           \leavevmode
           \includegraphics[width=9cm]{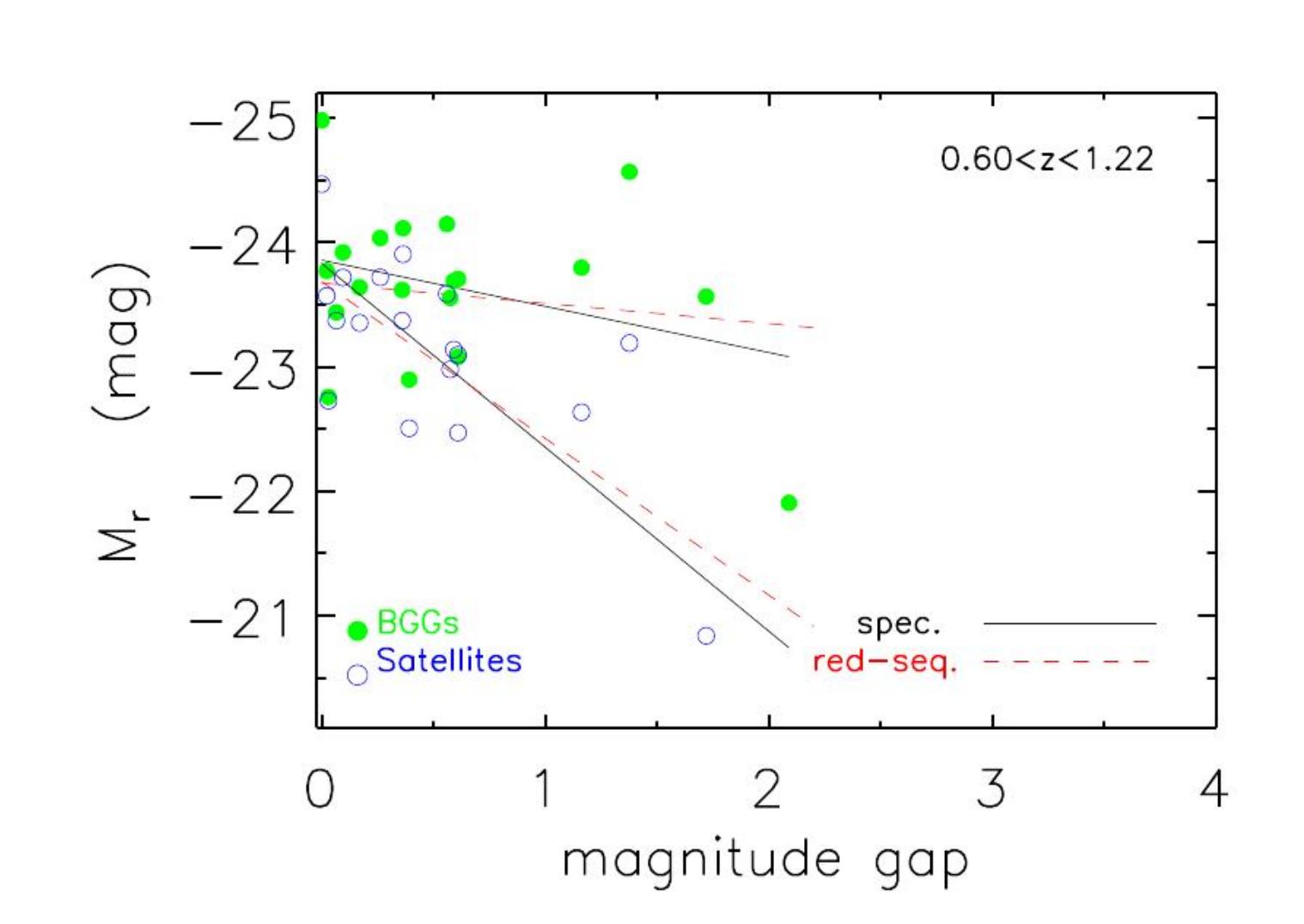}
          \includegraphics[width=9cm]{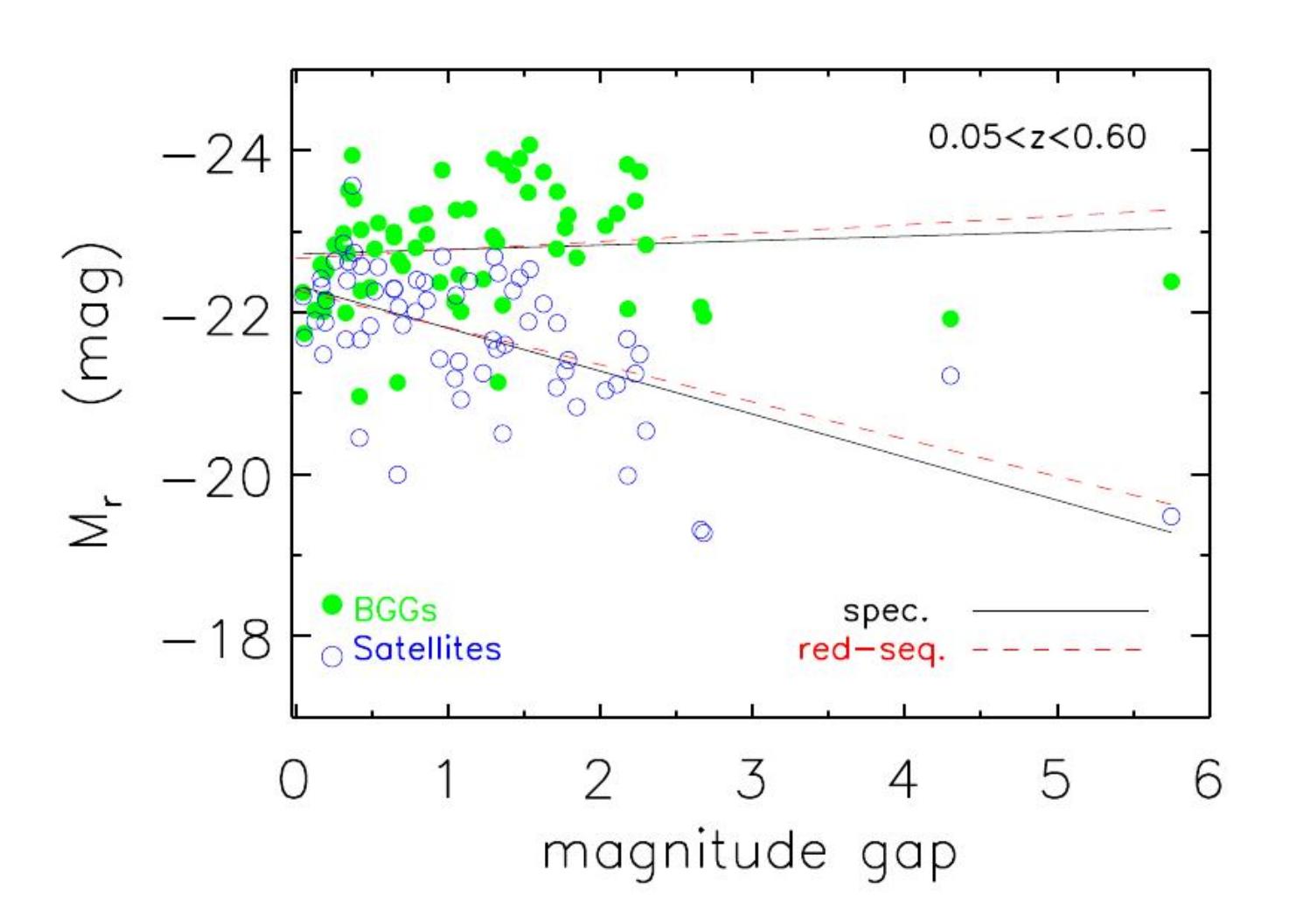}
           \end{center}
           \caption[flag]{The absolute r-band magnitudes of the first
             (filled green circles) and the second (open blue circles)
             brightest galaxies as a function of the magnitude gap for
             the COSMOS+AEGIS+XMM-LSS sample of groups, having spectroscopic
             members. The solid black and dashed red lines show linear
             fits for the spectroscopic and red sequence selections of
             group membership, respectively.  The {\it upper panel}
             and {\it lower panel} show the butterfly diagrams for
               high-z and low-z groups, respectively.  }
           \label{specbgg4}
           \end{figure}

\begin{figure}[ht!]
       \begin{center}  
       \leavevmode
       \includegraphics[width=9.5cm]{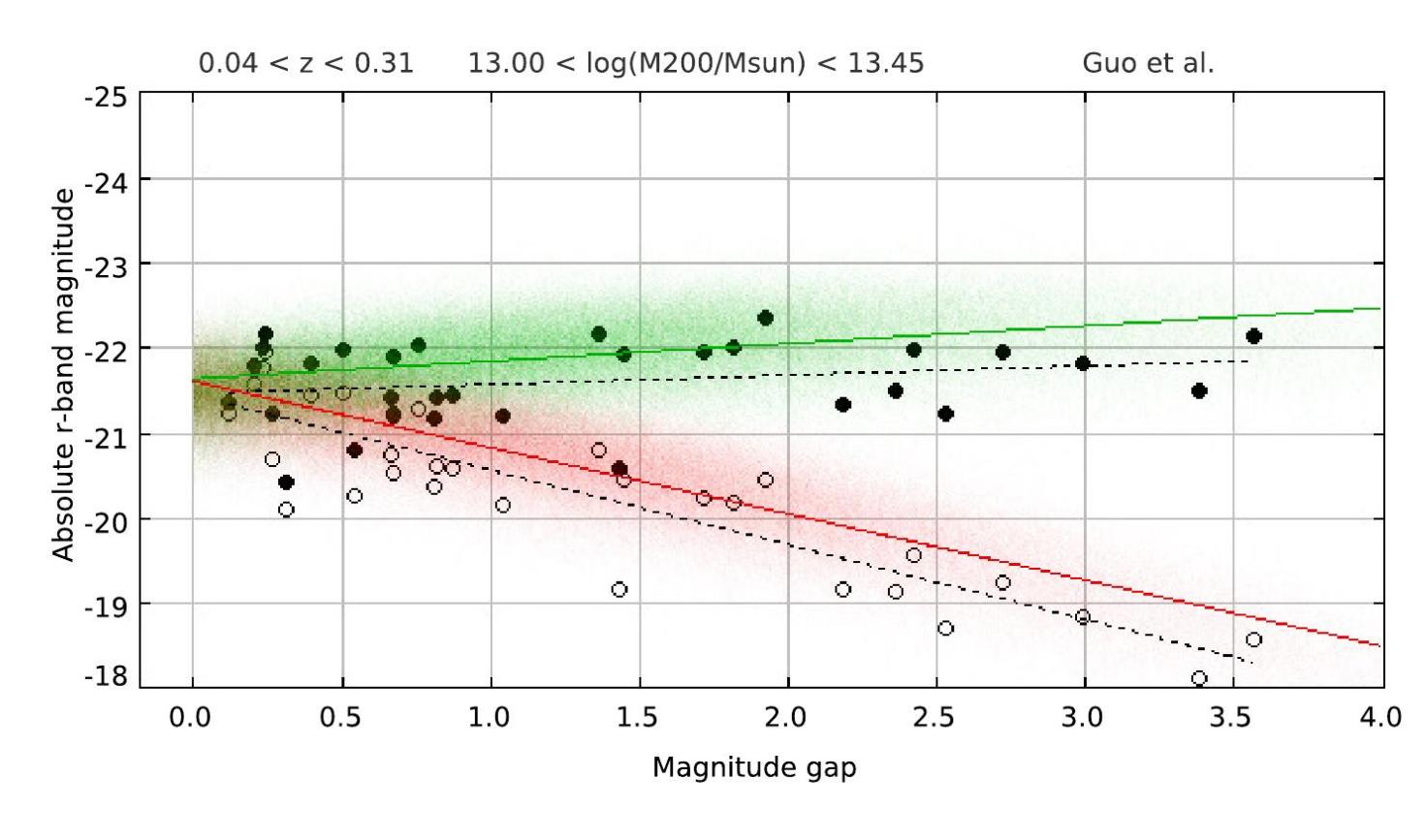}
       \includegraphics[width=9.5cm]{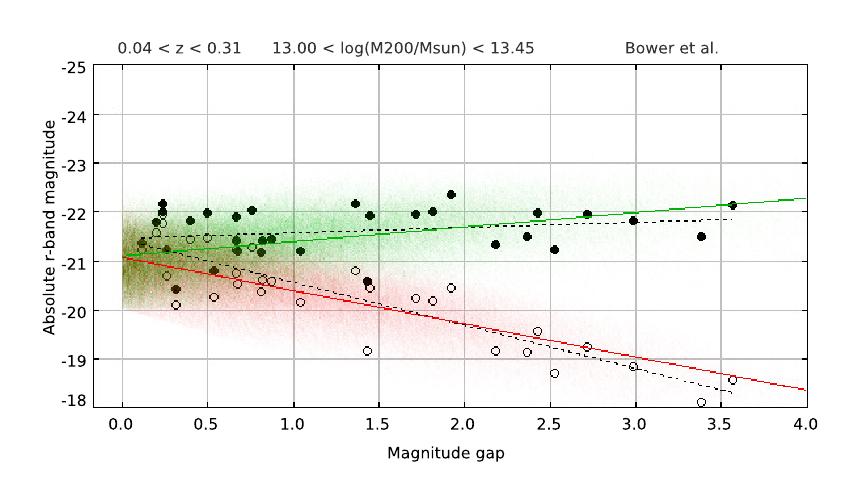}
       \includegraphics[width=9.5cm]{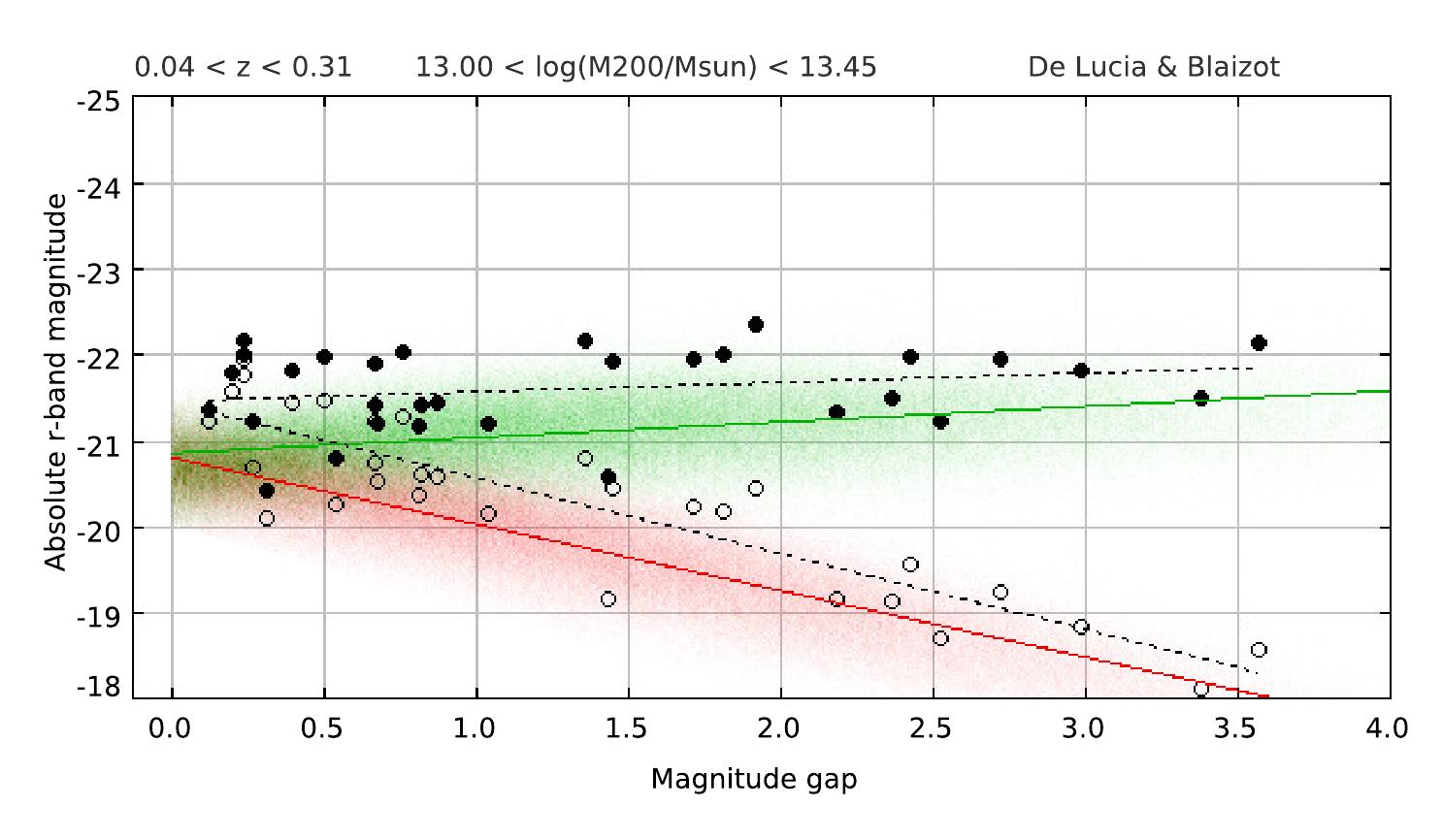}
       \end{center}
       \caption[flag]{The absolute r-band magnitudes of the first and
         the second brightest galaxies as a function of the magnitude
         gap. The S--I data are shown by filled (BCG) and open (2-nd
         ranked galaxy) black circles with the dashed black lines
         showing the linear fit. Each panel shows the results of one of
         the SAMs (G11--top, B06--middle, DLB07--bottom), with green
         (red) points and a line showing the individual points and the
         fit to the BCG (2-nd ranked galaxy).}
       \label{bgg1}
       \end{figure}

\begin{figure}[ht!]
      \begin{center}  
       \leavevmode                              
       \includegraphics[width=9.5cm]{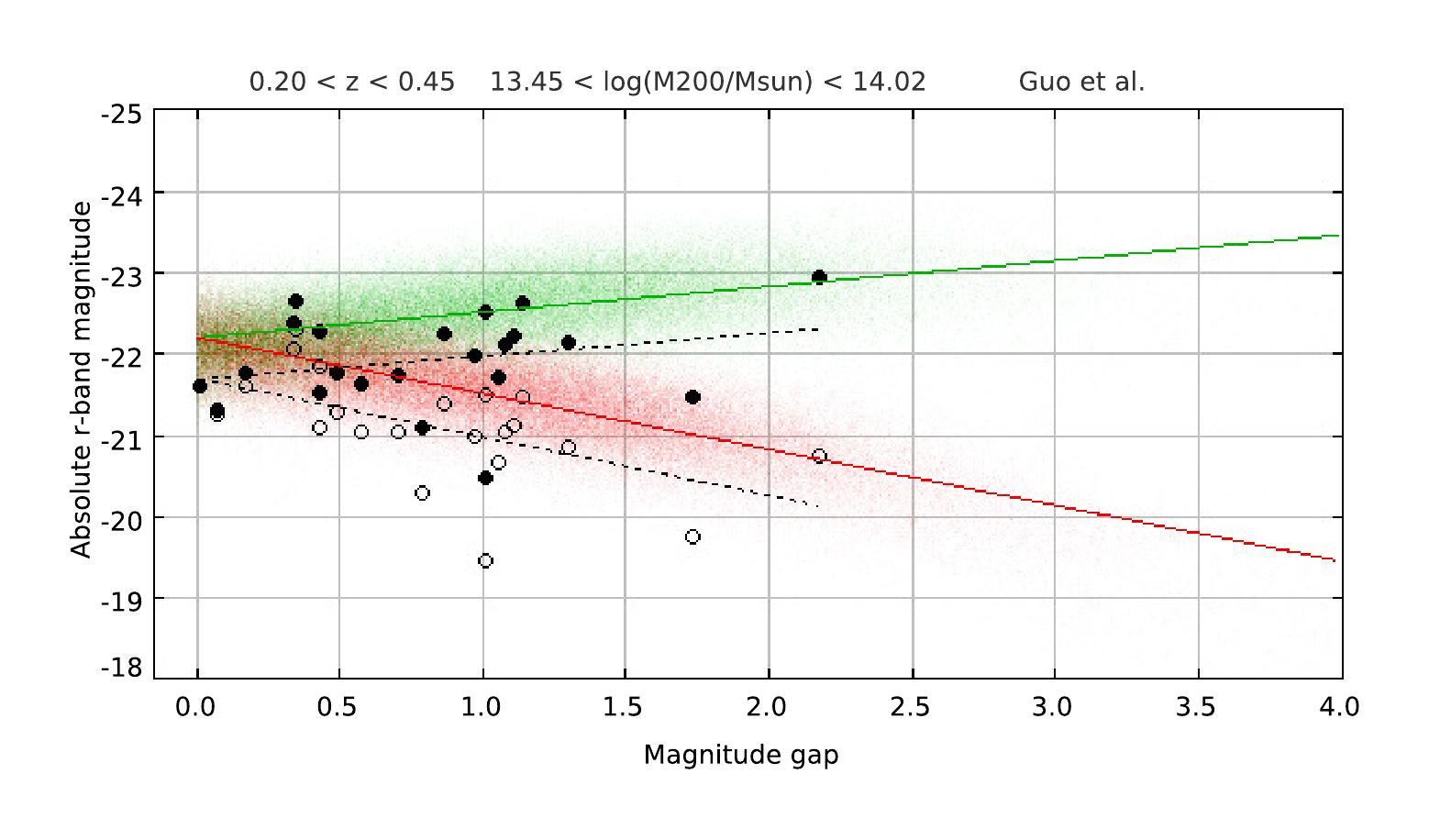}
       \includegraphics[width=9.5cm]{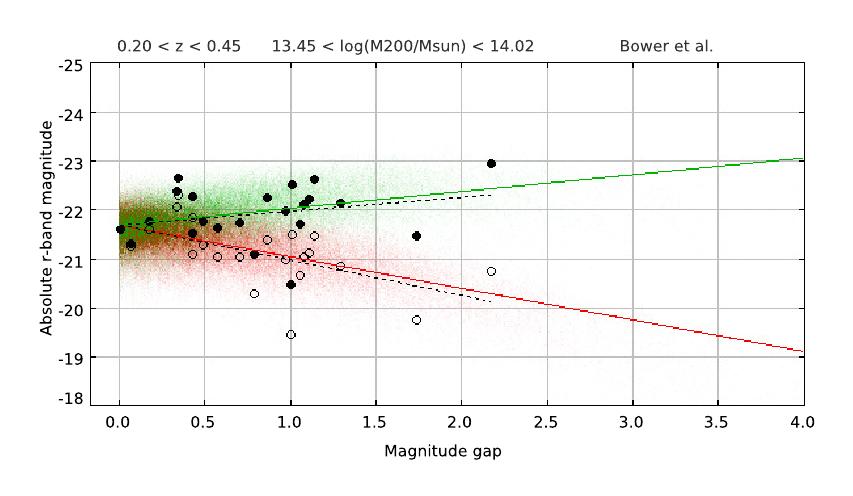}
       \includegraphics[width=9.5cm]{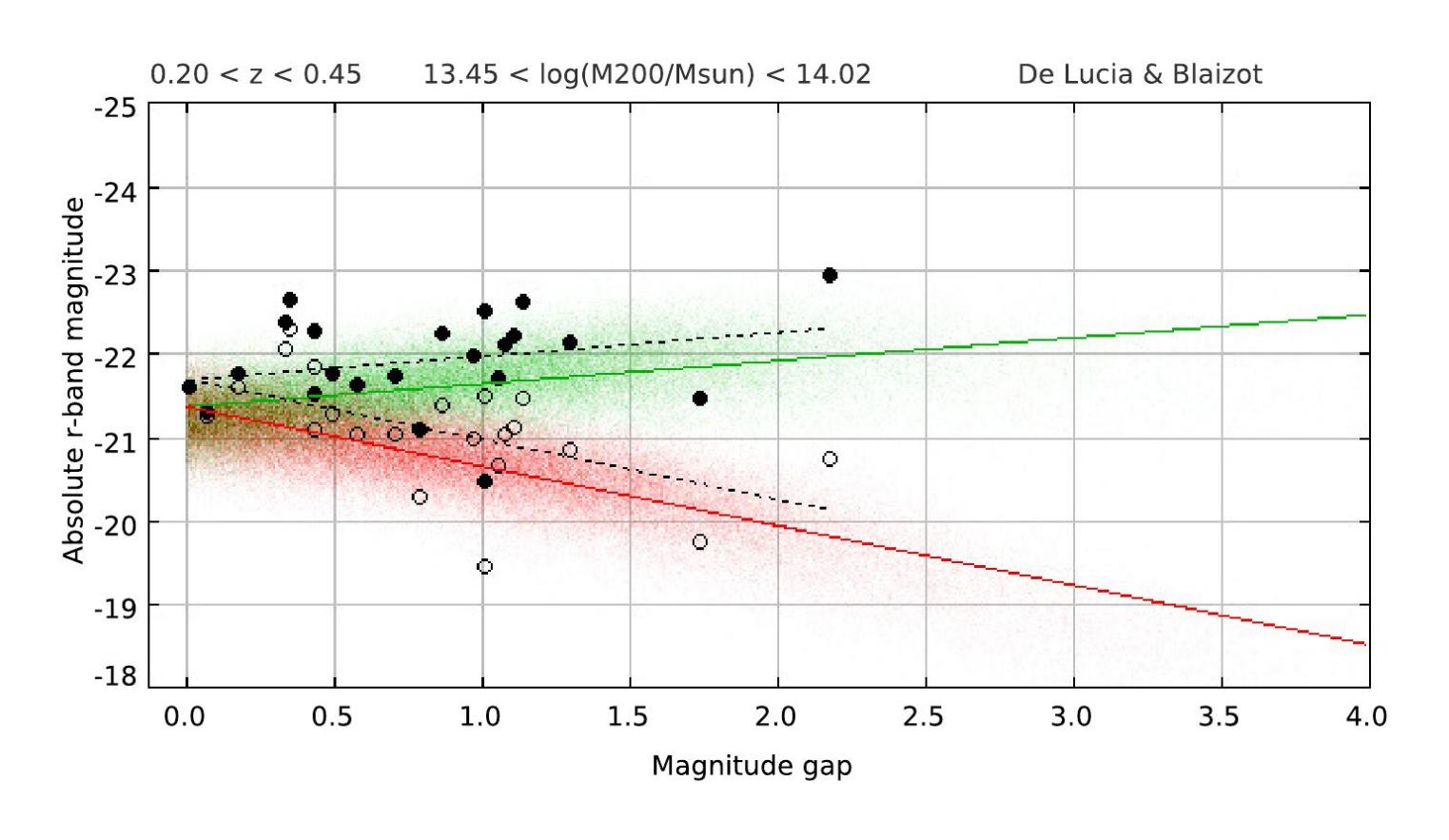}
       \end{center}
       \caption[flag]{Same as in Fig. \ref{bgg1} , but for the BCGs and second brightest satellites in  S--II.}
       \label{bgg2}
       \end{figure}

\begin{figure}[ht!]
     \begin{center} 
     \leavevmode \includegraphics[width=9.5cm]{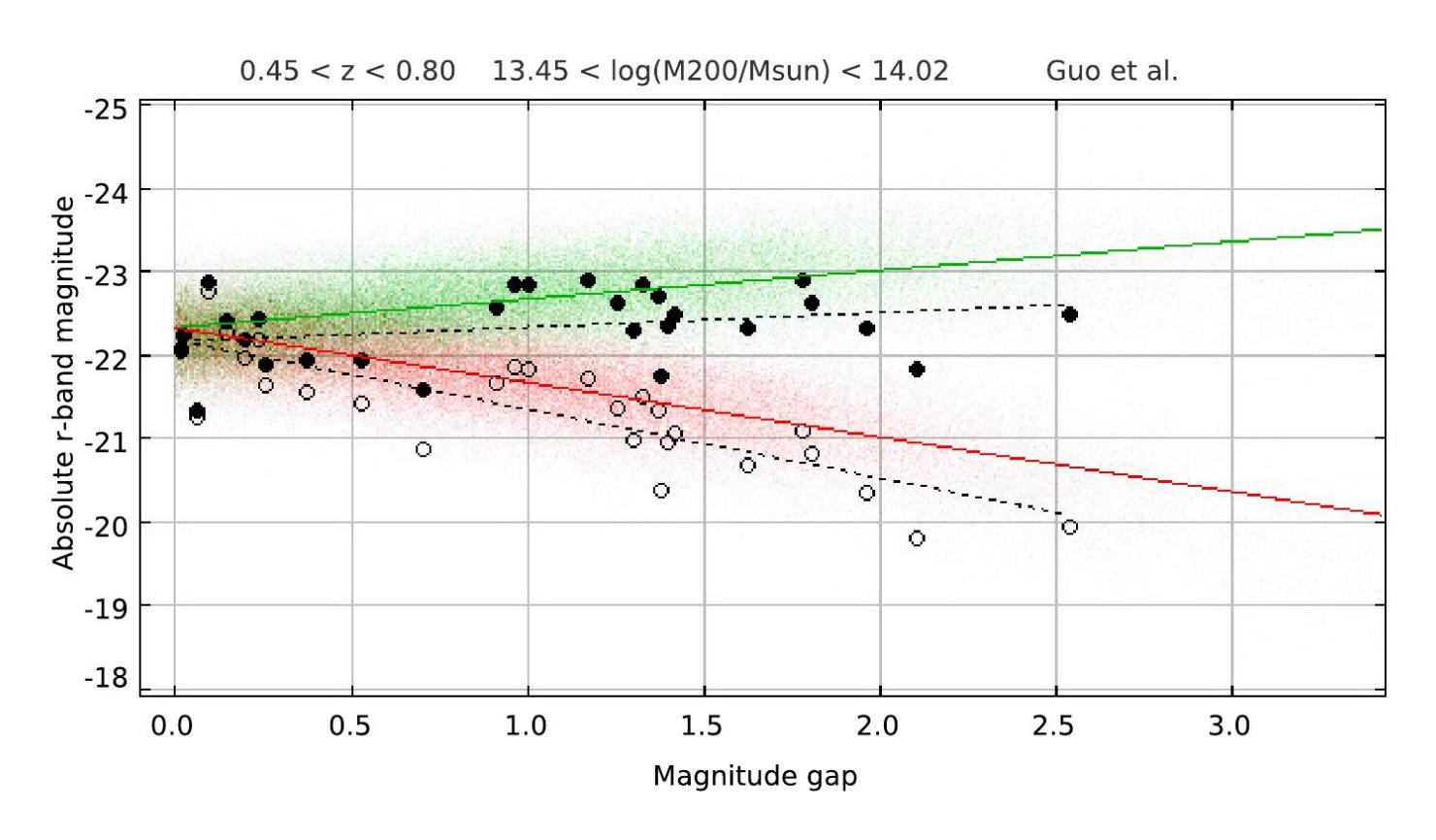}
     \includegraphics[width=9.5cm]{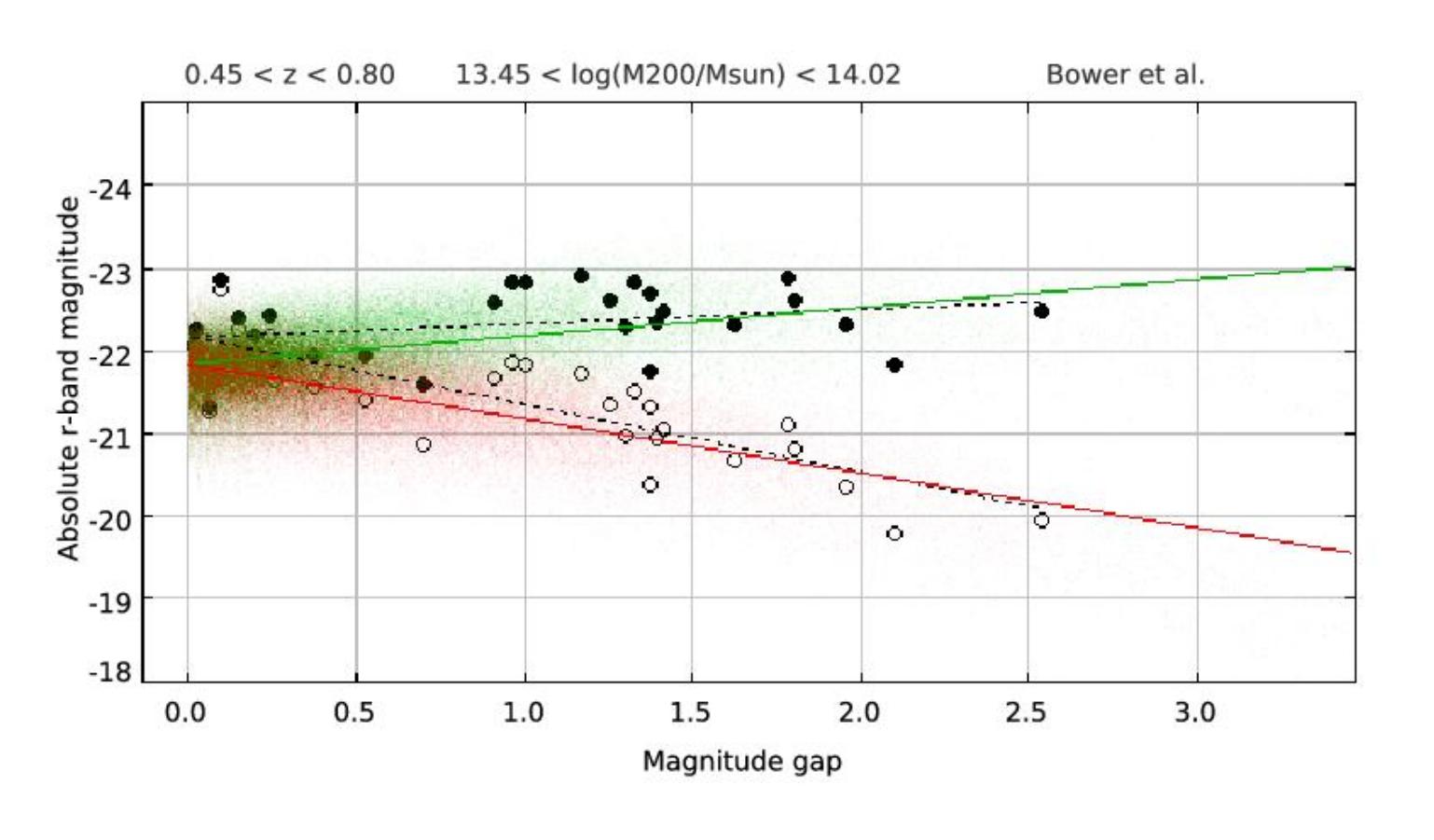}
     \includegraphics[width=9.5cm]{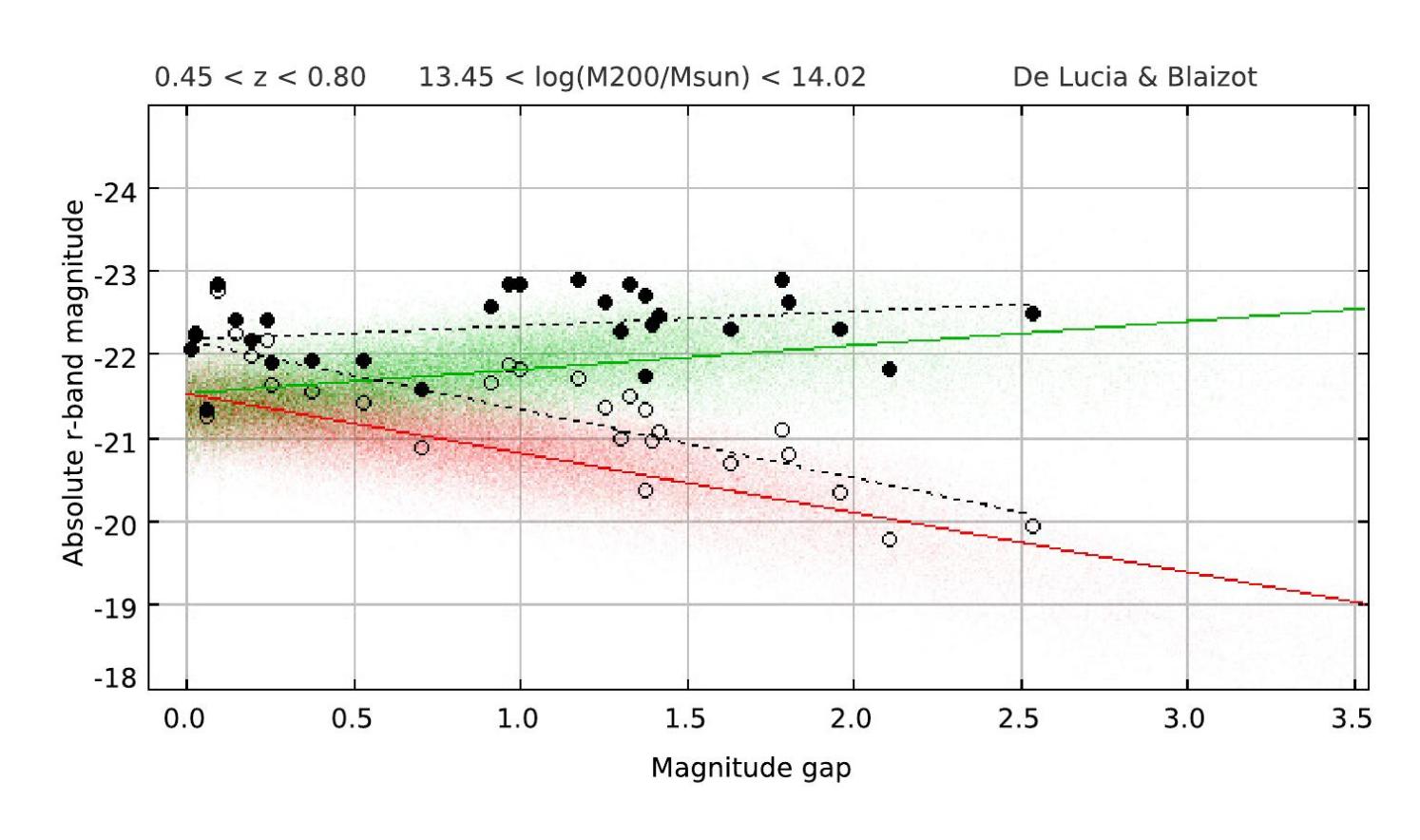} \end{center}
     \caption[flag]{Same as in Fig. \ref{bgg1}, but for the BCGs and second brightest satellites in S--III. } 
     \label{bgg3} 
\end{figure}

\begin{figure}[ht!]
\begin{center}  
\leavevmode
\includegraphics[width=9.5cm]{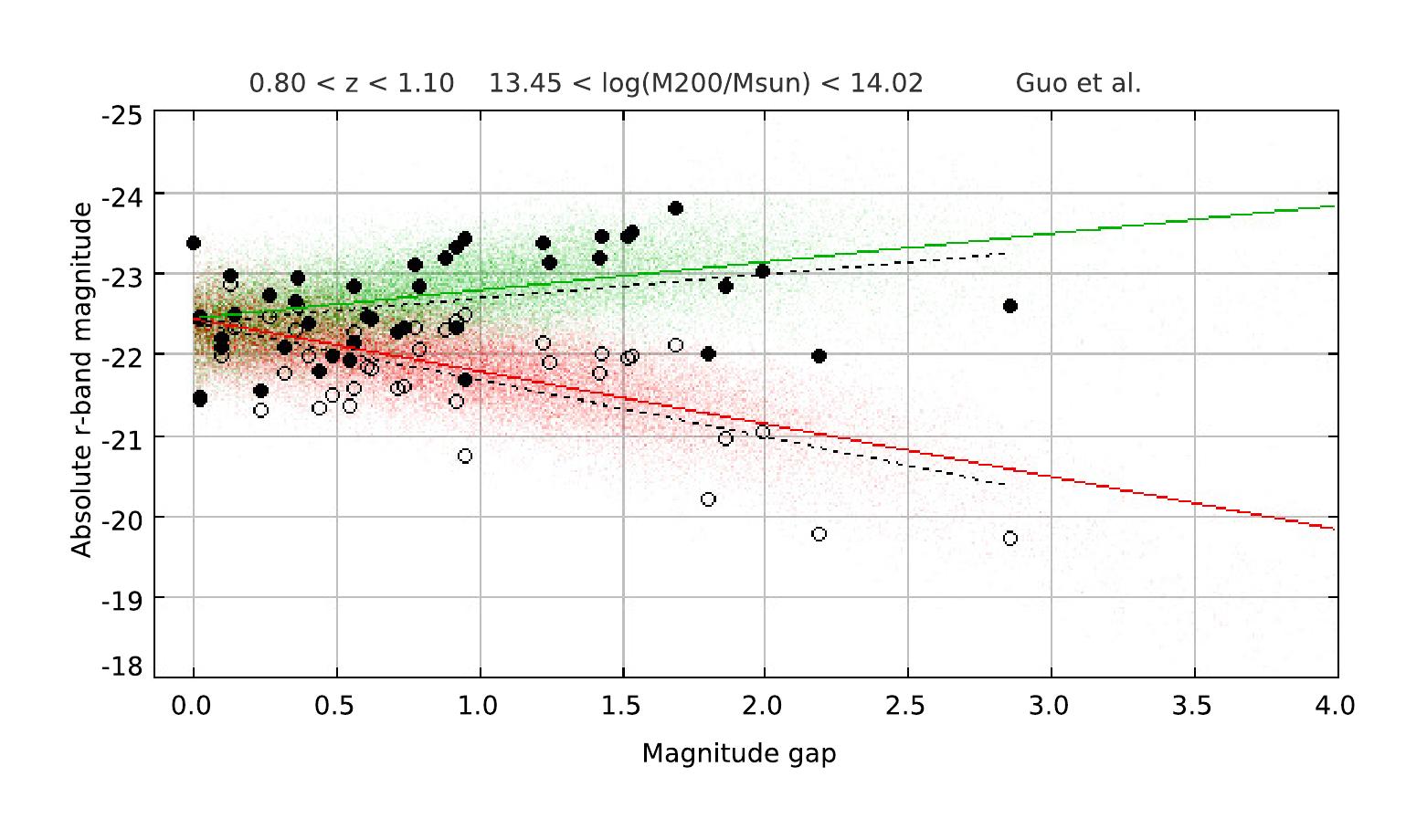}
\includegraphics[width=9.5cm]{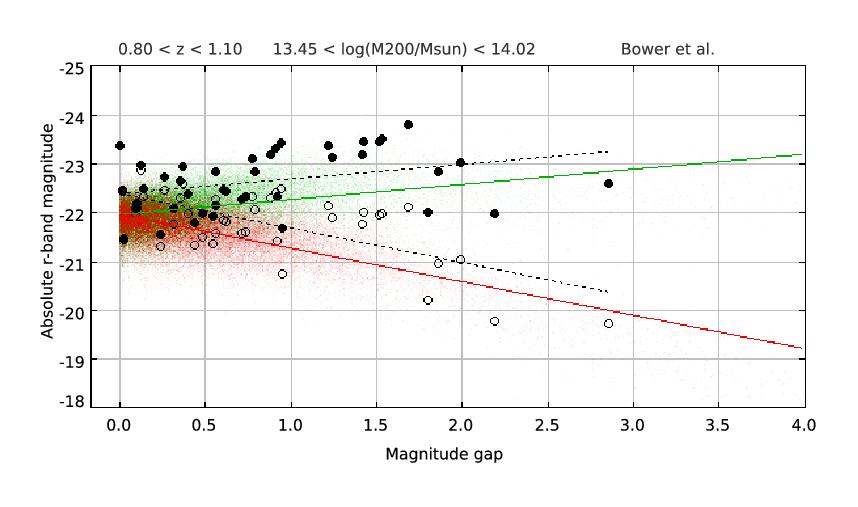}
\includegraphics[width=9.5cm]{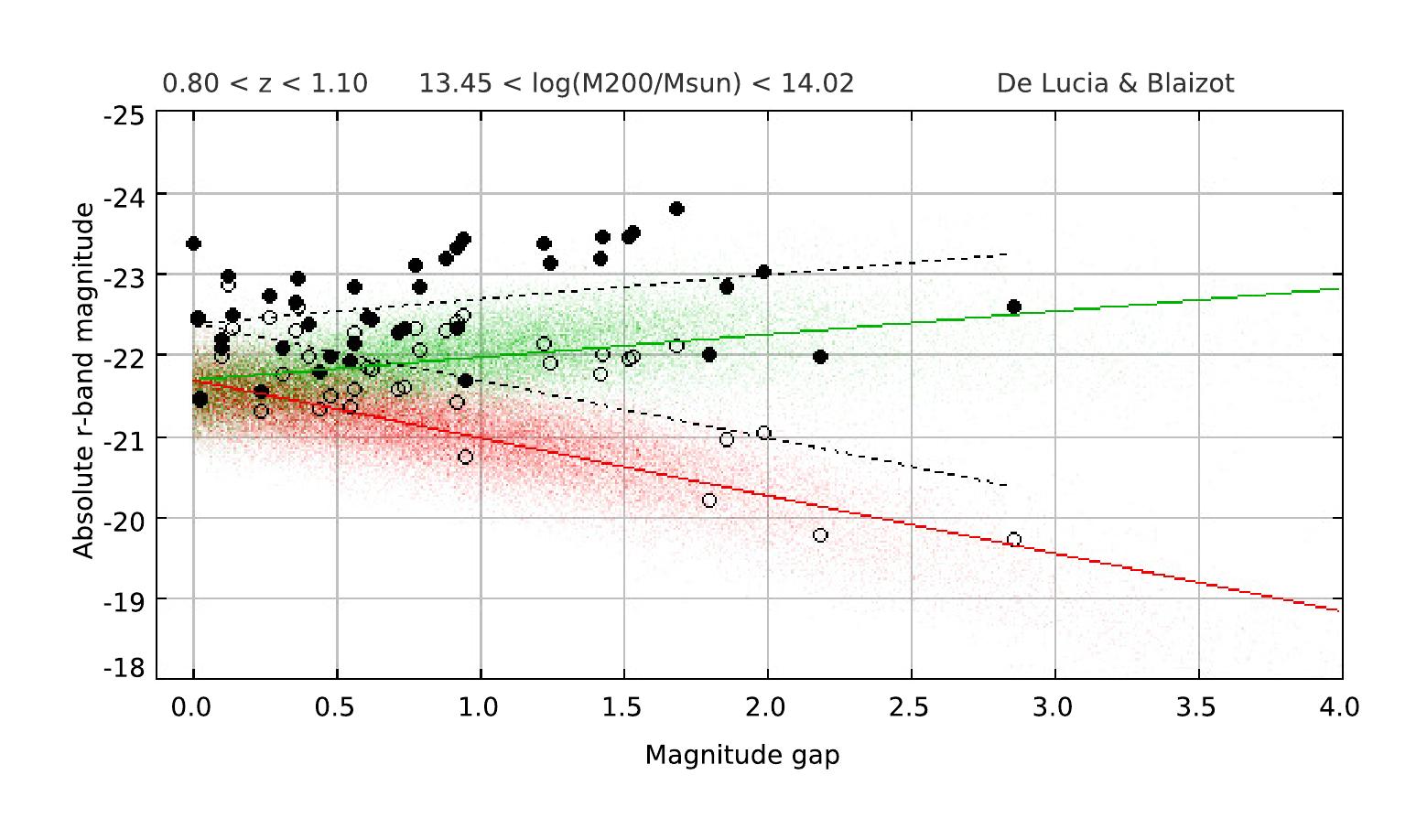} 
\end{center}
\caption[flag]{Same as in Fig. \ref{bgg1}, but for the BCGs and the second brightest satellites in S--IV. } 
\label{bgg4}
\end{figure}

 In Figs. \ref{specbgg4} to \ref{bgg4} we plot the absolute magnitudes
 of the first ($ M_{r,1} $) and the second ($M_{r,2}$) brightest
 galaxies against $\Delta M_{1,2}$. Previously, such butterfly
 diagrams have been used to quantify the comparison between
 observations and predictions of the SAMs below the redshift of
 $\sim$0.3 \citep{Smith10, Tavasoli11}. In this paper, we extend these
 studies to z$<$ 1.10.

 In \S 5.1 we examined whether contamination can affect the gap
 computations when the red-sequence method is applied for the
 selection of group membership versus the spectroscopic selection. In
 Fig. \ref{specbgg4} we examine this effect on the butterfly diagrams,
 which are constructed using the spectroscopic sample of the
 COSMOS+AEGIS+XMM-LSS groups at z$ < $0.6 (lower panel) and z$ > $0.6
 (upper panel). The redshift binning is limited by the sample size.
 In Fig. \ref{specbgg4} the filled green and open blue circles show
 the absolute r-band magnitudes of the spectroscopically selected BCGs
 and satellites as a function of $\Delta M_{1,2}$, respectively. These
 are quantified by linear regressions (solid black lines), $ M_{r,1} =
 a_{1}+b_{1}\Delta M_{1,2}$ and $ M_{r,2}=a_{2}+b_{2}\Delta
 M_{1,2}$. We have compared the results of these linear fits with
 those for the red-sequence selected BCGs and satellites (dashed red
 lines). Coefficients of fits are presented in Tab. \ref{bgg_param}.
 This comparison shows that the slopes of the red-sequence selection
 are slightly shallower than those for the spectroscopic membership
 and the membership contamination is a little more effective for the
 high-z groups compared to the low-z groups. However, the trends in
 the $M_{r,1} $ - $\Delta M_{1,2}$ and $M_{r,2} $ - $\Delta M_{1,2}$
 relations for both red-sequence and spectroscopic members at z$ <
 $0.6 and z$ > $0.6 are similar and differences between the
 slopes/intercepts lie within the corresponding uncertainties. Thus,
 we expect that our findings which are derived from butterfly diagrams
 of our XMM-LSS catalog are valid within observational errors.
                                                         
 In Fig. \ref{bgg1}, we compare the observed $ M_{r,1} $--$\Delta
 M_{1,2}$ and $ M_{r,2} $--$\Delta M_{1,2}$ relations with model
 predictions of G11 (top panel), B06 (middle panel) and DLB07 (bottom
 panel) for S--I. The observed luminosity of the first-ranked galaxy
 (filled black circles) increases very slowly with $\Delta M_{1,2}$,
 so the difference in the magnitude is due to the second-ranked
 galaxies (open black circles), which declines from $M_{r}\sim-21.5 $
 at $\Delta M_{1,2}\sim$0 to $M_{r}\sim$-18 at $\Delta
 M_{1,2}\sim$3.5. The values of the linear fit coefficients are given
 in Tab. \ref{bgg_param}. For S--I, the galaxies in the G11 model are
 more luminous than the observed galaxies by $\sim$0.3 mag. In
 contrast, galaxies in the B06 and DLB07 models tend to be less
 luminous than the observed galaxies by $\sim$0.5 to 0.7 mag.

 For the S--II, a similar comparison is presented in
 Fig.\ref{bgg2}. The increase in the absolute magnitude of the BCG
 with $\Delta M_{1,2}$ is $\approx$3 higher than in S--I, reaching
 -22.5 at $\Delta M_{1,2}\sim$3.5. The absolute magnitude of the
 satellite galaxy declines from $M_{r}\sim$-21.5 at $\Delta
 M_{1,2}\sim$0 to $M_{r}\sim$-19 at $\Delta M_{1,2}\sim$3.5. As in
 S--I, the galaxies in the G11 model tend to be 0.5 mag more luminous
 than the observed ones. The B06 model predicts reasonably well the
 observed luminosity of the first-ranked and second ranked galaxies as
 a function of $\Delta M_{1,2}$. The predicted luminosities of
 galaxies in the DLB07 model are less luminous than the observed ones.
     
We show the butterfly diagrams for S--III in Fig. \ref{bgg3}. As for the low-z
subsamples, the luminosity of the BCG rises very slowly with $\Delta
M_{1,2}$, spanning a range of -23$\lesssim M_{r}\lesssim$-21 with
$\Delta M_{1,2}$.  While, the luminosity of the satellite galaxy
decreases from $M_{r}\sim$-22 at $\Delta M_{1,2}\sim$0 to
$M_{r}\sim$-19 at $\Delta M_{1,2}\sim$ 3.5.

Finally, Fig. \ref{bgg4} illustrates the luminosities of BCG and 2-nd
ranked galaxy as a function of magnitude gap for S--IV. The BCG
absolute magnitudes span a range of -24$\lesssim M_{r}\lesssim$-21.5
 and increase slowly with $\Delta M_{1,2}$. The luminosity of the
second ranked galaxy declines from $M_{r}\sim$-22.5  at $\Delta
M_{1,2}\sim$0 to $M_{r}\sim$-20 at $\Delta M_{1,2}\sim$3.5.

   \setcounter{table}{2}
      \begin{table*}[ht!]
        \caption{\footnotesize
       The Best linear fit to the $ M_{1,r}$ and $ M_{2,r} $ as a function of $ \Delta M_{1,2} $. The four upper rows present coefficients of the linear fit to the butterfly diagrams of our catalog and four lower rows list those for the sample of 84 groups from COSMOS+AEGIS+XMM-LSS.   
        \label{bgg_param}}
        \tiny
        \centering
        \renewcommand{\arraystretch}{0.9}\renewcommand{\tabcolsep}{0.12cm}
         
        \begin{tabular}[tc]{ccccc ccccc cccc cc}
         \hline
        \hline\\
       Subsample   &  Mean redshift& a1 & b1 & a2 & b2     \\
       \hline \\
       XMM-LSS (our catalog)\\
                  S--I  & 0.175  &   $  -21.44 \pm   0.11 $ & $ -0.11 \pm  0.10 $ & $  -21.42 \pm 0.17 $ & $ 0.88\pm 0.11 $   \\
        S--II   &  0.325  &  $ -21.66 \pm  0.22 $ &  $  -0.28 \pm 0.15 $ & $ -21.67 \pm 0.22 $ & $ 0.72 \pm 0.20 $ \\
       S--III &   0.625  &   $   -22.13 \pm  0.15 $ &  $ -0.19\pm 0.12 $  & $ -22.14  \pm 0.15 $ &  $ 0.82 \pm 0.13 $  \\
        S--IV  &  0.95  &   $ -22.35 \pm  0.15 $ &  $  -0.30\pm 0.08 $ & $ -22.35  \pm 0.16 $ & $ 0.70 \pm 0.16 $ \\ 
       COSMOS+AEGIS+XMM-LSS  \\           
         red-sequence selection  & 0.3  &  $ -22.67\pm 0.15 $   & $ -0.10\pm0.10 $ & $ -22.28\pm0.14$ & $ 0.46\pm0.10 $\\
          spectroscopic selection &  0.3  &    $  -22.72\pm0.14 $ &  $ -0.05\pm0.09 $ &$ -22.34\pm0.14 $&   $ 0.53\pm0.09 $    \\ 
         
           red-sequence selection  & 0.9  &  $ -23.67\pm 0.23 $   & $ 0.16\pm0.34 $ & $ -23.69\pm0.23$ & $ 1.26\pm0.33 $\\
            spectroscopic selection &  0.9  &    $  -23.86\pm0.21 $ &  $ 0.37\pm0.26 $ &$ -23.83\pm0.20 $&   $ 1.48\pm0.25 $    \\

       \hline
       
       \end{tabular}
       \end{table*}

\subsubsection{Redshift evolution of the butterfly diagram } 

\begin{figure}[ht!]
  \begin{center}  
  \leavevmode
  \includegraphics[width=9cm]{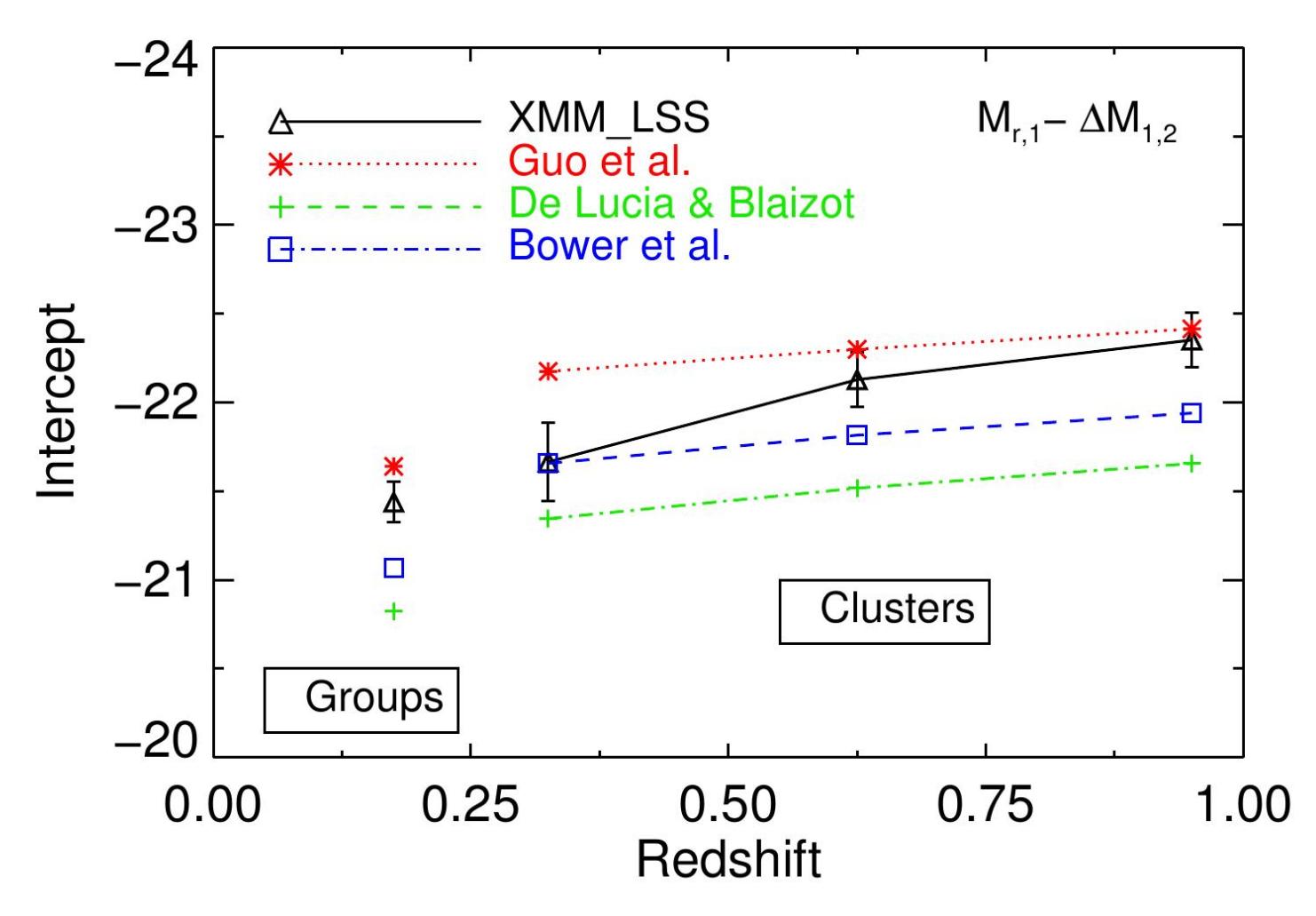}
  \includegraphics[width=9cm]{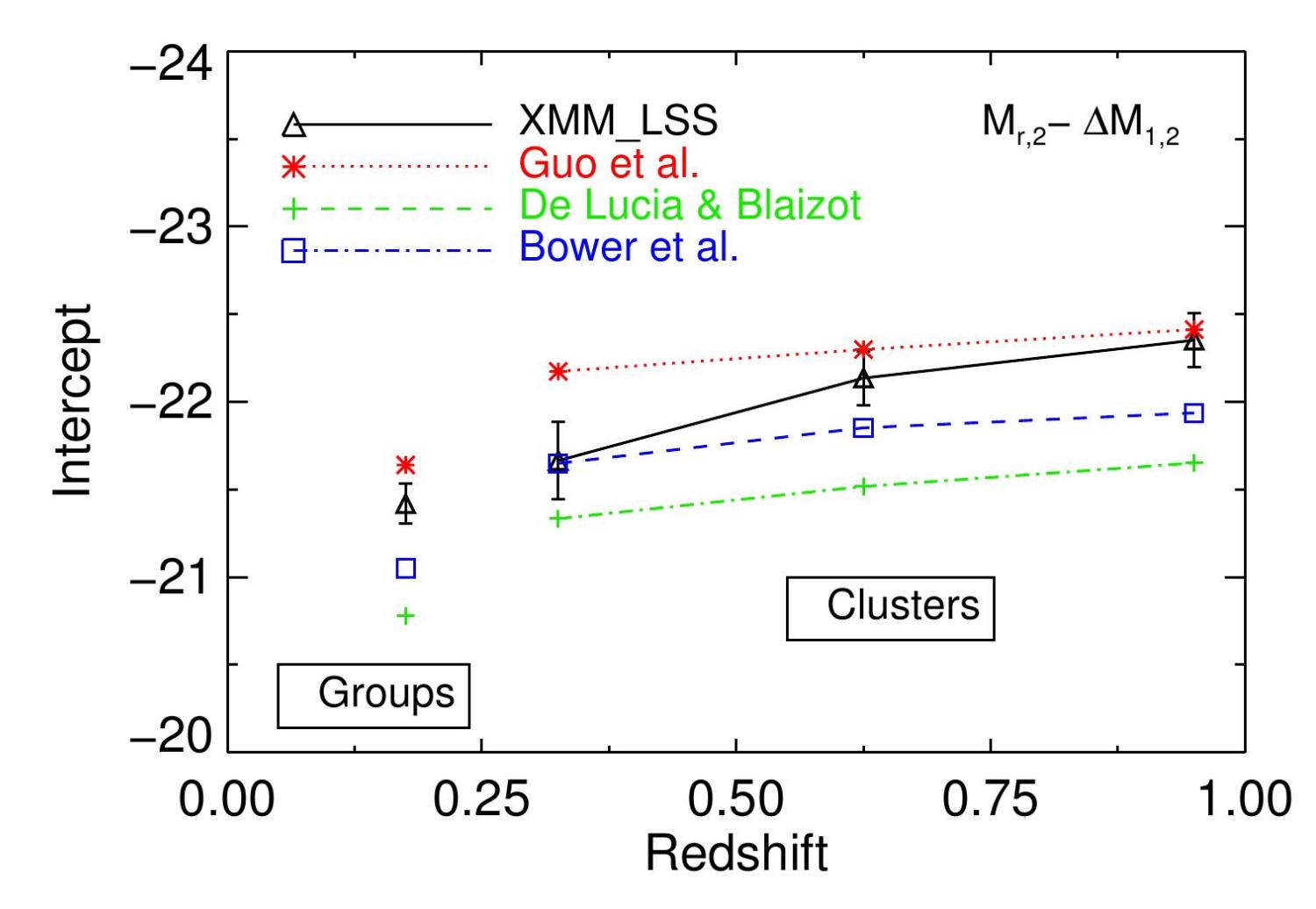}
  \end{center}
  \caption[flag]{ Evolution of the zero point of the $ M_{r,1}
    $--$\Delta M_{1,2}$ ({\it upper panel.})  and $ M_{r,2} $-- $\Delta
    M_{1,2}$ ({\it bottom panel.}) relations with redshift.  The solid
    black curve with error bars represents the data. Dotted red, dashed
    blue and dash-dotted green curves show the model predictions of G11, B06
    and DLB07, respectively. In contrast to the models, we observe a
    significant evolution with redshift.}
  \label{zerop}
  \end{figure}

In Fig.\ref{zerop}, we compare the redshift evolution of the intercept
of $ M_{r,1} - \Delta M_{1,2}$ (top panel) and $ M_{r,2} - \Delta
M_{1,2}$ (bottom panel) relations in our observations and the
models. We find a significant negative redshift evolution of the
intercepts for BCGs and their satellites by $ \sim$0.8 mag for
clusters at z$\gtrsim $0.25. All models exhibit a shallower evolution
of the intercepts with redshift compared to the data. A steeper
negative evolution in the BCG magnitudes in time, is most likely due
to younger stellar population age of BCGs, compared to the models. The
intercept of the predicted butterfly diagram for S--I at low redshifts
tend to be closer to G11 compared to two other models, indicating
importance of their advances in the modeling. A possible reason for
over-prediction of the BCG or BGG luminosity in G11 is that this model
incorporates a disruption mechanism which provides additional gas and
metal-rich material for the hot gas atmosphere of BCG, thus
increasing the cooling rate and star formation in BCG (see G11). To
achieve a good agreement with observations, it may still be needed to
increase slightly the strength of AGN feedback in this model.
 
\begin{figure}[ht!]
  \begin{center}  
  \leavevmode
  \includegraphics[width=9cm]{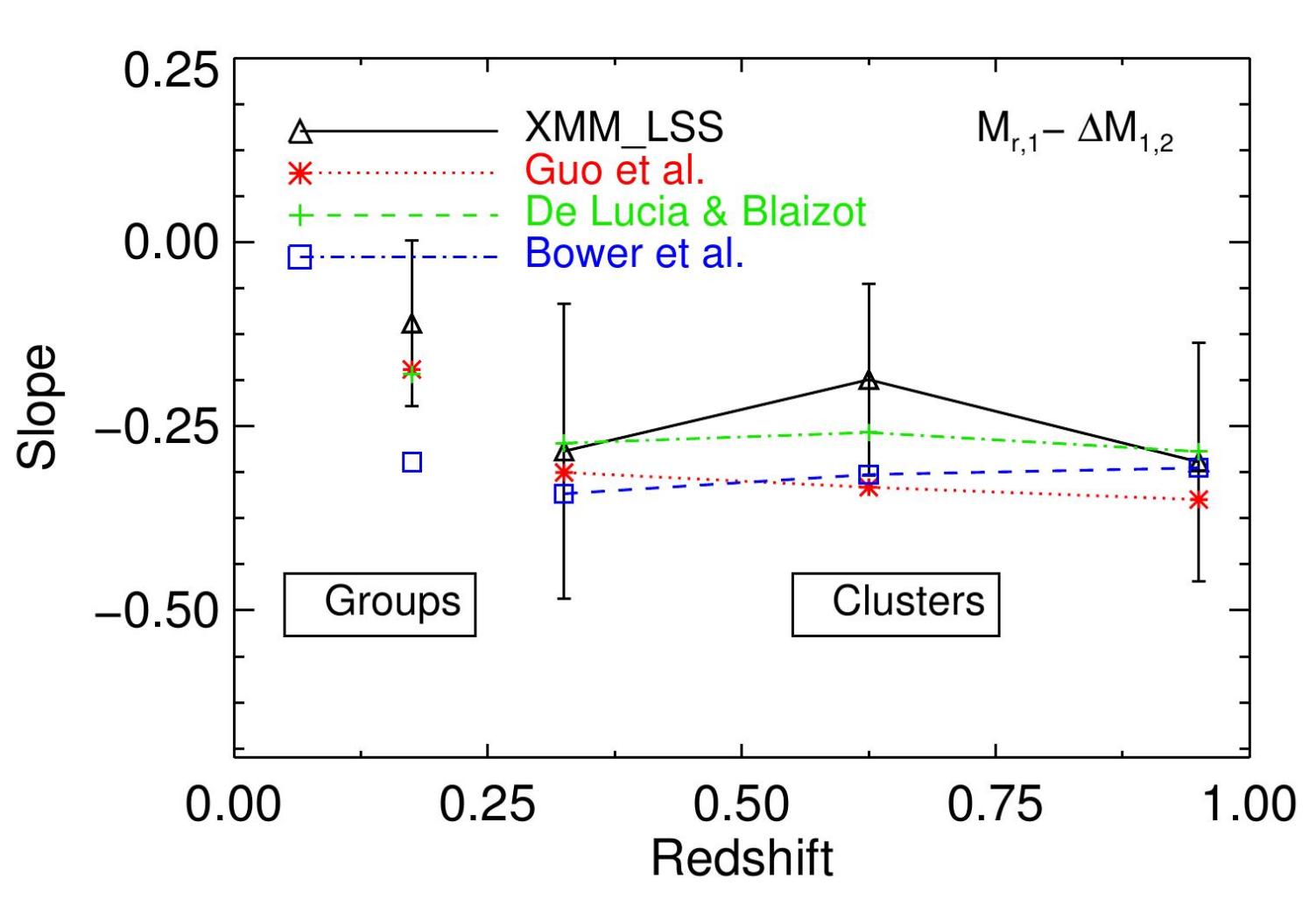}
  \includegraphics[width=9cm]{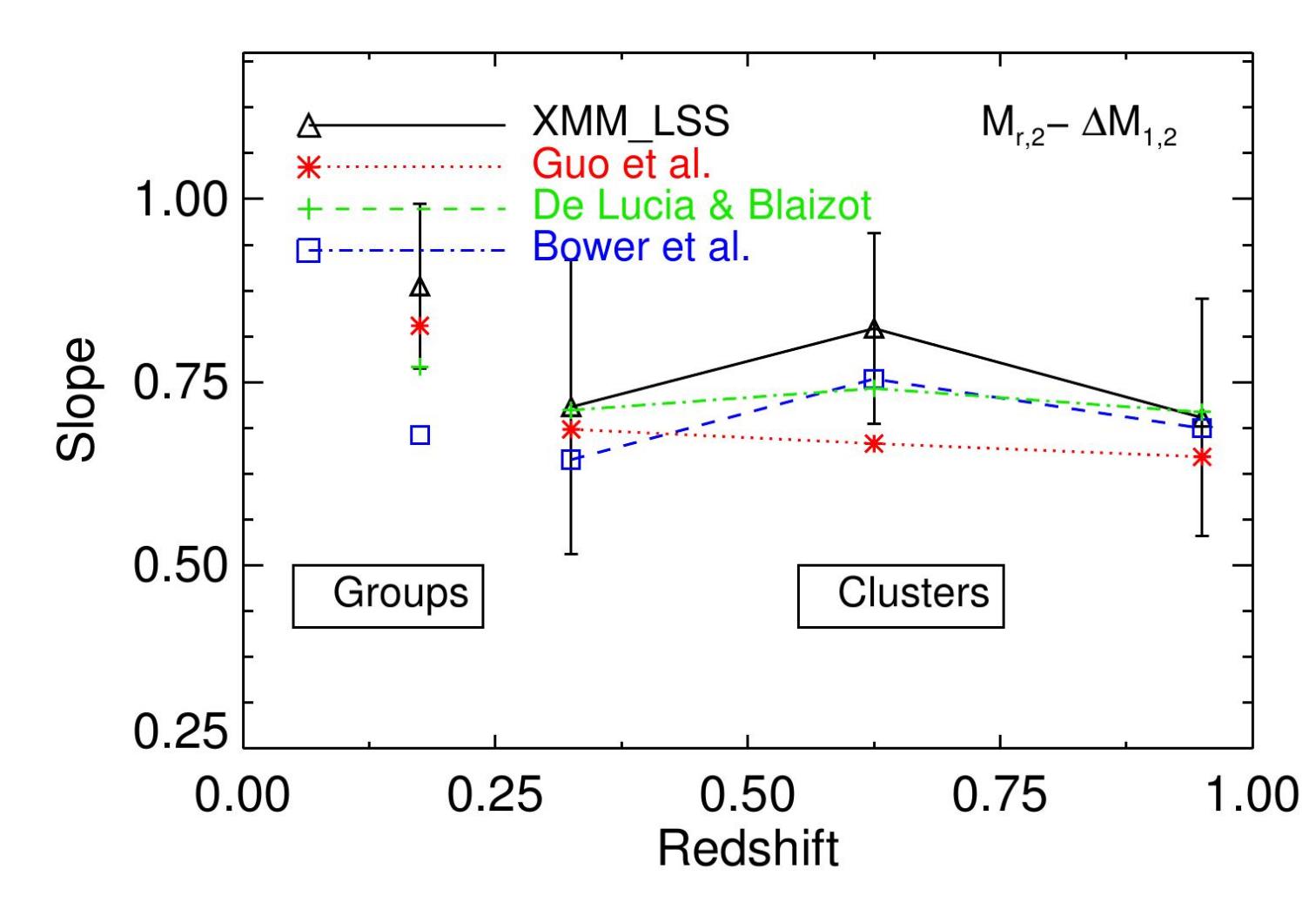}
  \end{center}
  \caption[flag]{ Evolution of the slopes of the $ M_{r,1} $--$\Delta
    M_{1,2}$ ({\it upper panel.}) and $ M_{r,2} $-- $\Delta M_{1,2}$
    ({\it bottom panel.})  relations with redshift. The models and the
    data reveal a nearly constant evolution with redshift.}
  \label{slope}
  \end{figure}

In Fig. \ref{slope}, we show a lack of the redshift evolution in the
slope of the butterfly diagram. No dependence on the absolute
magnitude of BCG and an agreement between all models, indicates that
the mechanism for creating large gaps is not strongly affected by the
star-formation and feedback, and is due to disruption of
galaxies. However, as we discussed above, the models show differences
in the fraction of high gap systems, which is controlled by the
efficiency of the transformation.
    
\subsection{Fossil groups }
 
Using the magnitude gap, we identify 23 fossil groups in our catalog.
The fossil groups constitute 22.2$\pm$6\% and 12.3$\pm $7.0\% of all
identified X-ray groups in our catalog at z$\leq$ 0.6 and z$>$0.6,
respectively. In this calculation we include the impact of the
contamination (when using the red-sequence method versus the
spectroscopic approach to select group membership) and completeness
effects on the gap measurements.  This misclassification causes the
fraction of fossil groups to be underestimated by $ \pm20 $\% at
high-z, which we add to the errors. Although the significance of the
redshift evolution of the fraction of the fossil groups is marginal,
it is also found in a nearly independent sample of the
COSMOS+AEGIS+XMM-LSS groups, which contains 23.1$ \pm $7\% and
10.5$\pm$8\% of all groups as fossil groups at z$\leq$ 0.6 and
z$>$0.6, respectively.
 
We mark the fossil groups in Fig. \ref{mass_z} with red circles to
illustrate their occurrence as a function of total mass and redshift.
Given the previous results on the prevalence of cool cores among the
fossil groups, in the following we verify that we do not have a
preferential selection of the fossil groups. Given the limited
statistics of the X-ray data, we select the extent of X-ray detection
as a parameter of the comparison. If the extends are much smaller, it
would indicate that the central part of the emission dominates the
detection and the X-ray selection of fossils is different to the bulk
of the groups.

\begin{figure}[ht!]
 \begin{center}  
 \leavevmode
 \includegraphics[width=9cm]{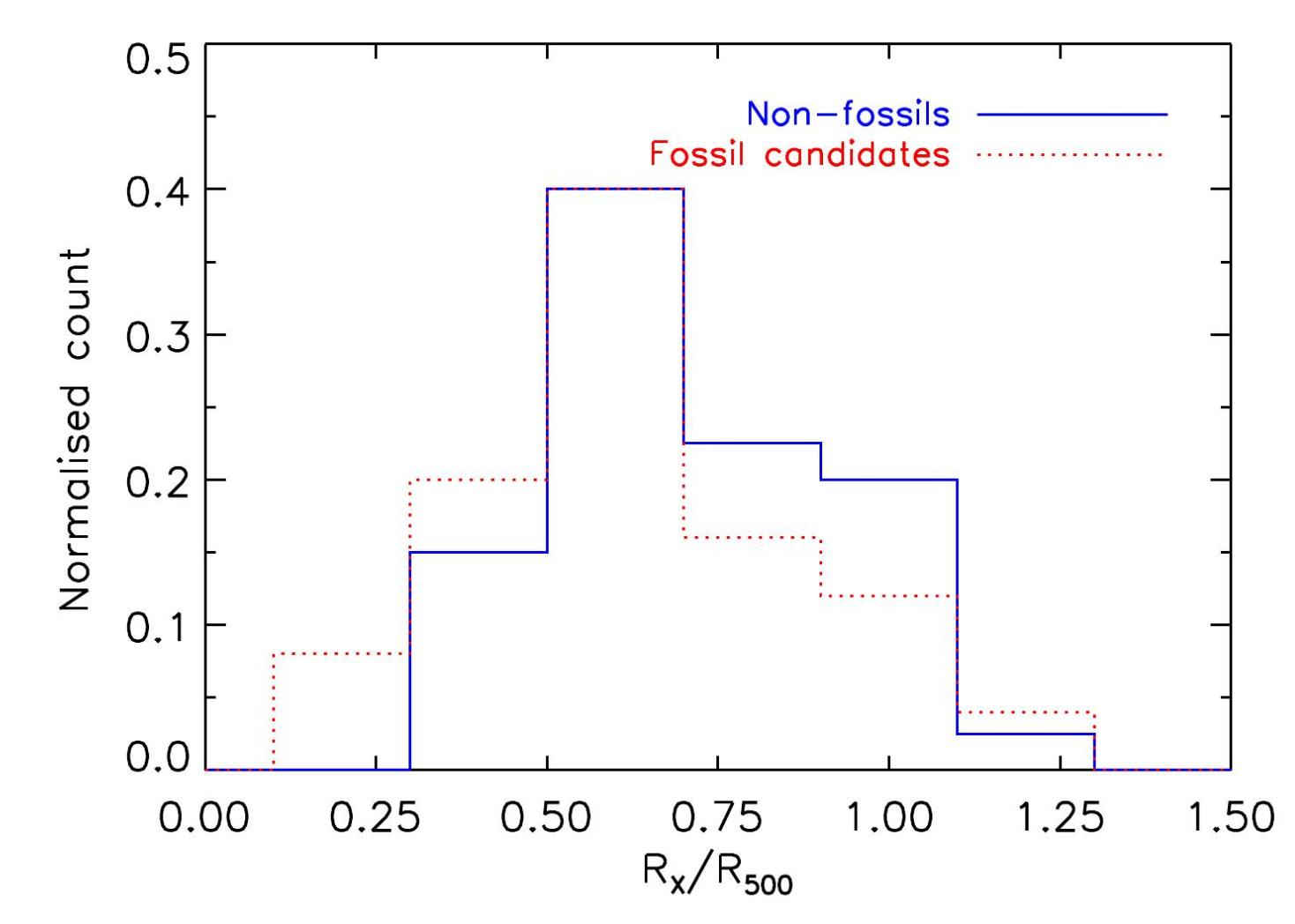}
 \includegraphics[width=9cm]{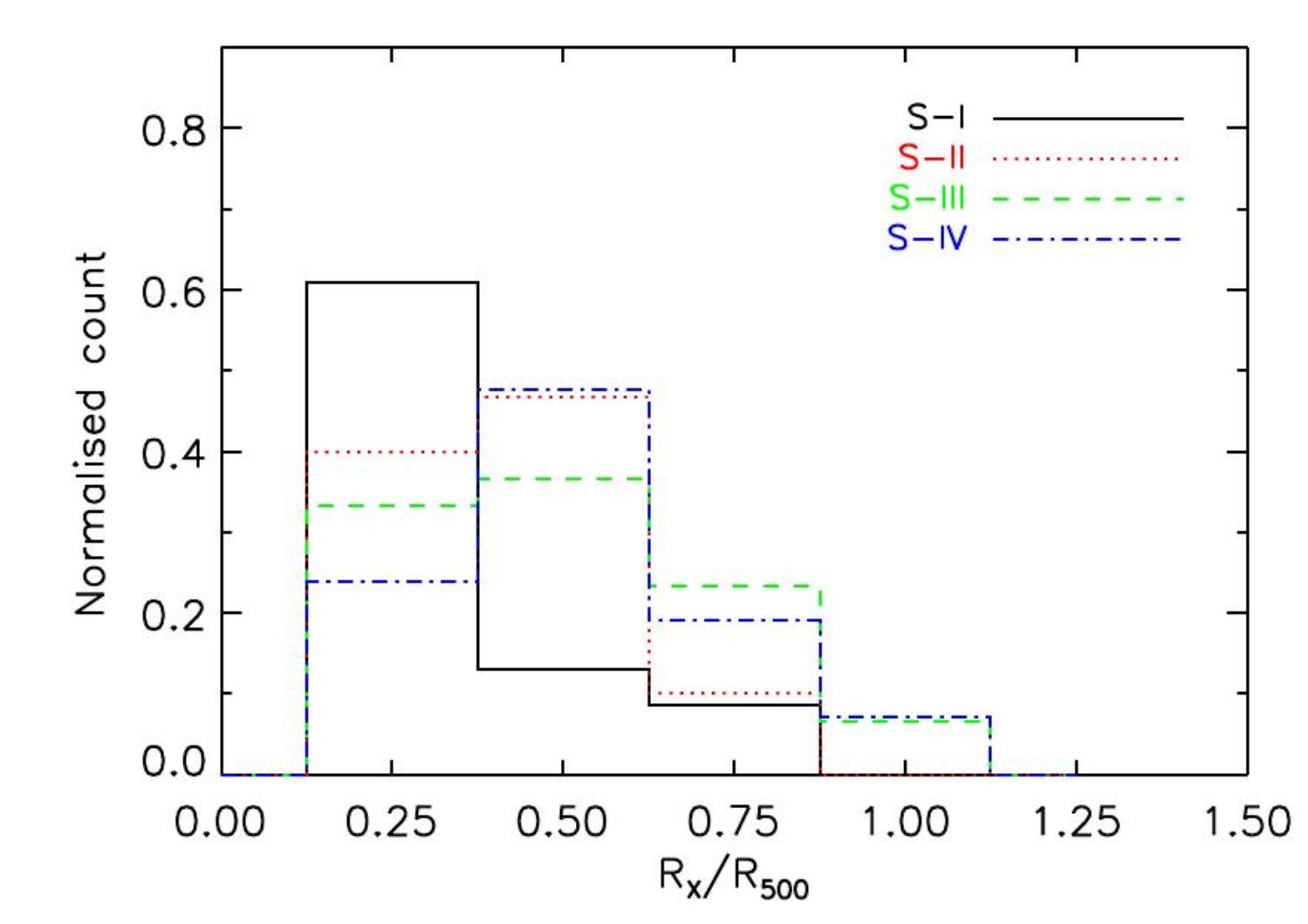}
 \includegraphics[width=9cm]{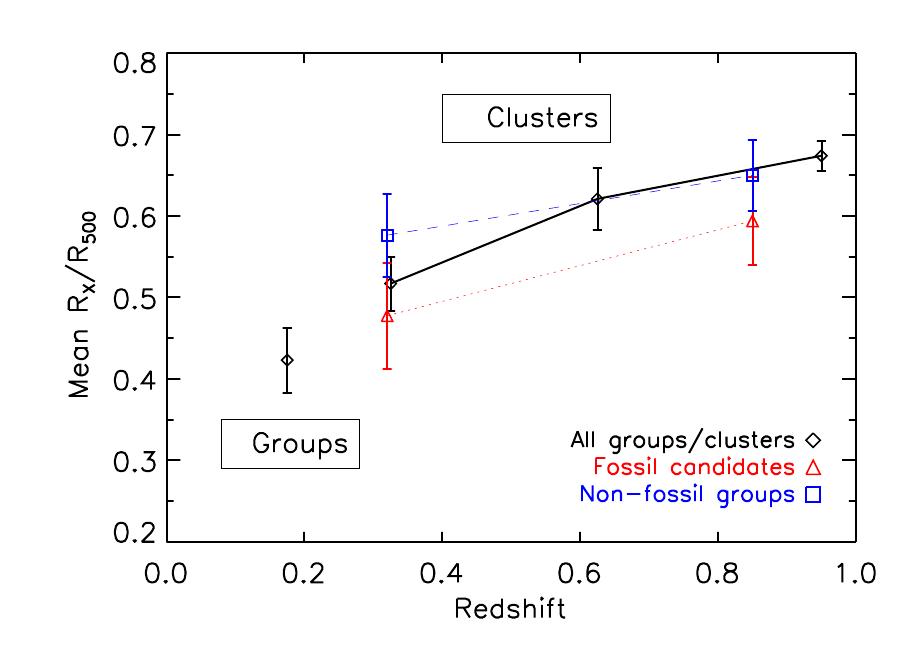}
 \end{center}
 \caption[flag]{{\it Upper panel.} Distribution of the ratio of the
   X-ray extent of the detection to the total extent of X-ray
   emission, $ \frac{R_{X}}{R_{500}} $ for the fossil group candidates
   ($\Delta M_{1,2} \ge$1.7) (dotted blue histogram) and non-fossils
   ($ \Delta M_{1,2} \le $0.5) (solid red histogram) in our X-ray
   group catalog. The fraction of fossil groups is skewed toward the
   lower values of $ \frac{R_{X}}{R_{500}} $ compared to the
   non-fossil groups.  {\it Middle panel.} The $ \frac{R_{X}}{R_{500}}
   $ distribution for X-ray groups in different subsamples. {\it Lower
     panel.} The redshift evolution of the extent of X-ray detection
   for galaxy groups/clusters. A redshift evolution in the ratio by
   0.25 in amplitude is seen for all groups.}
 \label{ks}
 \end{figure}

We compare the distributions of $\frac{R_{X}}{R_{500}}$ ratio for
fossils to non-fossils, selected by implying $\Delta
M_{1,2}\leq$0.5 in Fig. \ref{ks} (top panel). $R_X$ is the detected
extent of the X-ray emission, while $R_{500}$ is the extent of X-ray
emission anticipated for the given flux of the group. The distribution
of $\frac{R_{X}}{R_{500}}$ ratio is skewed toward the lower values
for fossils compared to the non-fossil groups, with 1\% probability of
chance occurrence, based on the K-S test. However, as also shown in
Fig. \ref{ks}, the observed extent of the X-ray emission is a strong
function of the redshift of the group. So, for a refined comparison of
the extent of X-ray detection for fossil candidates and non-fossils,
we separate the groups into two redshift bins: with z$>$0.6 and
z$\leq$0.6. We find that fossils exhibit only a marginally lower
values of $\frac{R_{X}}{R_{500}}$ ratio compared to non-fossils in
both redshift ranges, z$>$0.6 and z$\leq$0.6, which may also be due to
residual differences in the redshift distribution.  In our method, the
cores of groups are removed prior to detection, which as this test
shows, reduces the impact of X-ray selection.

In Fig. \ref{f1} to Fig. \ref{f25}  we show the X-ray
emission as contours on the CFHTLS RGB images for the fossils candidate
groups listed in Tab. \ref{fossil}. We show the color magnitude diagrams
of g$^{\prime}$-r$^{\prime} $ and r $^{\prime}$- i$^{\prime} $ versus
z$^{\prime} $ for the fossil candidates at z$ <$0.66, and r $^{\prime}$-i
$^{\prime} $ and i $^{\prime}$-z$^{\prime} $ versus z$^{\prime}$ for
those at z$\geq$0.66. Group members (dark circles) are selected
according to the method described in \S 4.1. We mark the BGG and the
second brightest galaxies on each RGB image and all color magnitude
diagrams.

\section{Summary and conclusions}

We search for the extended X-ray emission using the contiguous XMM
coverage of the CFHTLS W1 field by the XMM--LSS public data. We
provide a catalog of 129 X-ray galaxy groups including 44 groups
  with a spectroscopic redshift over $\sim$3 degrees$^{2}$, spanning
the redshift range 0.04$<$z$<$1.23, characterised by a rest frame
0.1--2.4 keV band luminosity range between $ 10^{41} $ and $10^{44} $
ergs s$ ^{-1} $. We utilize a two-color red-sequence finder for
the photometric selection of group members and to calculate the mean group
redshift. For the first time, we perform a statistical analysis on the
luminosity gap of groups over the wide redshift range of our catalog.
We compare our observational results with predictions of SAMs
presented in G11, DLB07 and B06. Our main results are as follows:

(i) We have compared the flux for the groups and clusters in common
with the C1 clusters of \cite{Adami11}. We show that clusters at
z$>0.6$ have important differences in flux estimates as a consequence of
point source contamination.

(ii) We show that two color selection of group members reduces the sample
contamination by DSFGs at $\lesssim$0.6 compared to a single color
selection. This contamination slightly affects the sample at z$\gtrsim$0.6
(by $ \approx$3\%).

In addition, to test the effect of a red sequence selection of members
on the magnitude gap measurements and the determination of the
fraction of fossil or non-fossil groups, we compare the magnitude gap
values computed using the spectroscopic and red sequence members for a
sample of 84 spectroscopic X-ray groups from our catalogs in the
AEGIS, COSMOS and XMM-LSS fields. We find that the systematic effect
on the magnitude gap computation is limited to $<6$ \% at 95\% Poisson
confidence level, which is low compared to our statistical
uncertainties. We apply this effect on our estimate of the fraction of
fossil groups and number of groups in each $\Delta M_{1,2}$ bin in
Fig. \ref{gap}. Using this sample we show that our results which are
derived by combining groups with red-sequence and spectroscopic
membership are valid and we quantified the corresponding
uncertainties.

(iii) We demonstrate that the fraction of groups as a function of
$\Delta M_{1,2}$ exhibits a flatter slope for low-z and low-mass
groups in S--I compared to higher redshift subsamples, revealing the
build-up of high magnitude gap groups at low redshift. We derive a
marginally higher fraction of groups with $\Delta M_{1,2} \geq$2
compared to all models for S--I. We find that the B06 model is not
able to reproduce the observed trends for $ \Delta M_{1,2} $
distributions. On the other hand, G11 and DLB07 models predict
reasonably well $ \Delta M_{1,2} $ distributions for groups and
clusters.

It appears that the magnitude gap distribution of groups with $\Delta
M_{1,2}\lesssim$2 is well modeled by these two models. However the
consistency between them and observations is reduced for groups and
clusters with large magnitude gap, such as fossil groups, indicating
that the fraction of these objects have not been understood
completely. We report a good agreement between distributions of the
magnitude gap for our groups catalog with red sequence selection of
group members and that of our spectroscopic sample of 84 groups within
the redshift range of 0.05$<$z$<$1.22 (see Fig. \ref{specgap}).

(iv) We show that the model abundance by volume of galaxy groups is in
agreement with low-z X-ray groups in S--I. However, at higher
redshifts the observed number density of groups with higher halo
masses is below the predicted model values which we attribute to
cosmic variance and sensitivity of the abundance of high-mass systems
to cosmological parameters. We demonstrate that the observed mean
fraction of X-ray clusters with $\Delta M_{1,2}\geq$1 evolves slowly
with redshift in agreement with all models, with the exception of B06,
which under-predicts the fraction of clusters with $\Delta
M_{1,2}\le$1.  A larger fraction of groups with $ \Delta M_{1,2}\le$1
is detected at lower redshifts, in agreement with the models.

(v) We investigate the absolute r-band magnitude of the first and the
second brightest galaxies as a function of the magnitude gap. We find
a significant negative evolution of the intercept ($ \sim$0.8 mag) of
the $ M_{r,1} $ - $\Delta M_{1,2}$ and $ M_{r,2} $ - $\Delta M_{1,2}$
relations for clusters at 0.2$\lesssim$z$\lesssim$1.10. In comparison,
all the models predict a flatter redshift evolution for the
intercepts. We attribute the steeper negative evolution in the BCG
magnitudes to a more recent build-up of stellar mass in these galaxies
(in agreement with the models), suggestive of a stronger redshift
dependence of AGN feedback. The G11 model predicts well the intercepts
for low-z and low-mass groups, indicating the importance of the
improvements which have been made to this model. Conversely, B06 and
DLB07 models under-predict the intercepts for the BGGs and their
satellites. The observed slope of the $ M_{r,1}-\Delta M_{1,2}$ and $
M_{r,2}-\Delta M_{1,2} $ relations for clusters show a nearly constant
redshift evolution, consistent with all models. We conclude that the
mechanisms for creating magnitude gaps is well modeled in the
SAMs. For galaxy groups, we find that the G11 and DLB07 models predict
reasonably well the slopes of the butterfly diagrams. We conclude that
satellite distribution in galaxy groups is too efficient in the B06
model.

(vi) We have selected 22 fossil group candidates using the magnitude
gap criterion. Similar to other studies we find that fossil groups
constitute 22.2$\pm$6\% of all groups at z$<$0.6. We report a redshift
evolution in this fraction which drops to 12.3$\pm$7\% at
z$>$0.6. These numbers and their evolution are consistent with those
of our spectroscopic sample of COSMOS+AEGIS+XMM-LSS groups where we
classify 23.1$\pm$7\% and 10.5$ \pm $8 \% of all groups as fossils at
z$\leq$0.6 and z$>$0.6, respectively.  We show that some fossil group
candidates include few galaxies, located outside the 0.5R$_{200} $,
which are brighter than the second bright galaxy which used for the
magnitude gap estimate, supporting a suggestion of
\cite{VonBendaBeckmann08} that large magnitude gaps can be refilled by
infall.

Finally, this study demonstrates that magnitude gap can be used to
differentiate between models. We find that the recent improvements in
the SAM of G11 brings it closer to the observations compared to the
early version of the SAMs, DBL07 and B06. However, it still fails to
reproduce all the results. For current SAMs only a single model output
is available. However, in order to improve the understanding further,
a grid of model predictions fully sampling the parameter space of
amplitude and redshift evolution of the relevant physical mechanisms
such as AGN feedback is essential.

    \section{Acknowledgements }
    
    Initial stages of this work have been partially funded by the
    German Deutsche Forschungsgemeinschaft, DFG Leibniz Prize $ (FKZ
    HA 1850/28-1)$. This work also has been partially supported by the
    grant of Finish Academy of Sciences to the University of Helsinki,
    decision number 266918.  We thank Guenther Hasinger, Hannu
    Koskinen, Kirpal Nandra, Christophe Adami, Birgit Boller,
    Marguerite Pierre, Peter Johansson and Kimmo Kettula for helpful
    assistance and comments. We thank the anonymous referee for the
    patience in reviewing the manuscript and many insightful
    comments. We used the Millennium Simulation databases in this
    paper and the web application providing online access to them were
    constructed as part of the activities of the German Astrophysical
    Virtual Observatory.

\newpage
   \appendix 
 
\section{Fossil group candidates}    

\begin{figure}[H]
\begin{center}  
\leavevmode
\resizebox{\hsize}{!}{\includegraphics{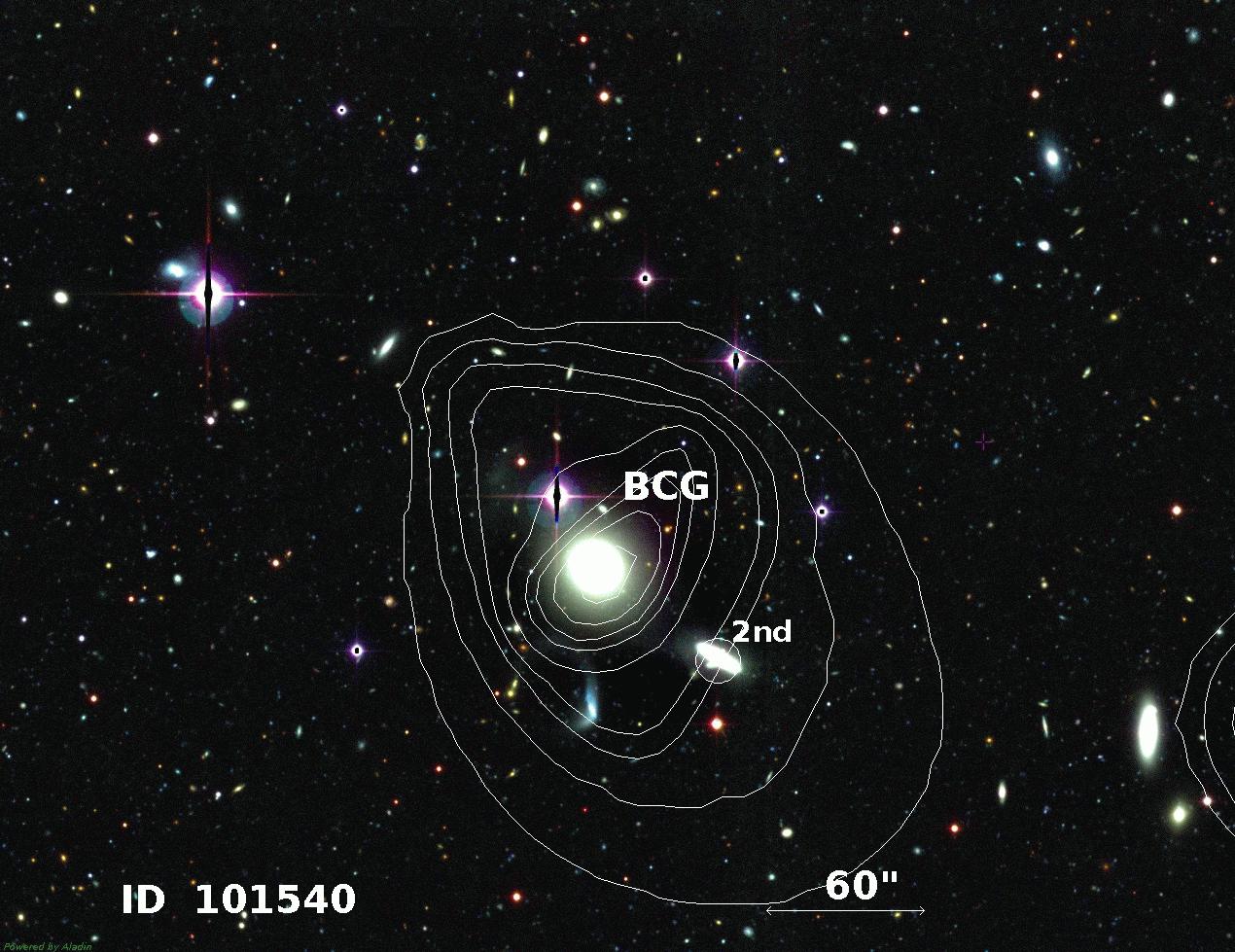}}
\resizebox{\hsize}{!}{\includegraphics{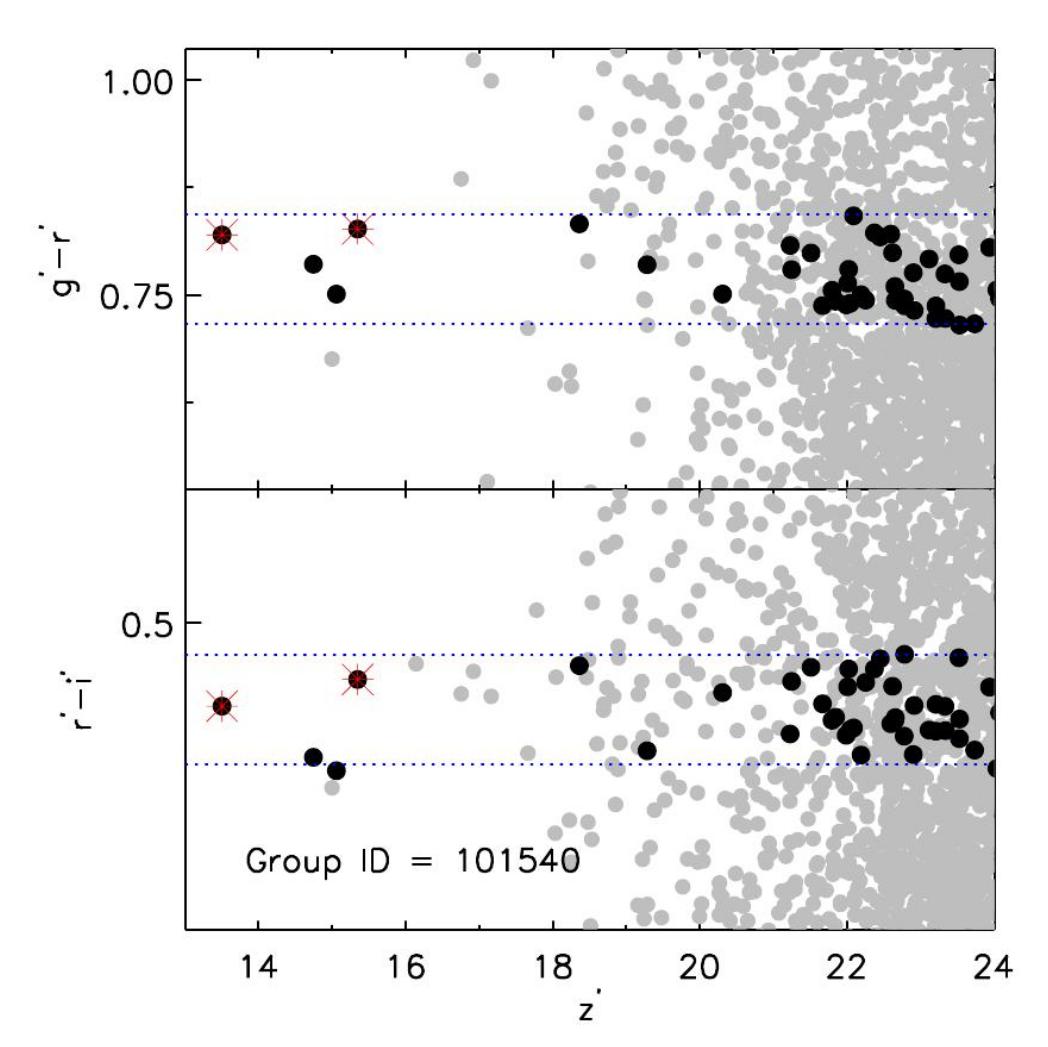} }
\end{center}
\caption[flag]{{\it Upper panel.} Contours of the extended X-ray
    emission overlaid on the CFHTLS RGB image of the fossil group
    101540 at z=0.07.  {\it Middle panel.} g$^{\prime}$- r$^ {\prime}
    $ versus z$^{\prime}$. {\it Lower panel.} r$^{\prime}$ -
    i$^{\prime} $ versus z$^{\prime} $. Filled black circles
    illustrate group members selected by the method described in \S
    4.1. The BGG and second brightest satellite galaxy within $
    0.5R_{200} $ have been marked with red asterisks within each color
    magnitude diagram.  The upper and lower limits of colors have been
    shown by horizontal dashed lines.}
\label{f1}
\end{figure}

\begin{figure}[H]
\begin{center}  
\leavevmode
\resizebox{\hsize}{!}{\includegraphics{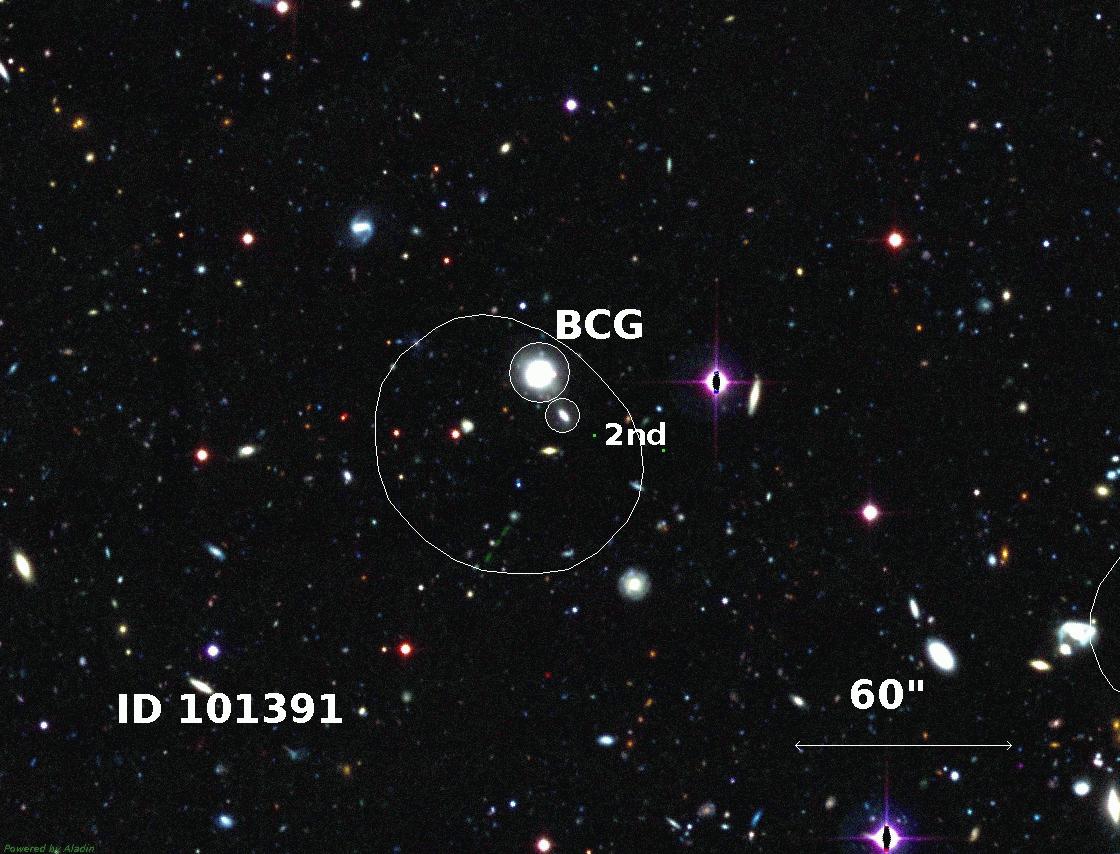}}
\resizebox{\hsize}{!}{\includegraphics{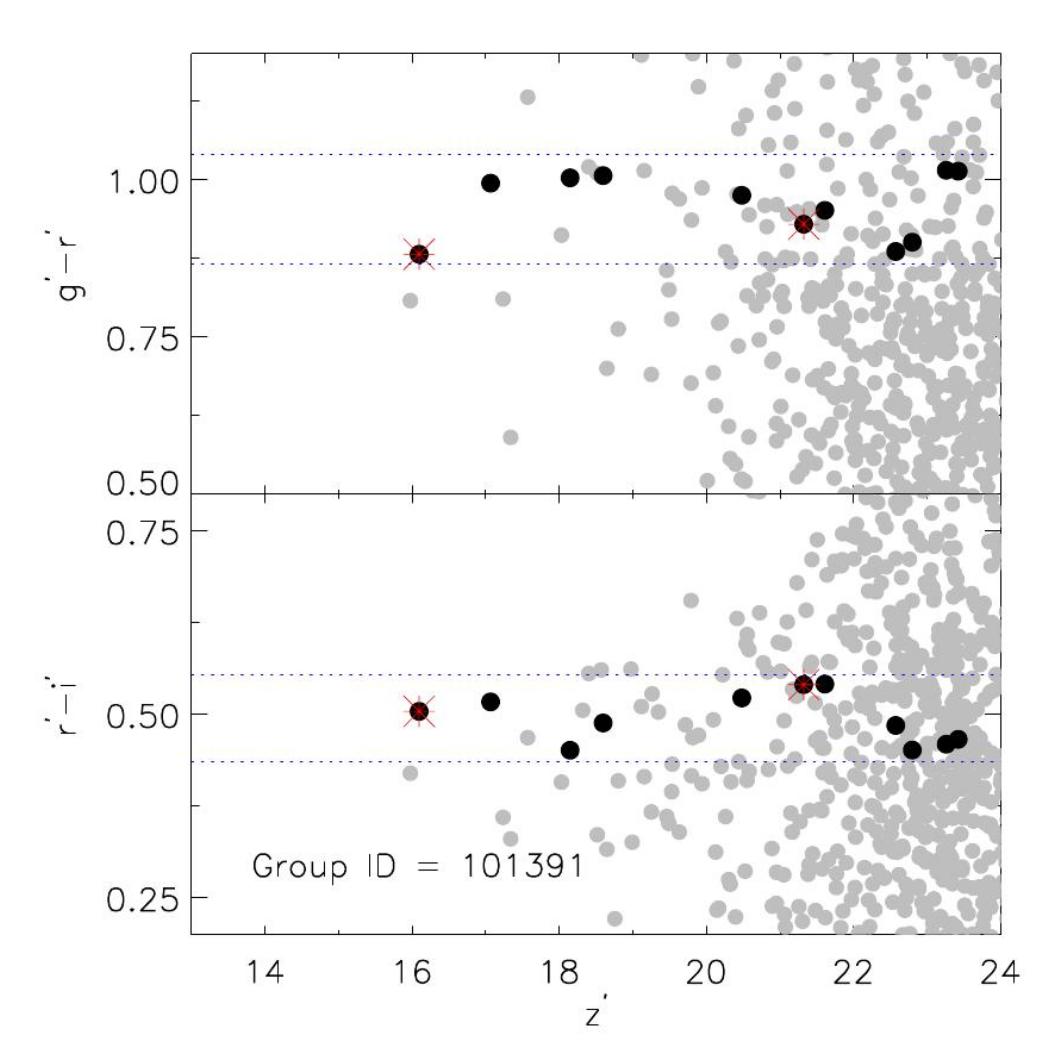}} 
\end{center}
\caption[flag]{ Same as in Fig. \ref{f1}, but for the group 101391 at z= 0.07.}
\label{f2}
\end{figure}

\begin{figure}[H]
\begin{center}  
\leavevmode
\resizebox{\hsize}{!}{\includegraphics{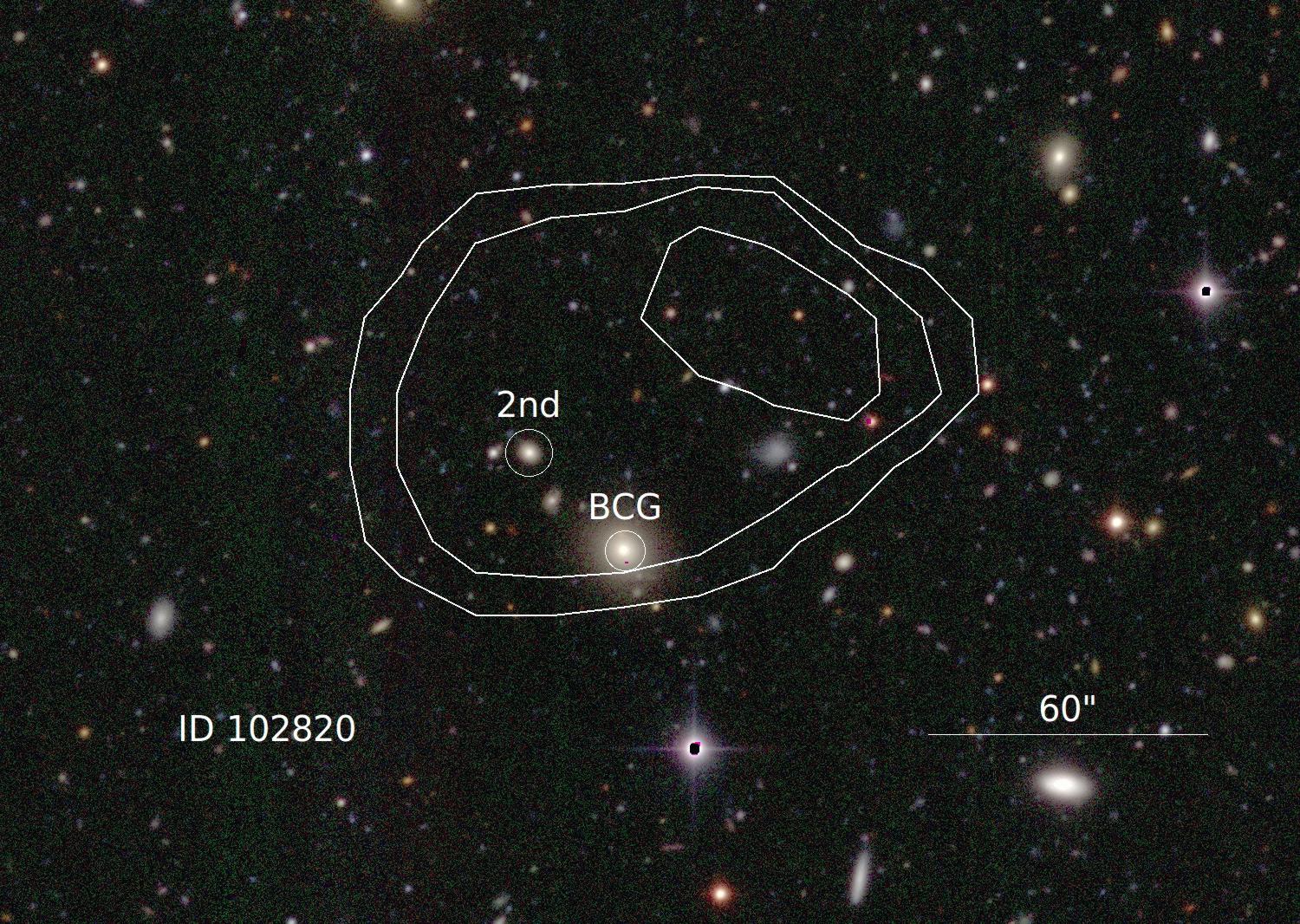}}
\resizebox{\hsize}{!}{\includegraphics{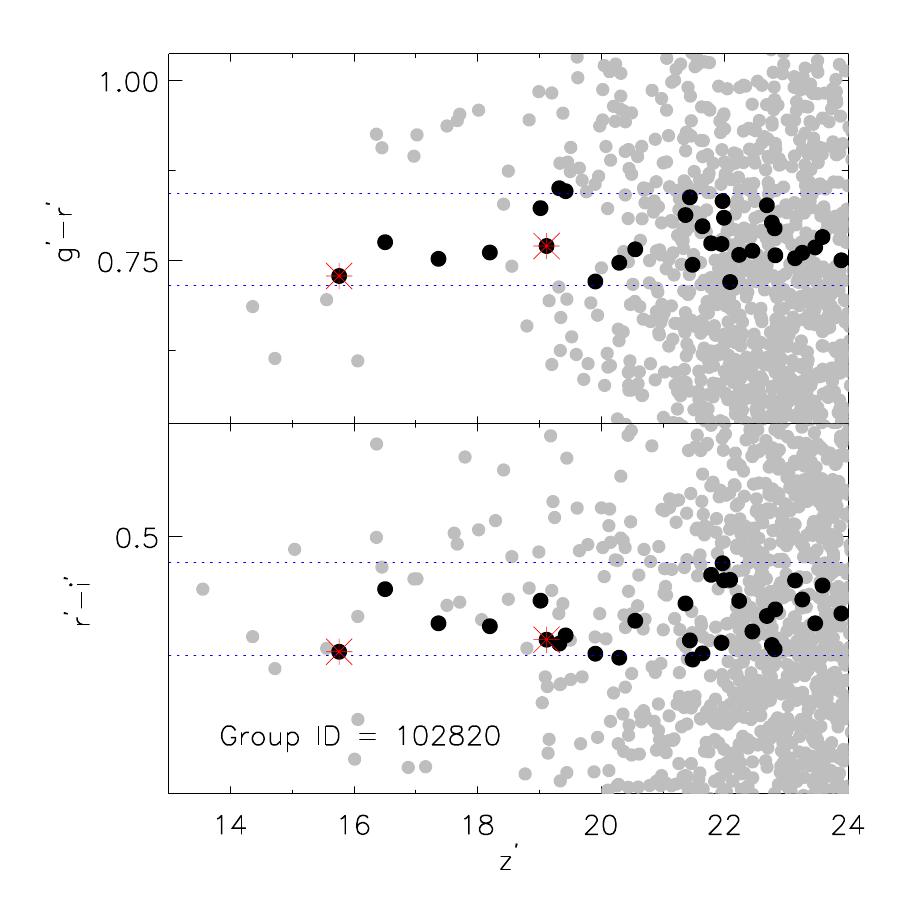}}
\end{center}
 \caption[flag]{ Same as in Fig. \ref{f1}, but for the group 102820 at  z=0.11.}
\label{f3}
\end{figure}

\begin{figure}[H]
 \begin{center}  
\leavevmode
\includegraphics[width=9cm]{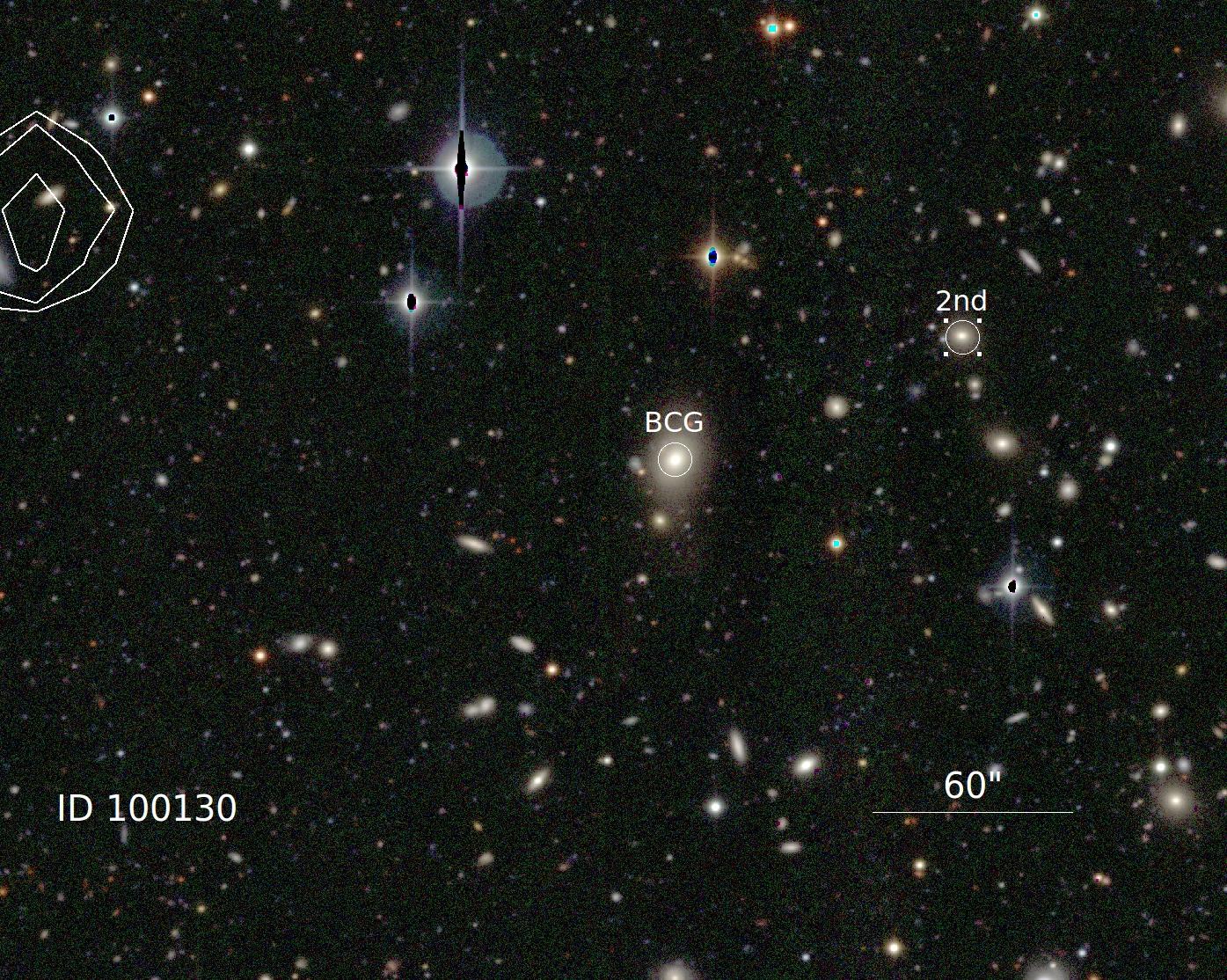}
\resizebox{\hsize}{!}{\includegraphics{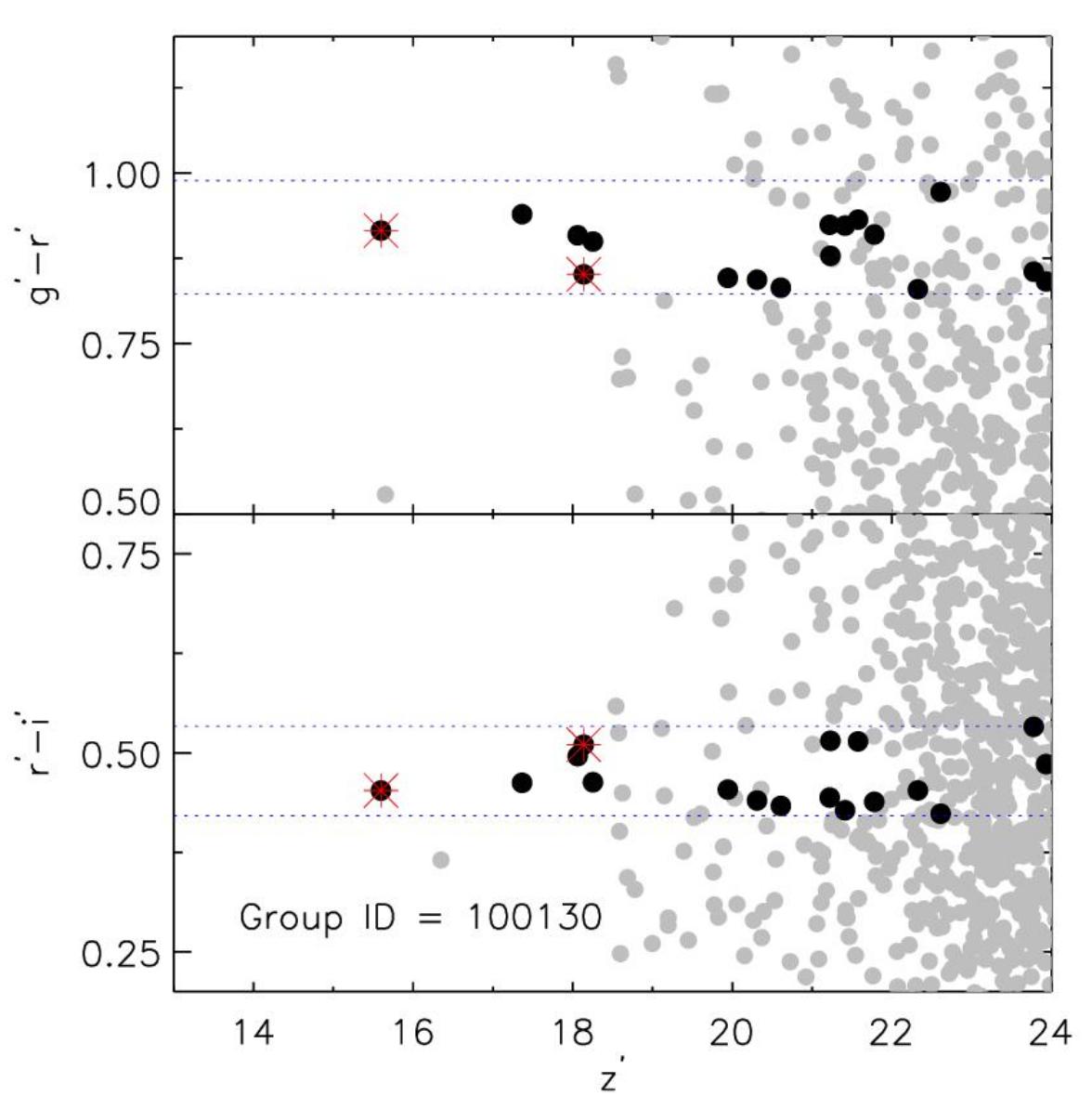}} 
 \end{center}
\caption[flag]{Same as in Fig. \ref{f1}, but for the group 100130, at  z= 0.13.}
\label{f4}
\end{figure}

\begin{figure}[H]
\begin{center}  
\leavevmode
\resizebox{\hsize}{!}{\includegraphics{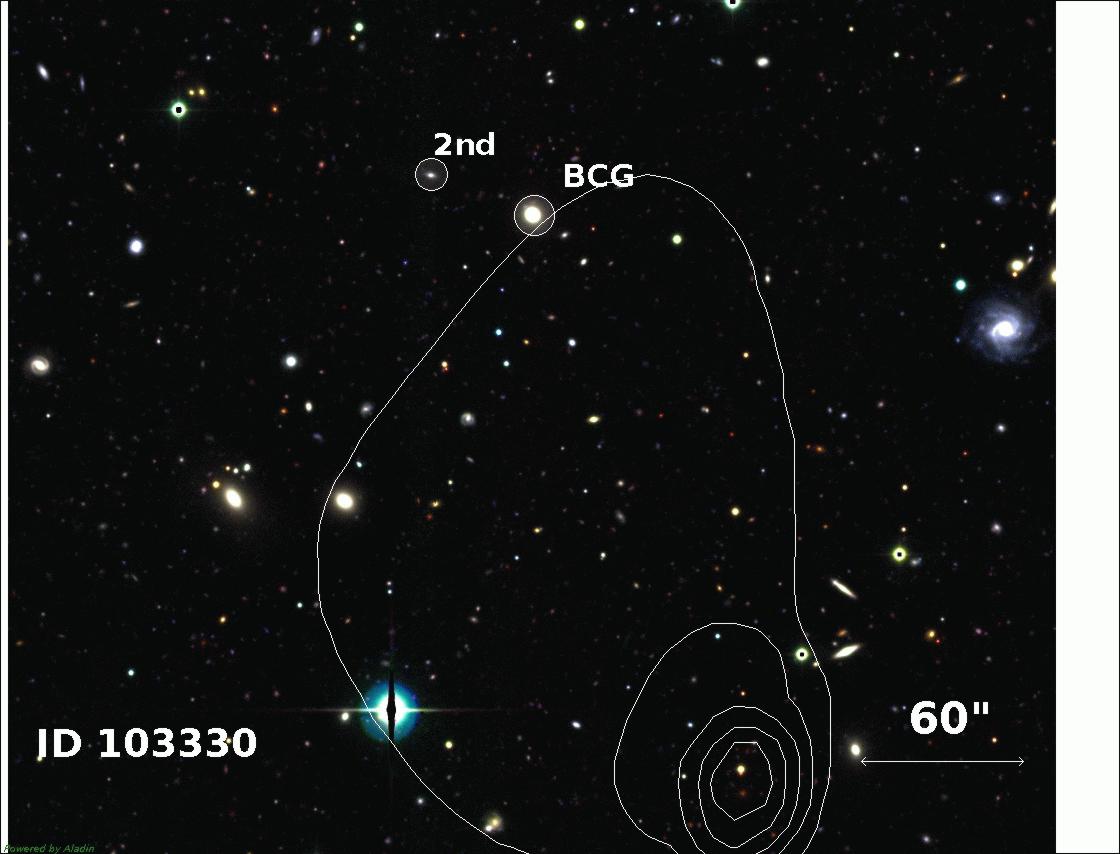}}
\resizebox{\hsize}{!}{\includegraphics{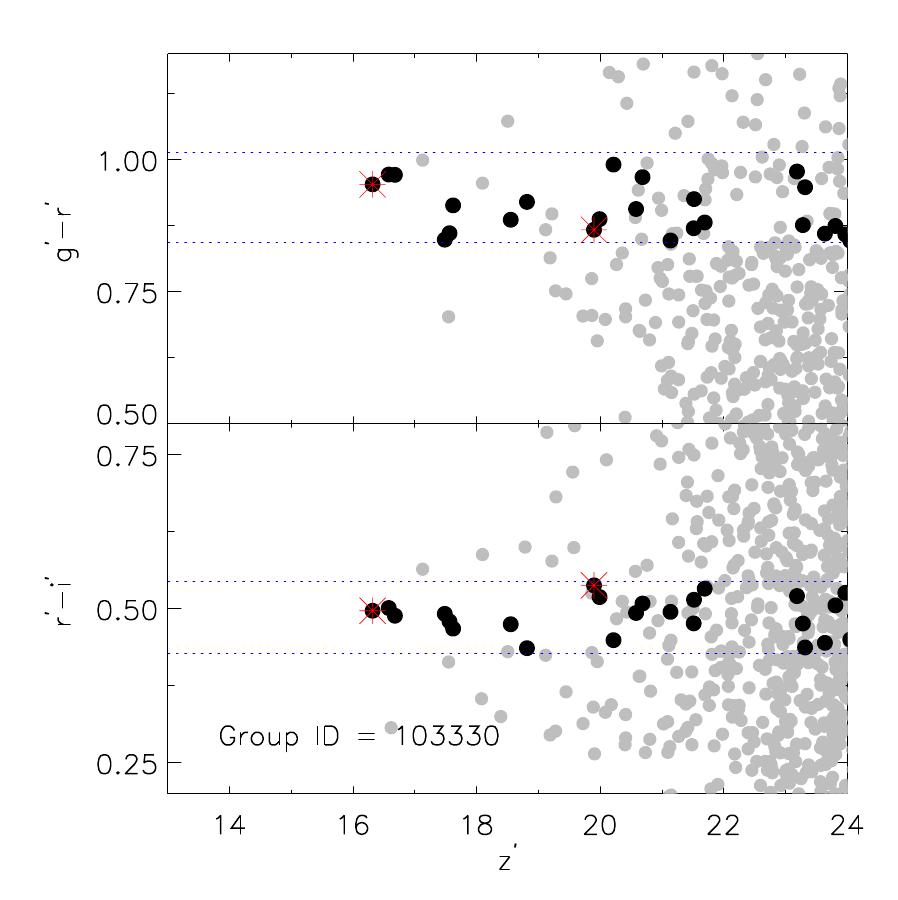}} 
\end{center}
\caption[flag]{Same as in Fig. \ref{f1}, but for the group 103330 at  z= 0.14.}
\label{f6}
\end{figure}

\begin{figure}[H]
\begin{center}  
\leavevmode
\resizebox{\hsize}{!}{\includegraphics{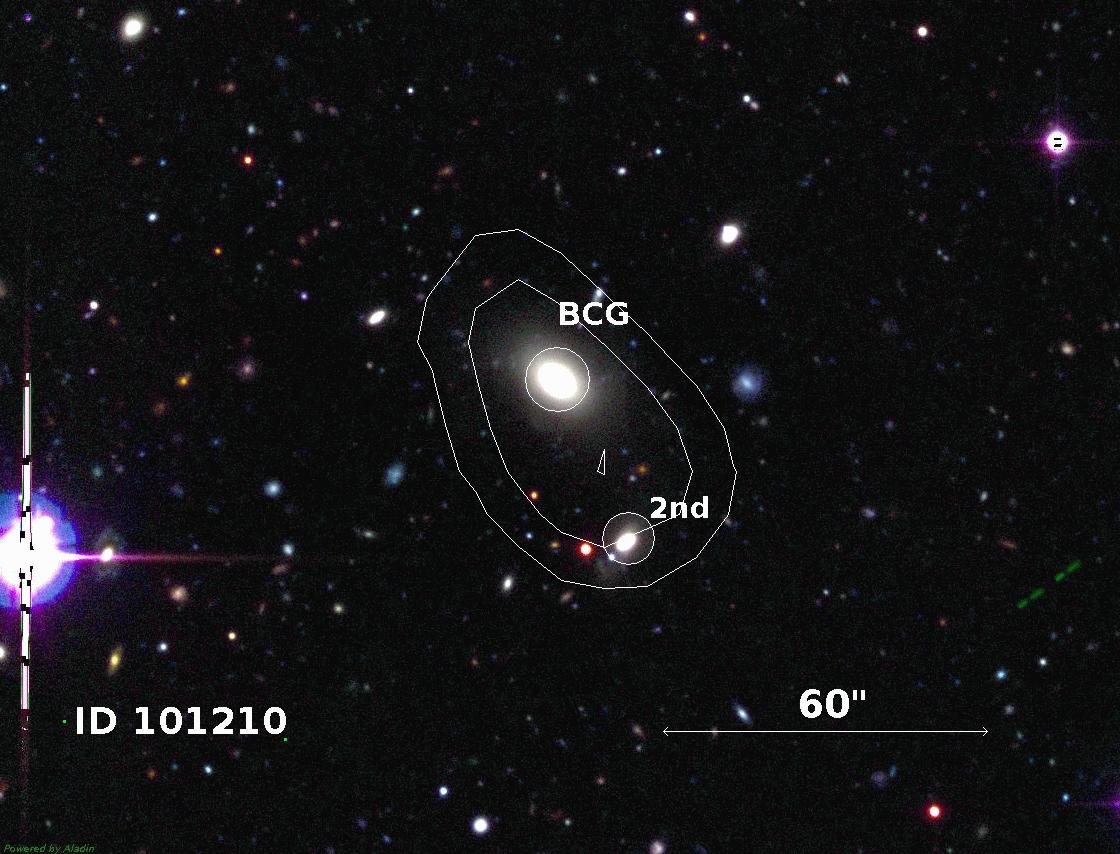}}
\resizebox{\hsize}{!}{\includegraphics{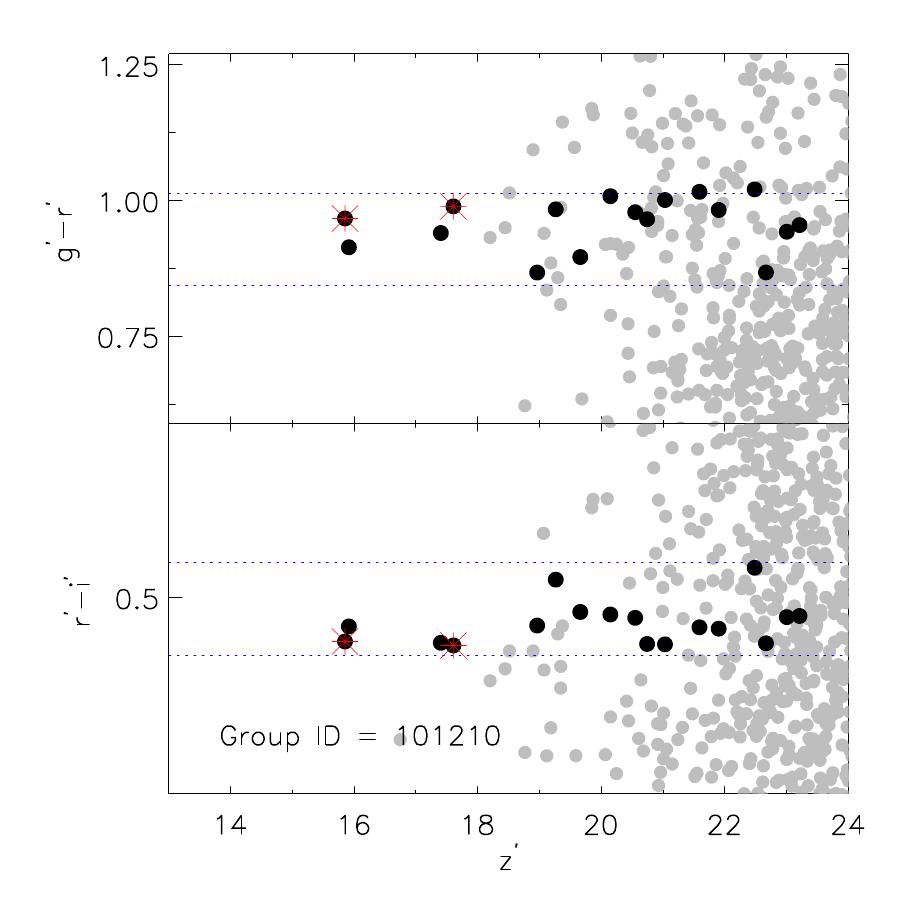}} 
\end{center}
\caption[flag]{Same as in Fig. \ref{f1}, but for the group 101210 at  z= 0.15.}
 \label{7}
\end{figure}

\begin{figure}[H]
\begin{center}  
\leavevmode
\resizebox{\hsize}{!}{\includegraphics{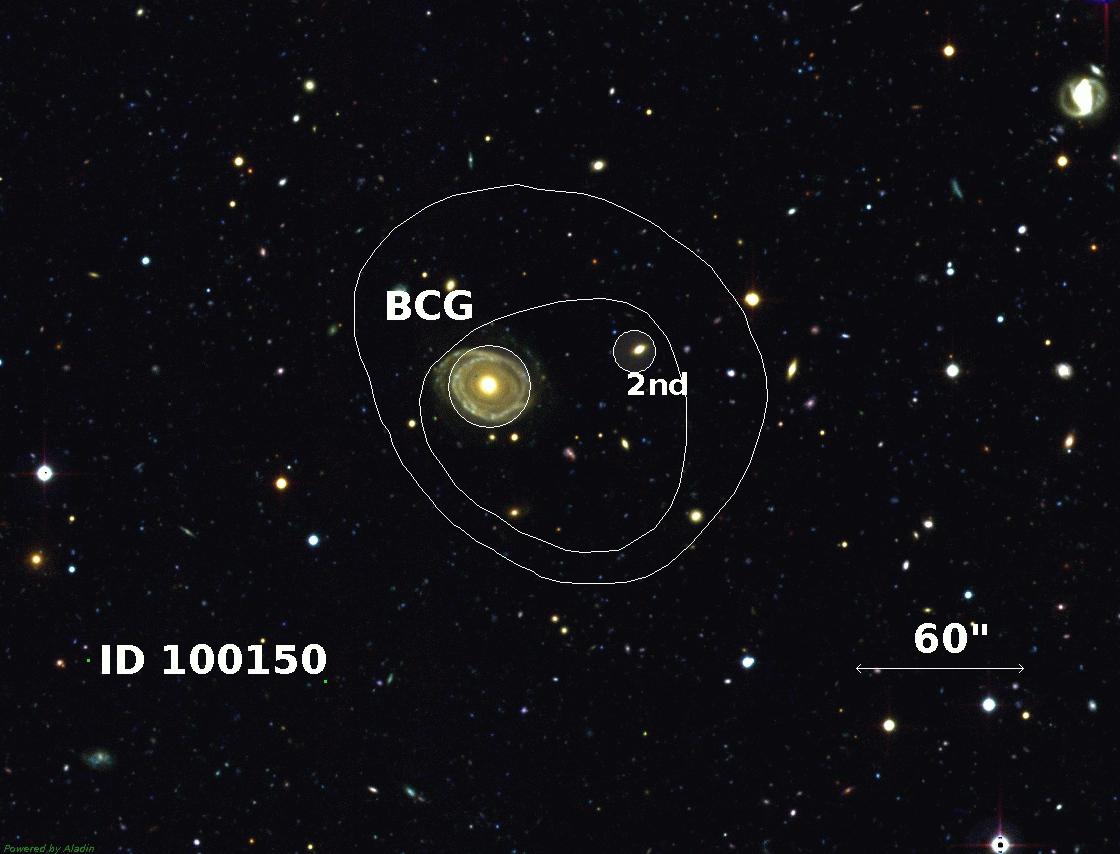}}
\resizebox{\hsize}{!}{\includegraphics{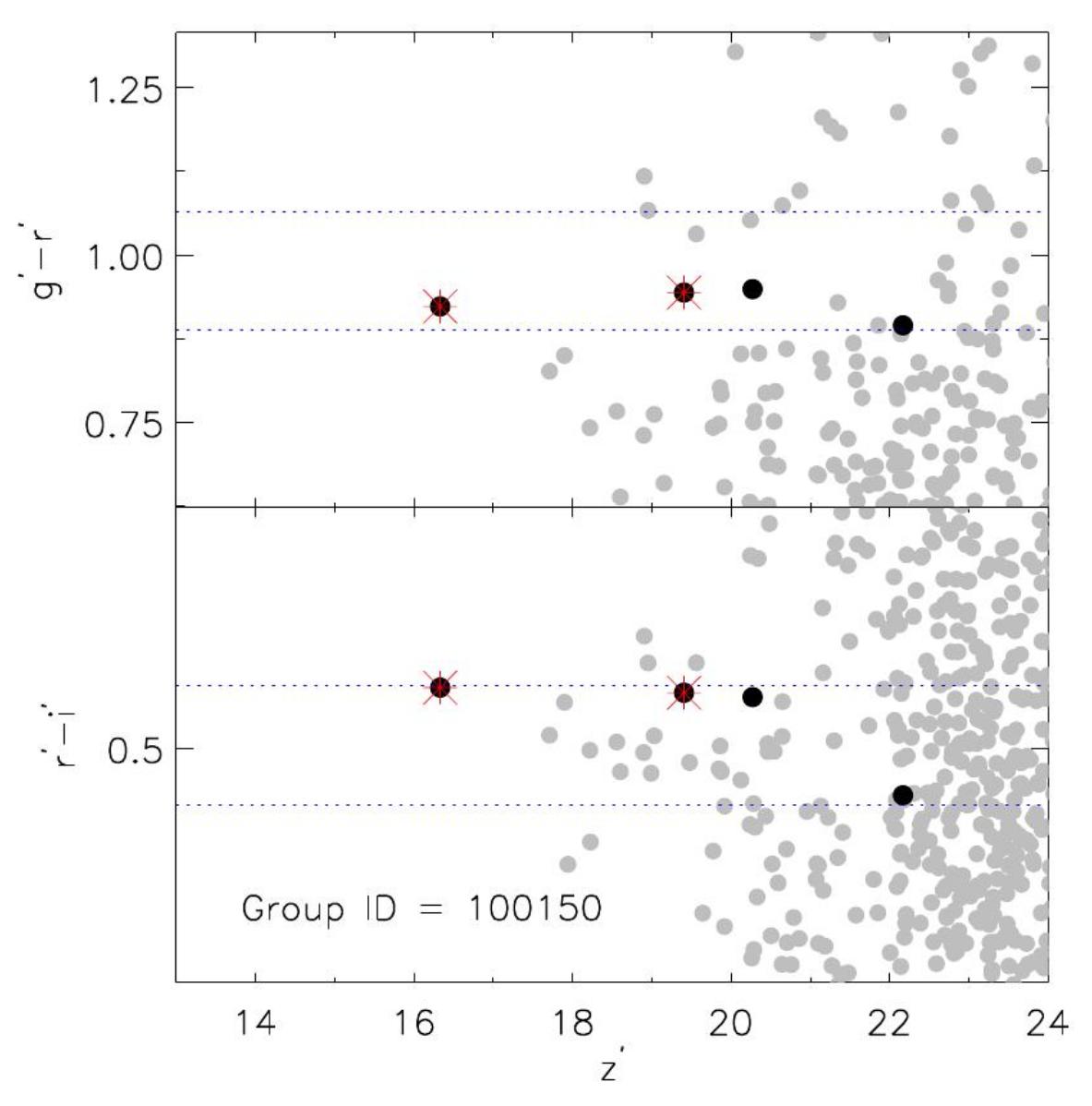}} 
\end{center}
\caption[flag]{Same as in Fig. \ref{f1}, but for the group100150 at  z= 0.16.}
\label{f8}
\end{figure}

\begin{figure}[H]
 \begin{center}  
\leavevmode
\resizebox{\hsize}{!}{\includegraphics{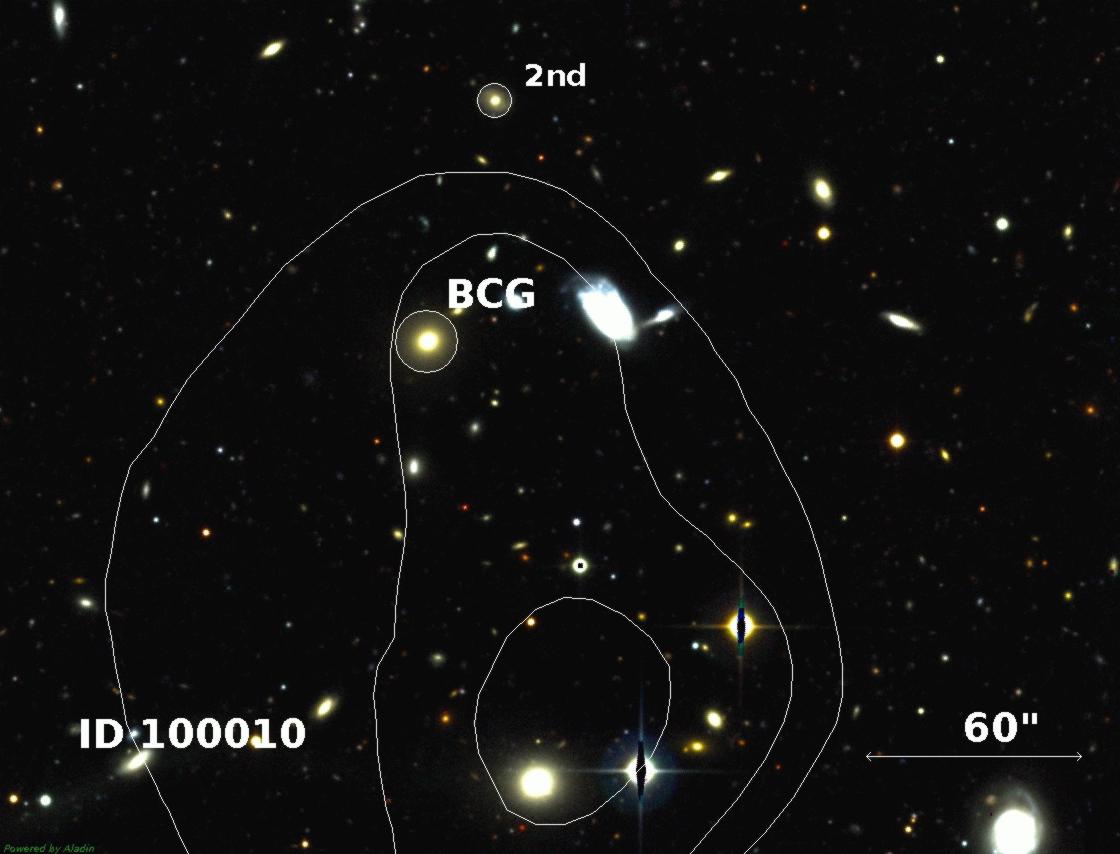}}
\resizebox{\hsize}{!}{\includegraphics{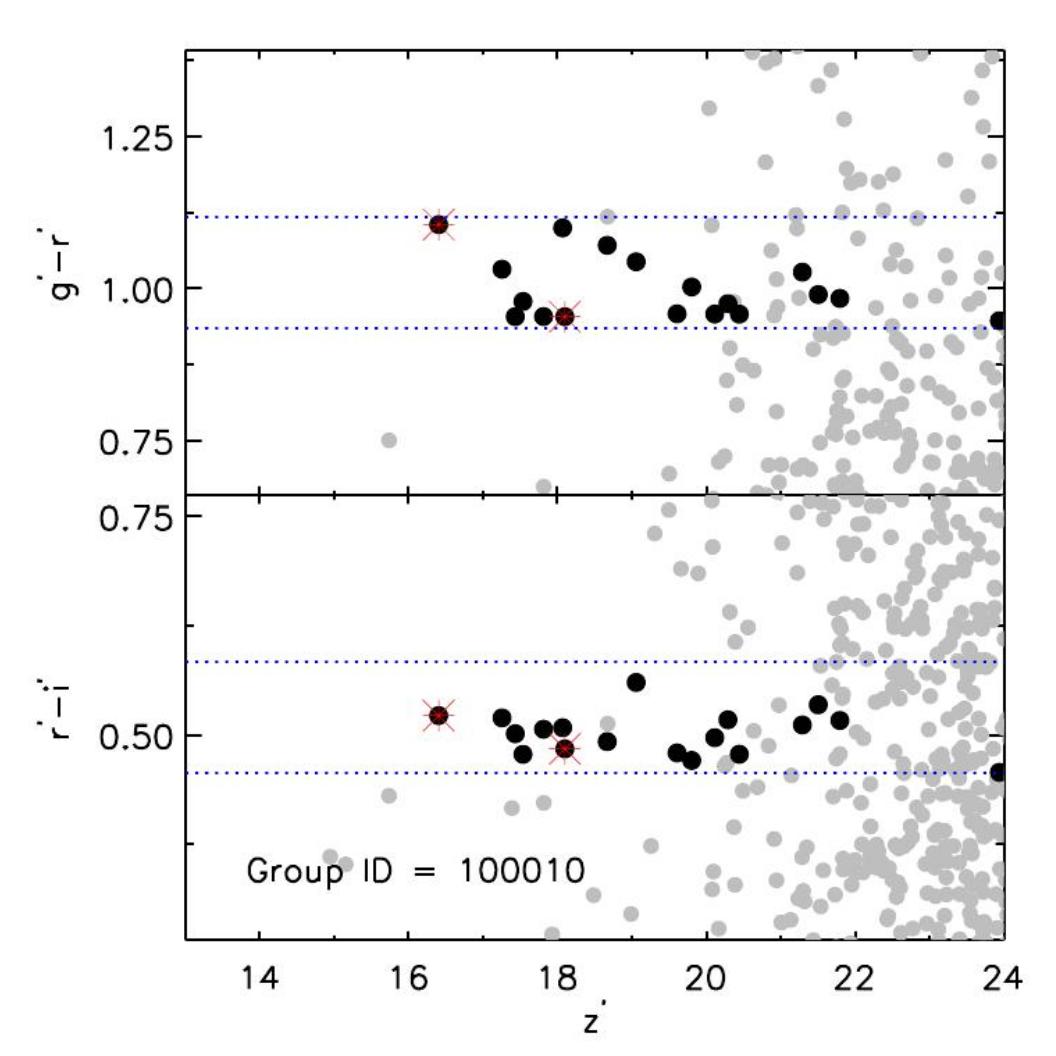}} 
\end{center}
\caption[flag]{Same as in Fig. \ref{f1}, but for the group 100010, at  z=0.22.}
\label{f10}
\end{figure}

\begin{figure}[H]
 \begin{center}  
\leavevmode
\resizebox{\hsize}{!}{\includegraphics{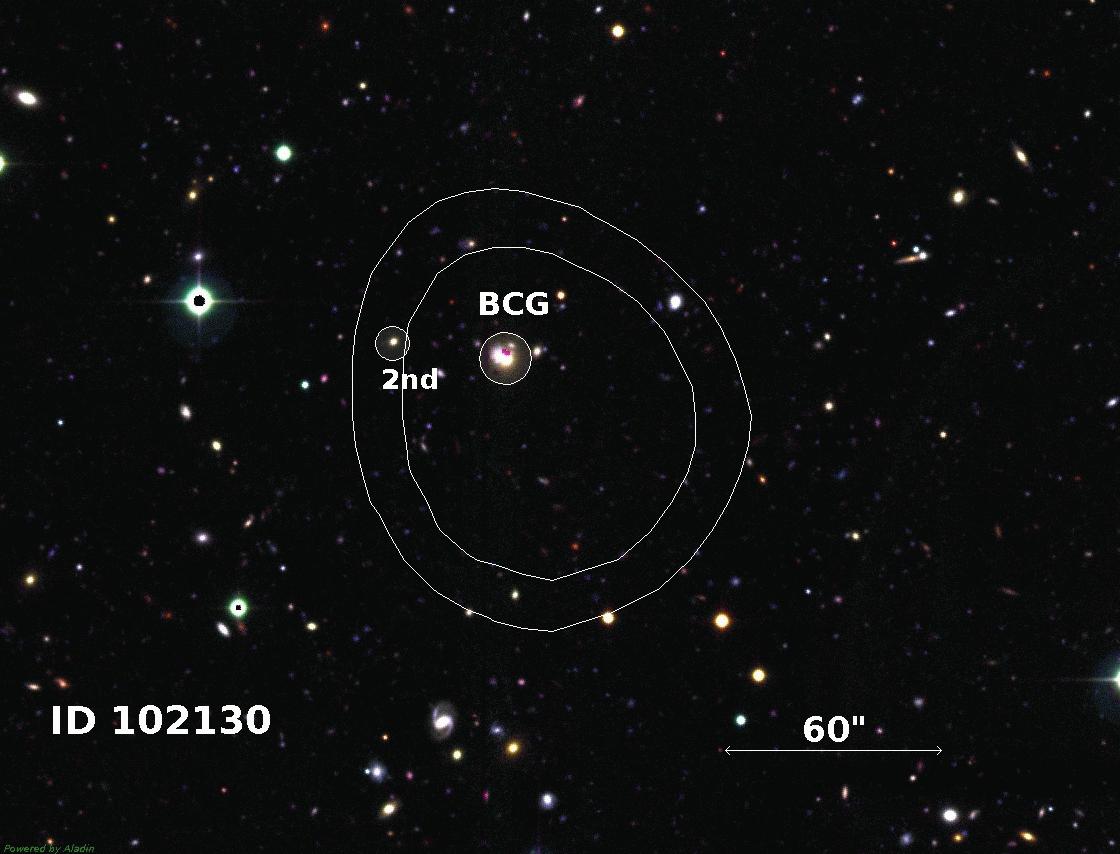}}
\resizebox{\hsize}{!}{ \includegraphics{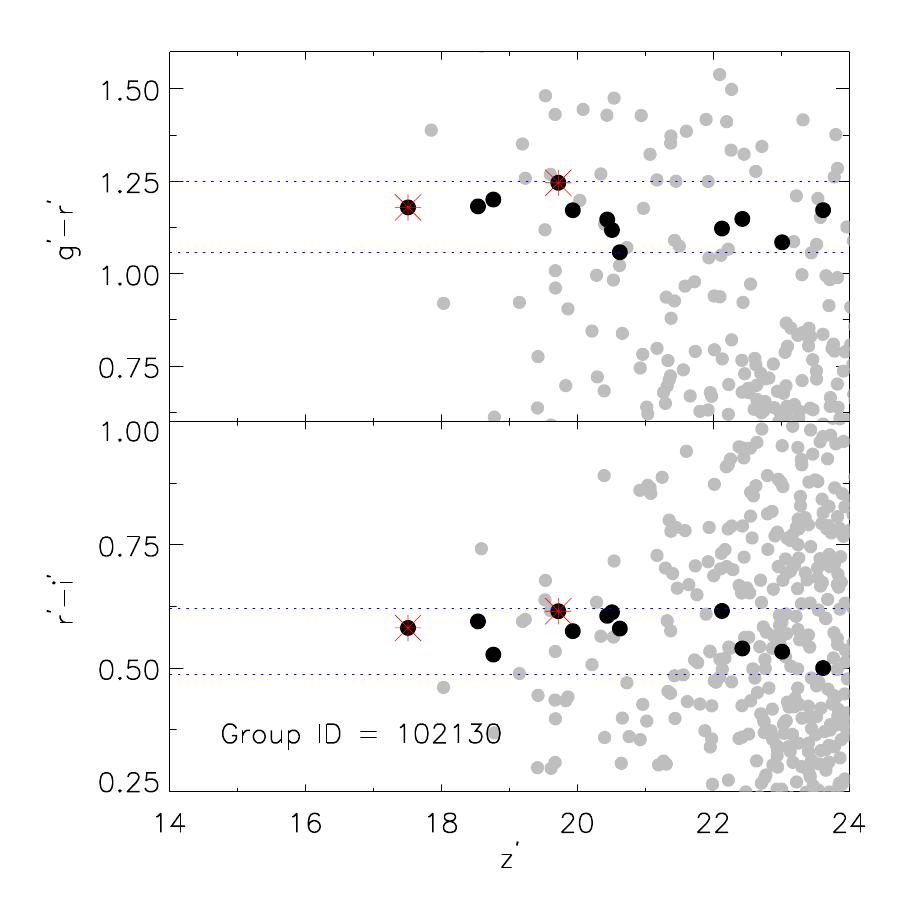} }
\end{center}
\caption[flag]{Same as in Fig. \ref{f1}, but for the group 102130 at z=0.23. }
\label{f11}
\end{figure}

\begin{figure}[H]
\begin{center}  
\resizebox{\hsize}{!}{\includegraphics{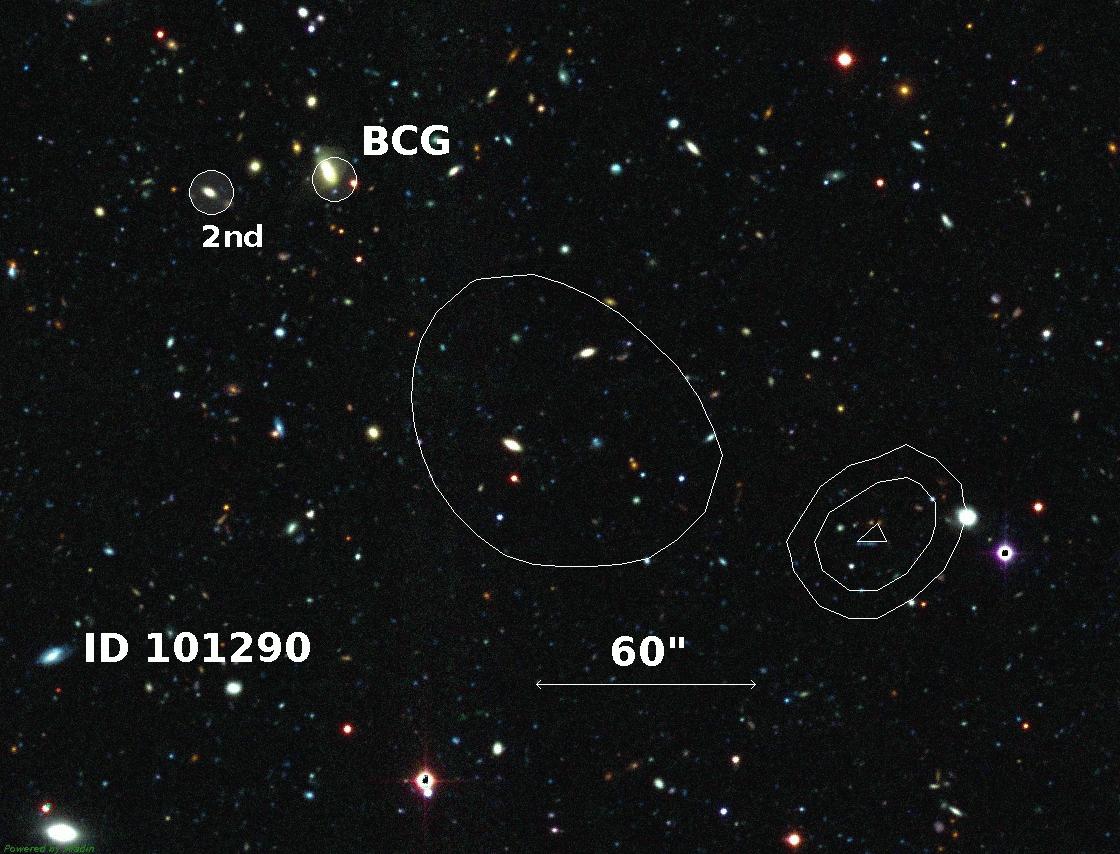}} 
\resizebox{\hsize}{!}{\includegraphics{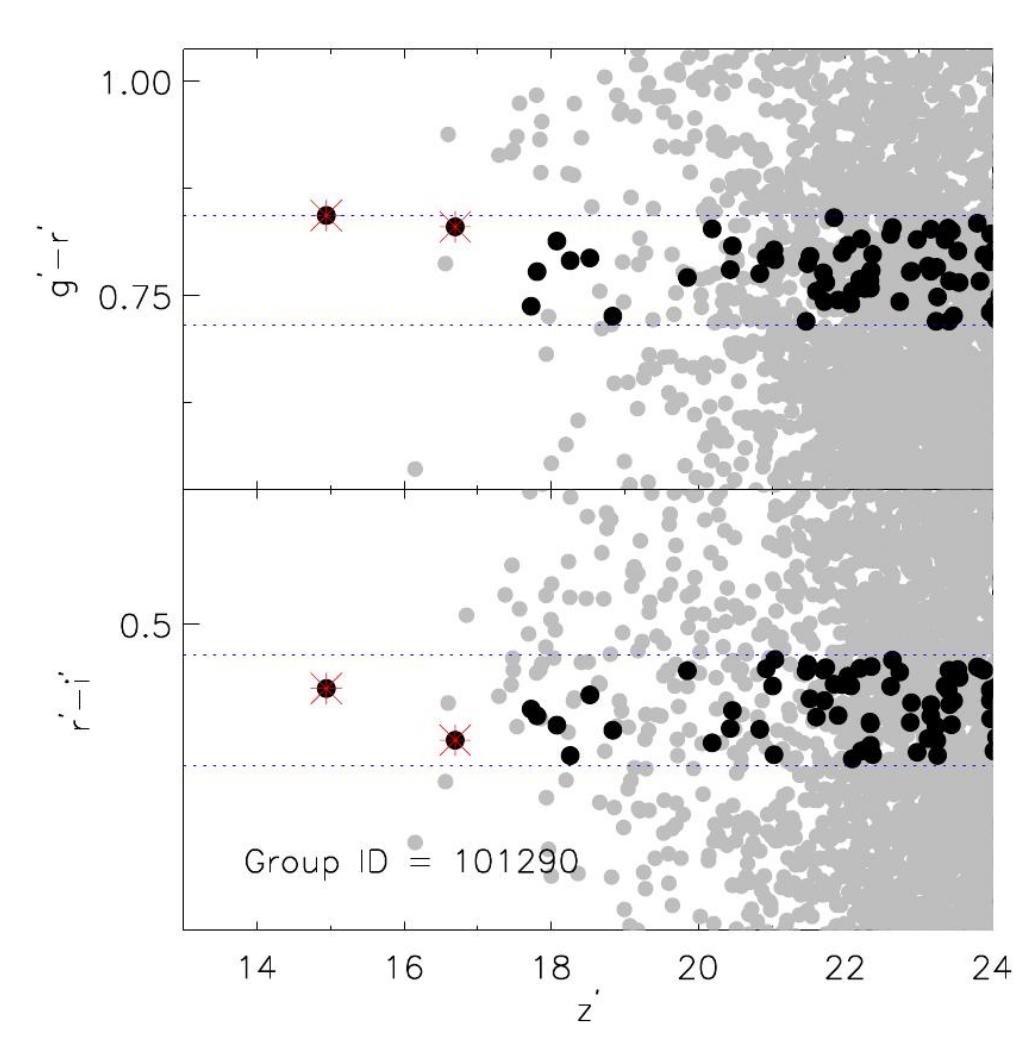}}                     
\end{center}
\caption[flag]{Same as in Fig. \ref{f1}, but for the group 101290 at  z=0.25.}
\label{f12}
\end{figure}

\begin{figure}[H]
\begin{center}  
\leavevmode
\resizebox{\hsize}{!}{\includegraphics{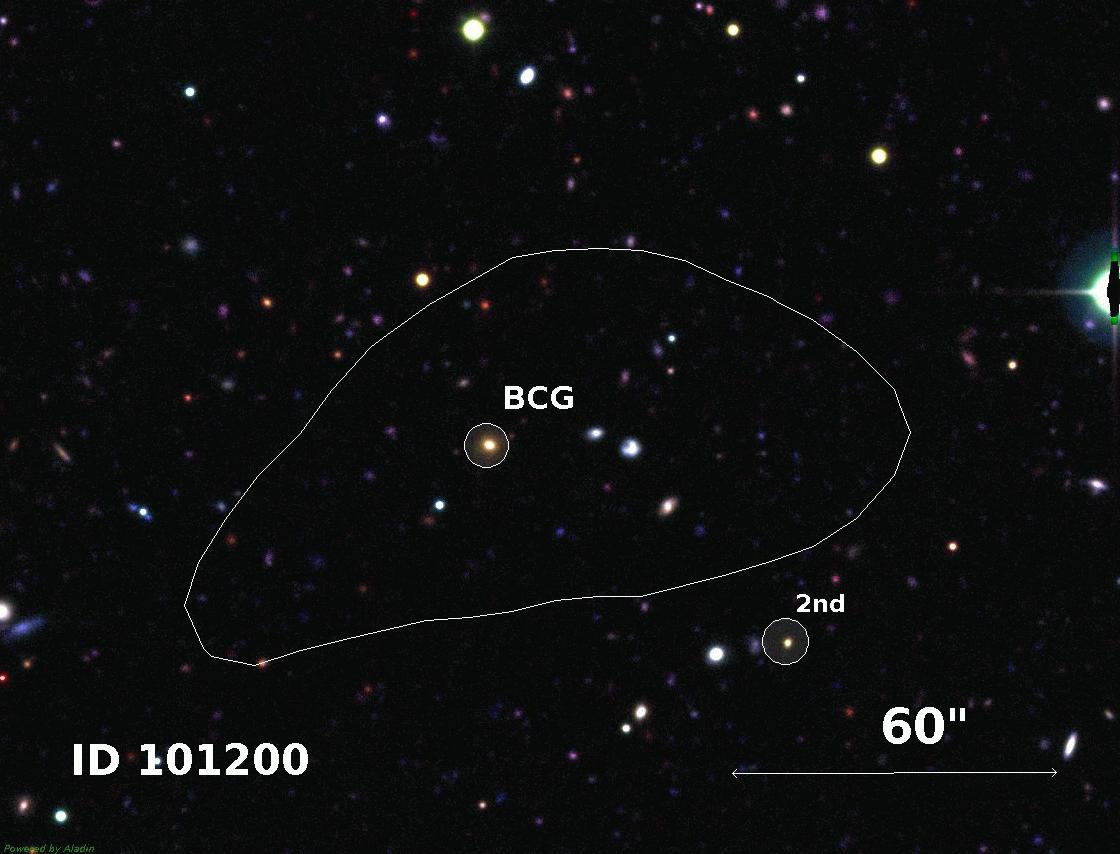}}
\resizebox{\hsize}{!}{\includegraphics{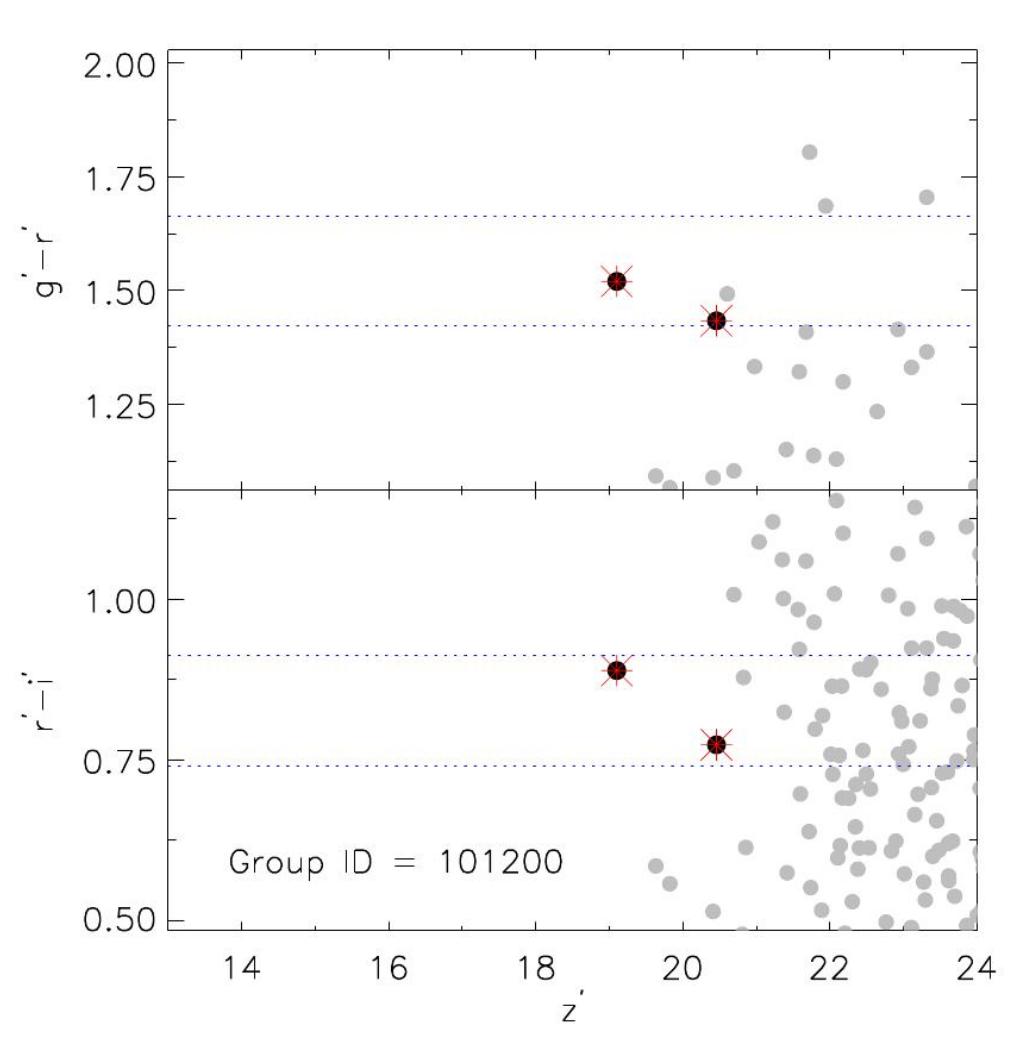}}
\end{center}
\caption[flag]{ Same as in Fig. \ref{f1}, but for the group 101200 at  z=0.47. The 2-nd brightest galaxy cannot be found inside 0.5R$ _{200} $ due to completeness effect and the marked galaxy is at 0.6R$_{ 200} $ distance to the center.}
\label{f14}
\end{figure}

\begin{figure}[H]
\begin{center}  
\resizebox{\hsize}{!}{\includegraphics{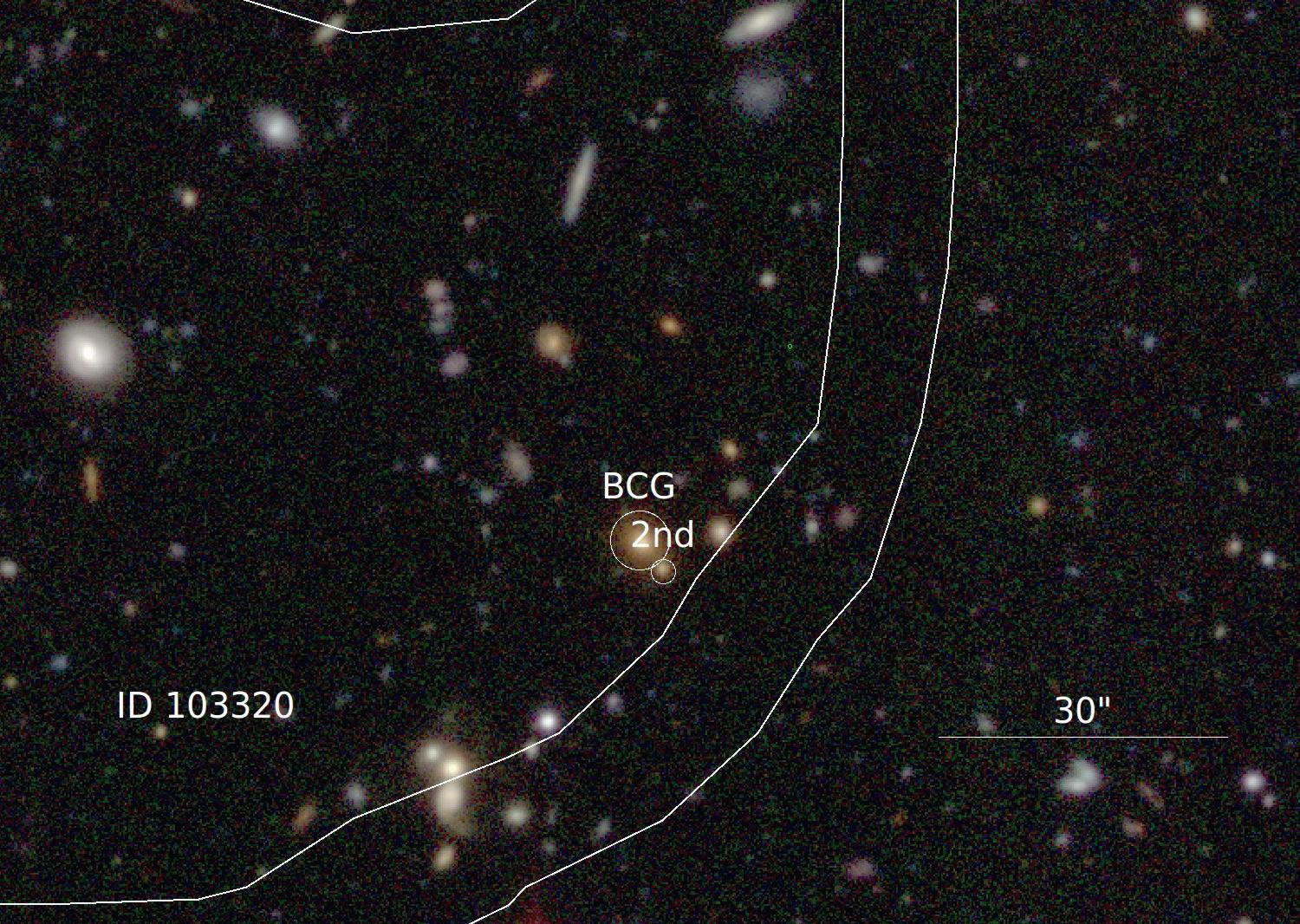}}
\resizebox{\hsize}{!}{\includegraphics{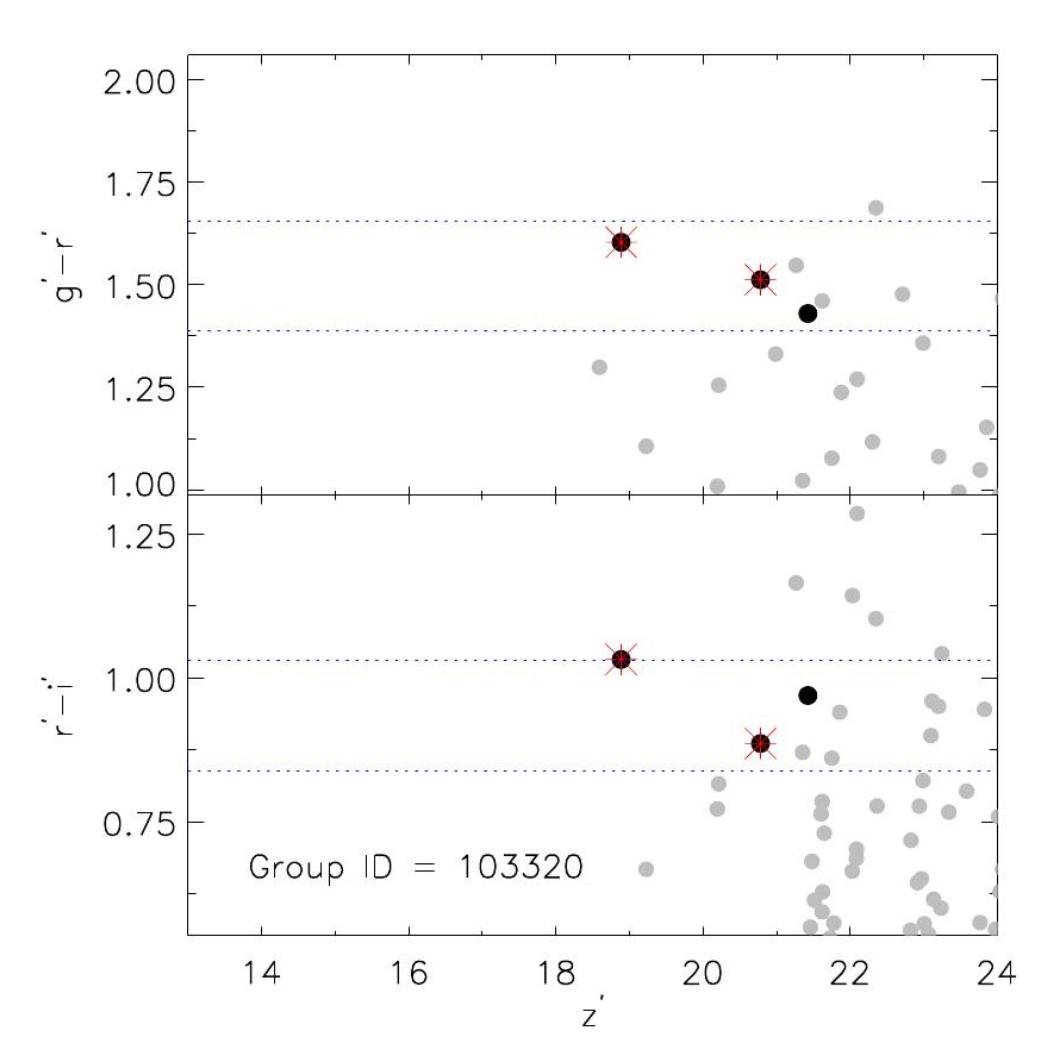}} 
 \end{center}
\caption[flag]{Same as in Fig. \ref{f1}, but for the group  103320 at  z=0.53.} 
\label{f16}
\end{figure}

\begin{figure}[H]
\begin{center}  
\leavevmode
\resizebox{\hsize}{!}{ \includegraphics{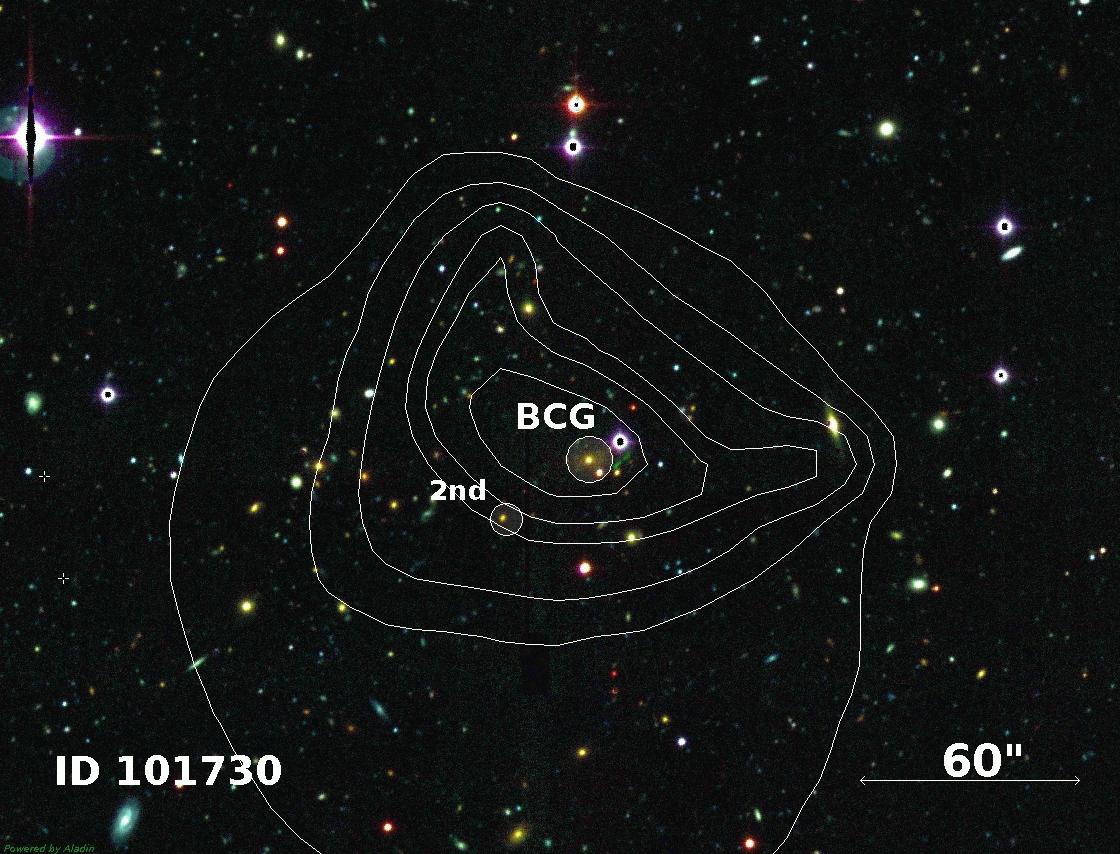} }
\resizebox{\hsize}{!}{\includegraphics{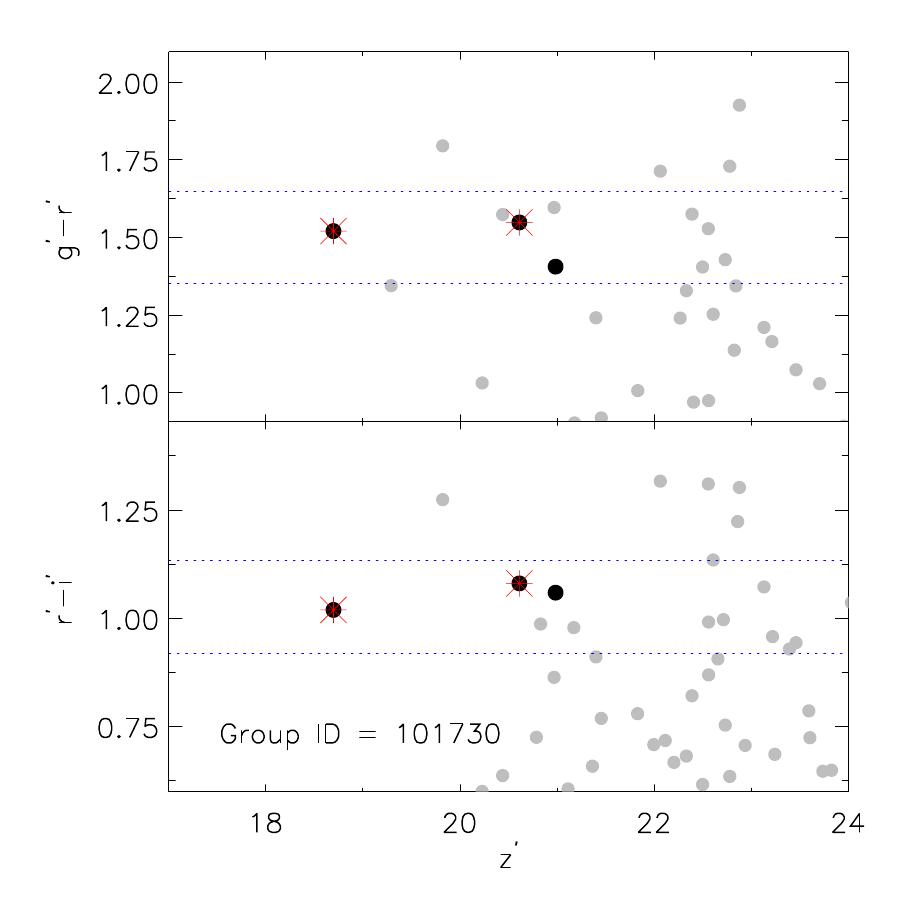}} 
\end{center}
\caption[flag]{Same as in Fig. \ref{f1}, but for the group 101730, at z 0.60. }
\label{f17}
\end{figure}

\begin{figure}[H]
  \begin{center}  
  \leavevmode
  \resizebox{\hsize}{!}{\includegraphics{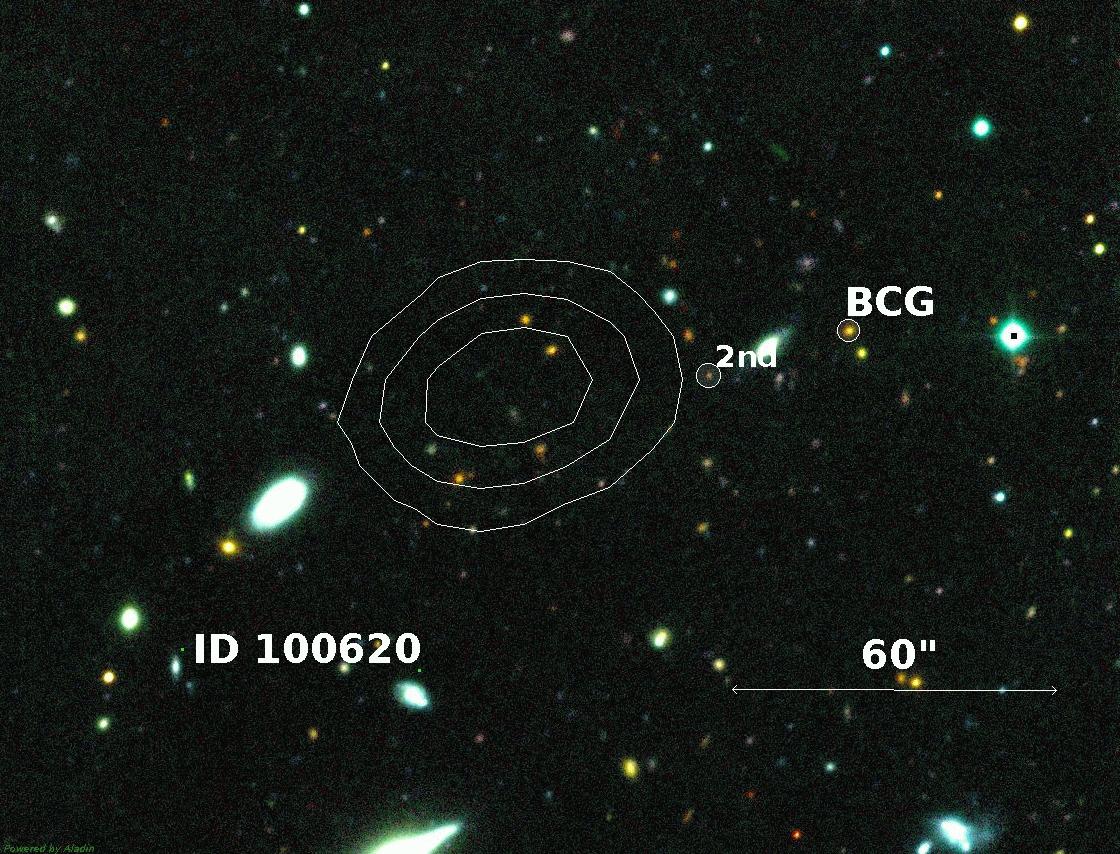} }
  \resizebox{\hsize}{!}{\includegraphics{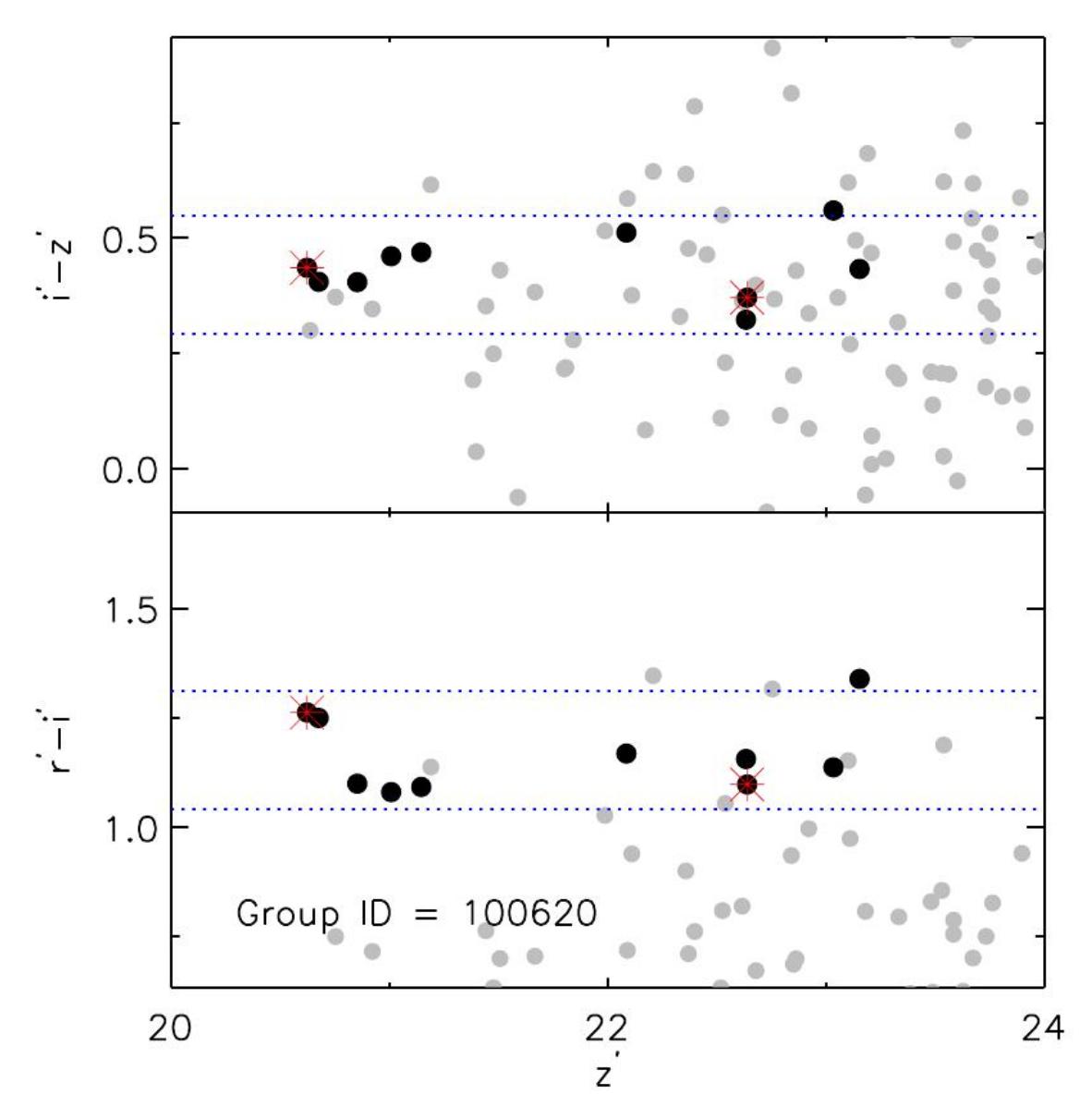}} 
  \end{center}
  \caption[flag]{{\it Upper panel.} Contours of extended X-ray
      emission overlaid on the CFHTLS RGB image of the fossil group
      100620 at z=0.66.  {\it Middle panel.} i$^{\prime}$- z$^
      {\prime} $ versus z$^{\prime}$. {\it Lower panel.} r$^{\prime}$
      - i$^{\prime} $ versus z$^{\prime} $. Filled black circles
      illustrate group members selected by the method described in \S
      4.1.  The BGG and second brightest satellite galaxy within $
      0.5R_{200} $ have been marked with red asterisks within each
      color magnitude diagram.  The upper and lower limits of colors
      have been shown by horizontal dotted blue lines.}
   \label{f19}
  \end{figure}

\begin{figure}[H]
 \begin{center}  
  \leavevmode
  \resizebox{\hsize}{!}{\includegraphics{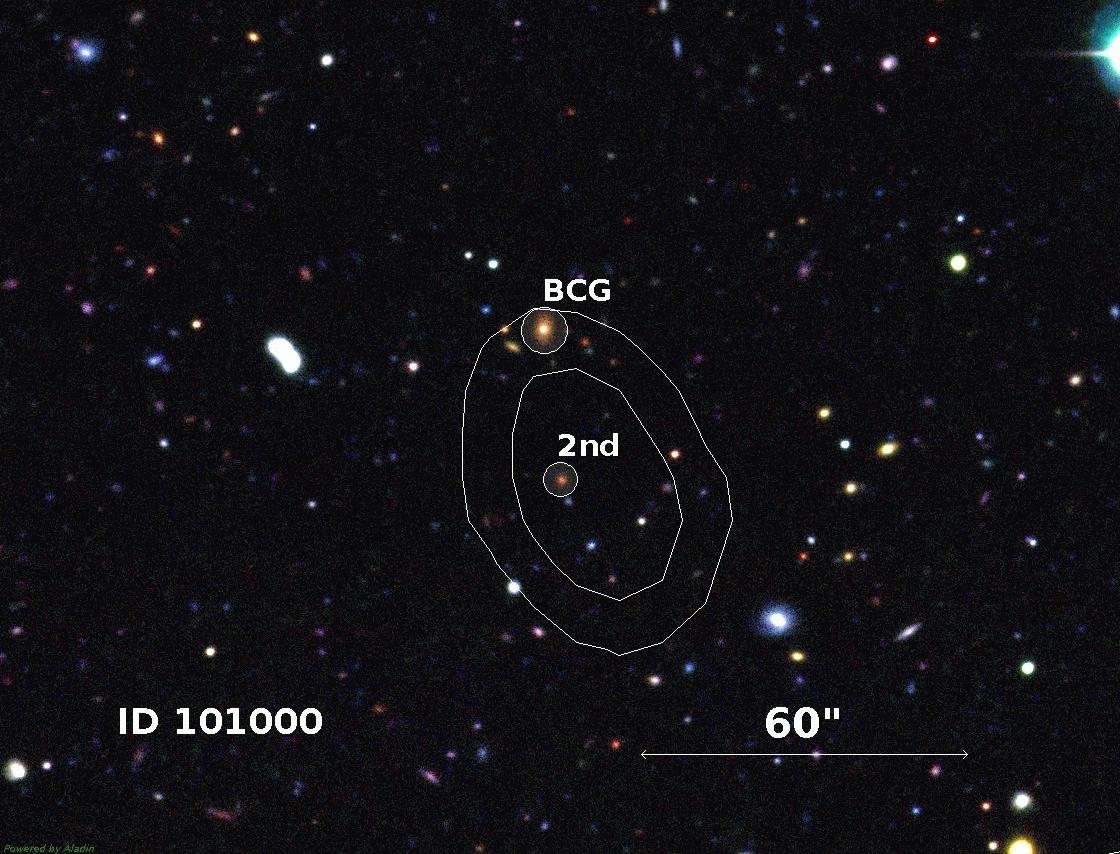}}
  \resizebox{\hsize}{!}{\includegraphics{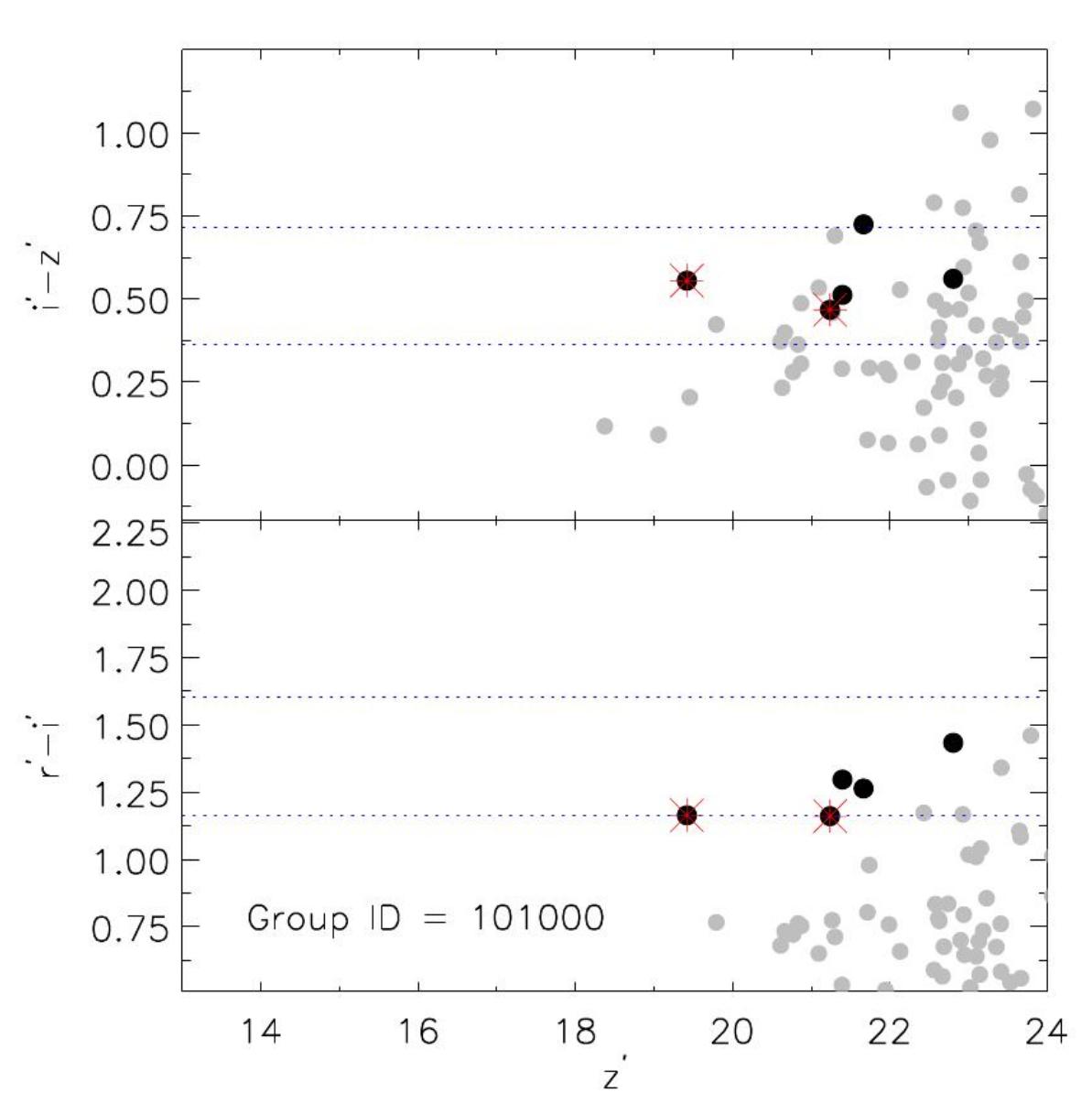}} 
 \end{center}
\caption[flag]{Same as in Fig. \ref{f19}, but for the group 101000 at  z=0.76.}
 \label{f20}
 \end{figure}

\begin{figure}[H]
\begin{center}  
\leavevmode
\resizebox{\hsize}{!}{\includegraphics{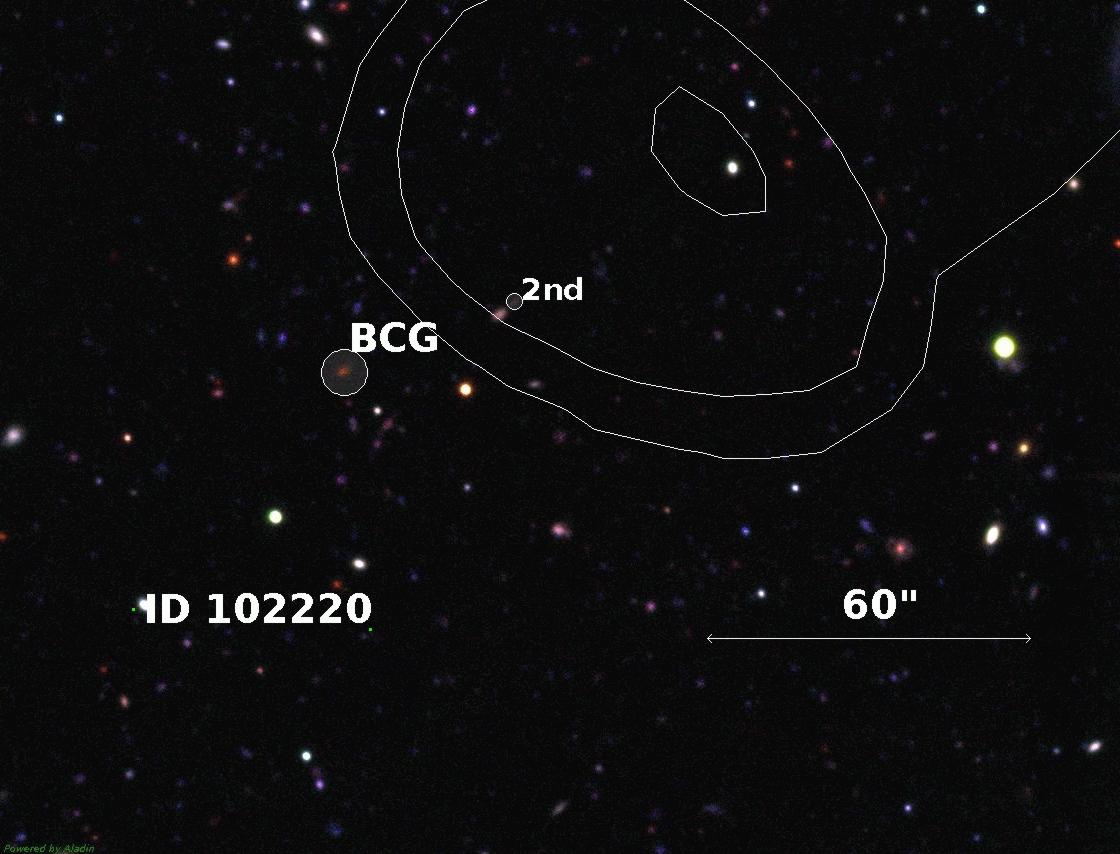} }
\resizebox{\hsize}{!}{\includegraphics{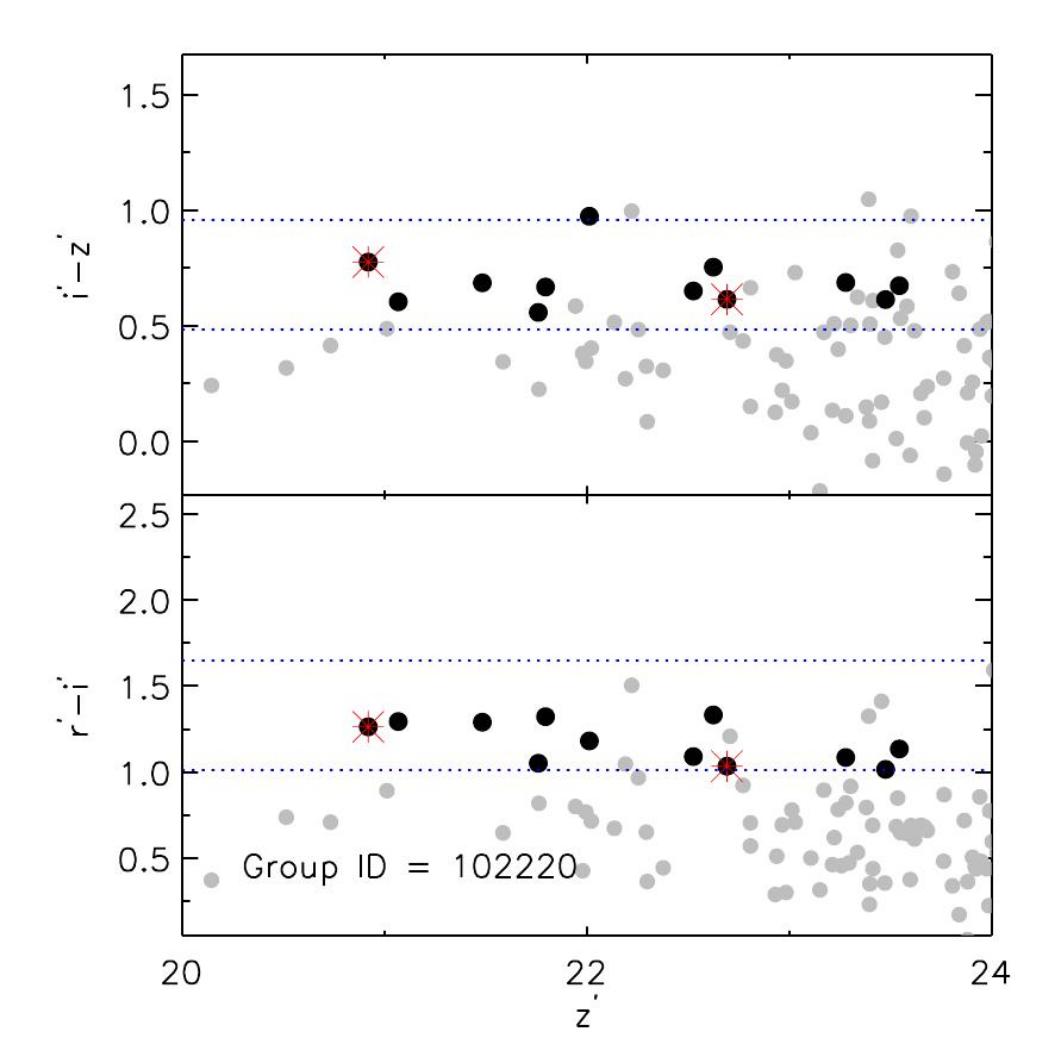}} 
\end{center}
\caption[flag]{Same as in Fig. \ref{f19}, but for the group 101120 at  z=0.86.}
 \label{f21}
\end{figure}

\begin{figure}[H]
\begin{center}  
 \leavevmode
\resizebox{\hsize}{!}{ \includegraphics{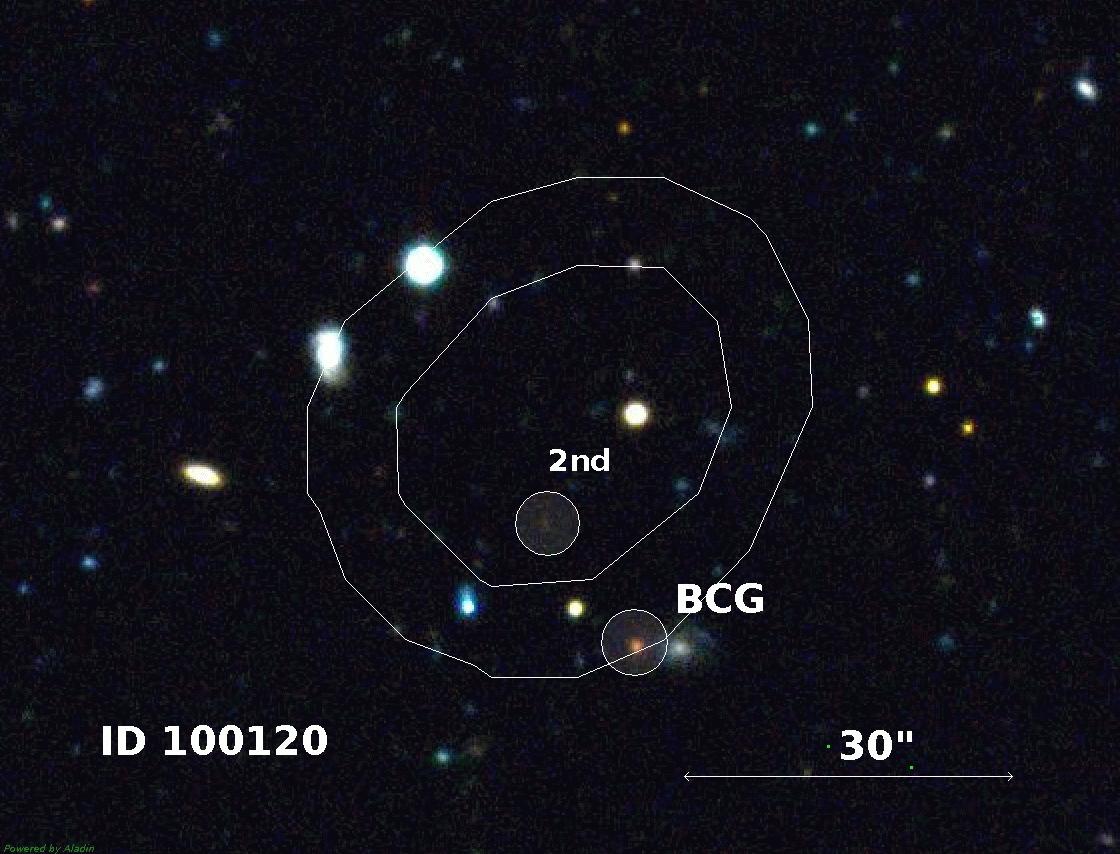}}
  \resizebox{\hsize}{!}{ \includegraphics{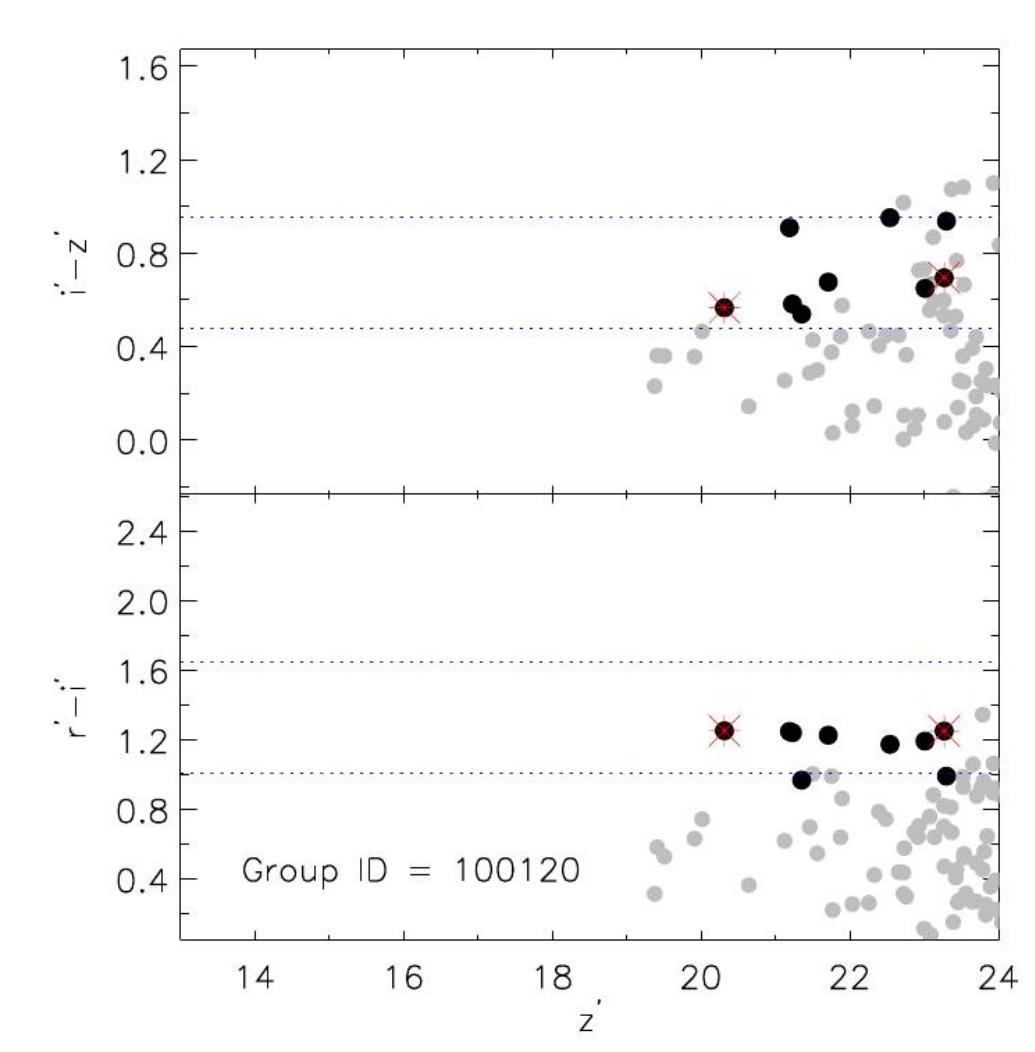} }
\end{center}
 \caption[flag]{Same as in Fig. \ref{f19}, but for the group 100120 at  z=0.88.}
\label{f22}
\end{figure}

\begin{figure}[H]
\begin{center}  
\leavevmode
\resizebox{\hsize}{!}{\includegraphics{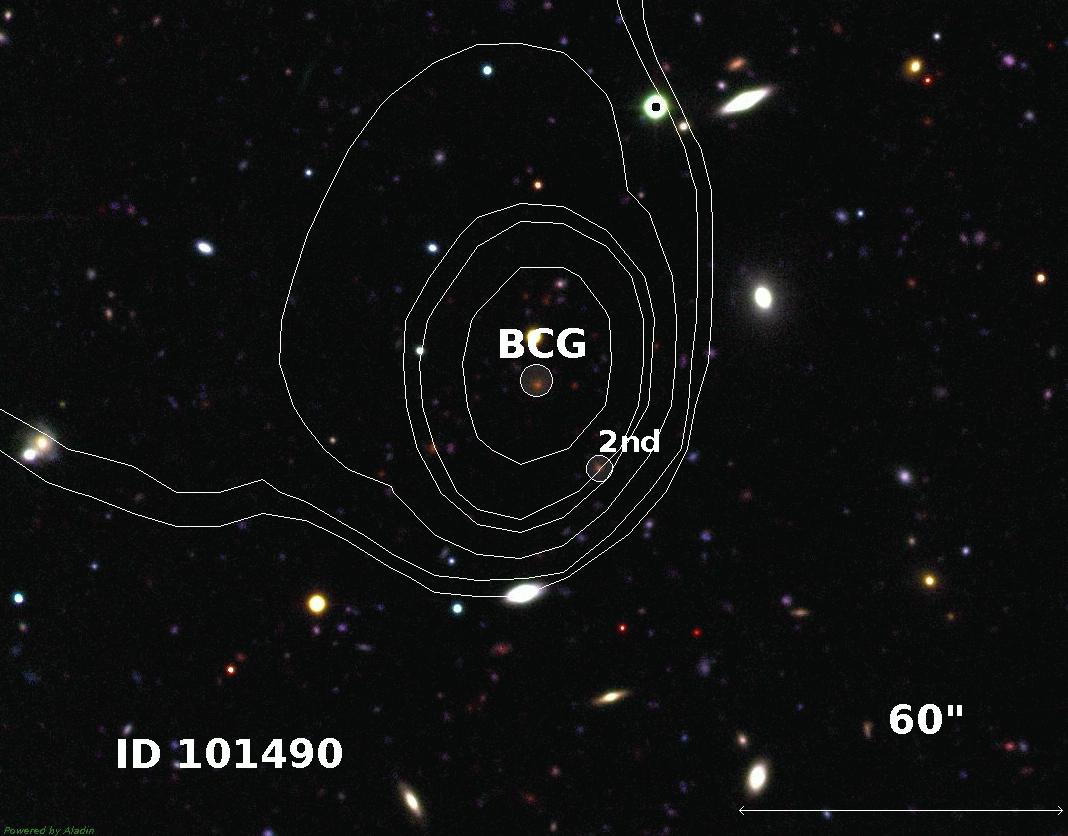} }
\resizebox{\hsize}{!}{\includegraphics{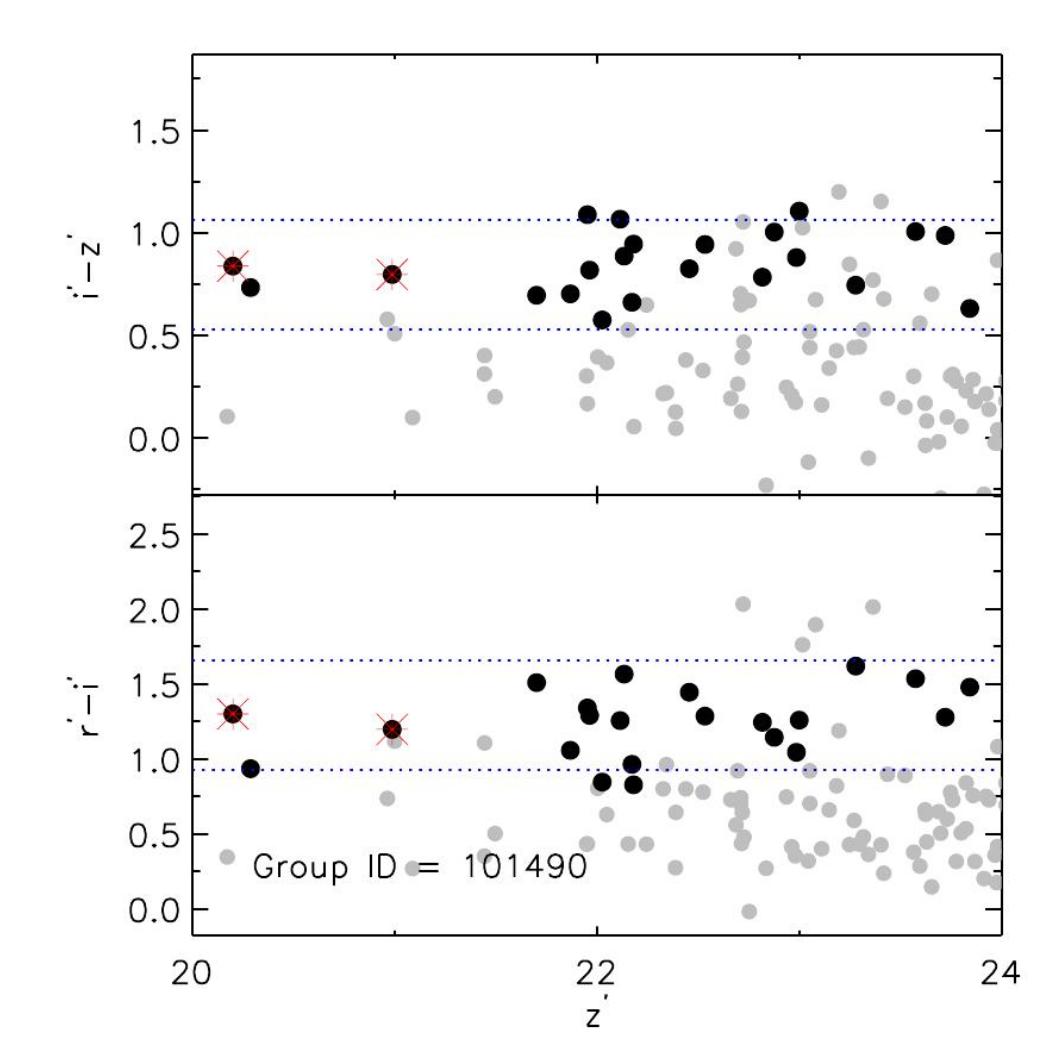}} 
\end{center}
 \caption[flag]{Same as in Fig. \ref{f19}, but for the group 101490, at  z=0.94.}
 \label{f23}
\end{figure}

\begin{figure}[H]
\begin{center}  
\leavevmode
\resizebox{\hsize}{!}{\includegraphics{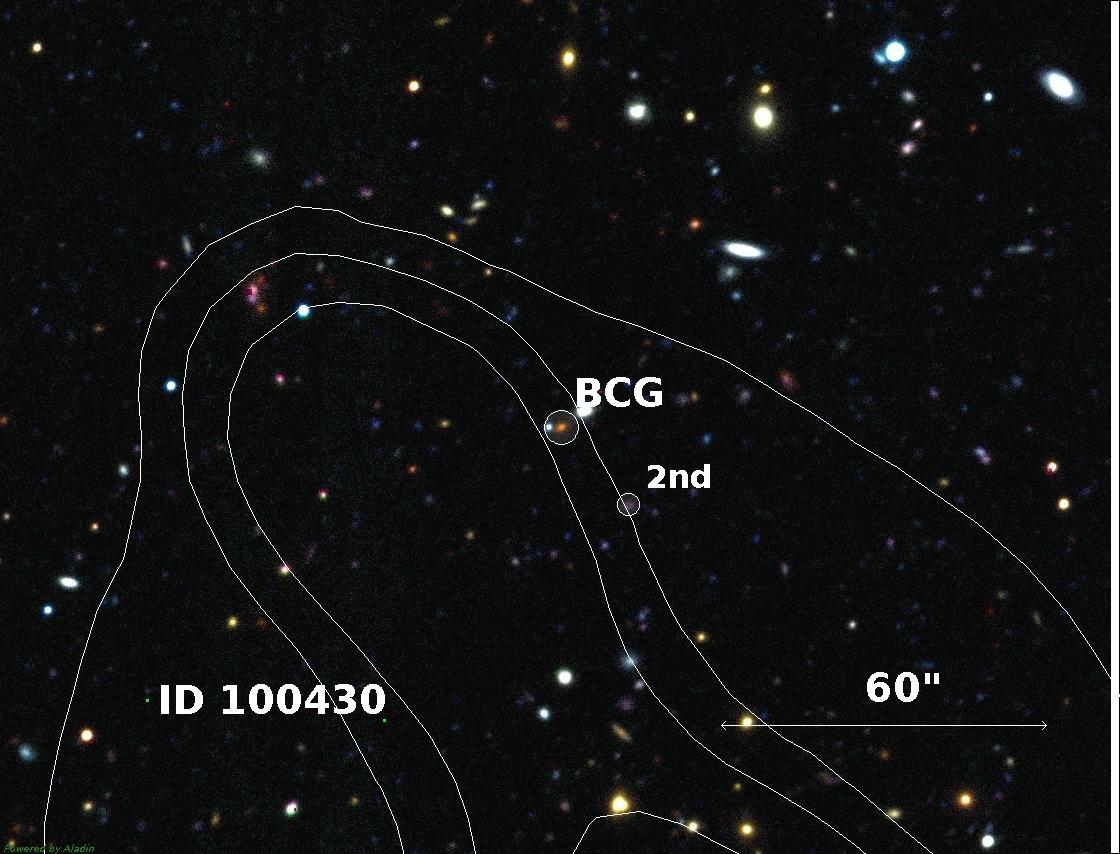} }
\resizebox{\hsize}{!}{\includegraphics{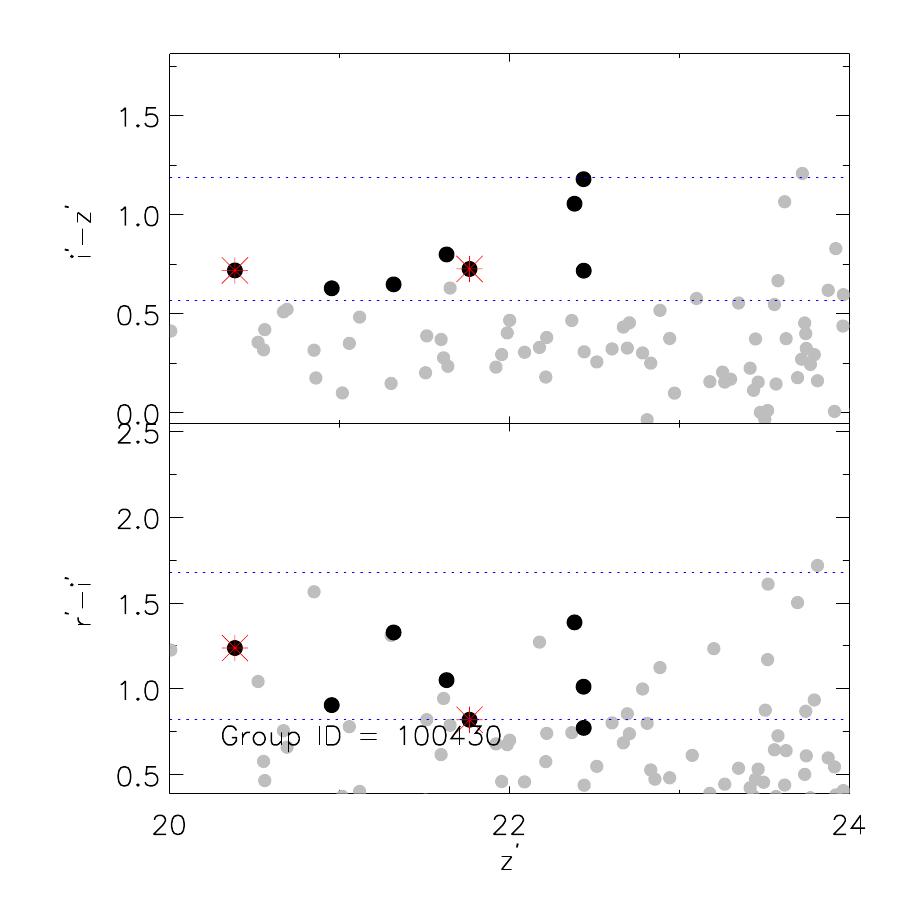}} 
\end{center}
 \caption[flag]{Same as in Fig. \ref{f19}, but for the group 100430, at  z=0.98.}
 \label{f24}
\end{figure}

\begin{figure}[H]
\begin{center}  
  \leavevmode
 \resizebox{\hsize}{!}{\includegraphics{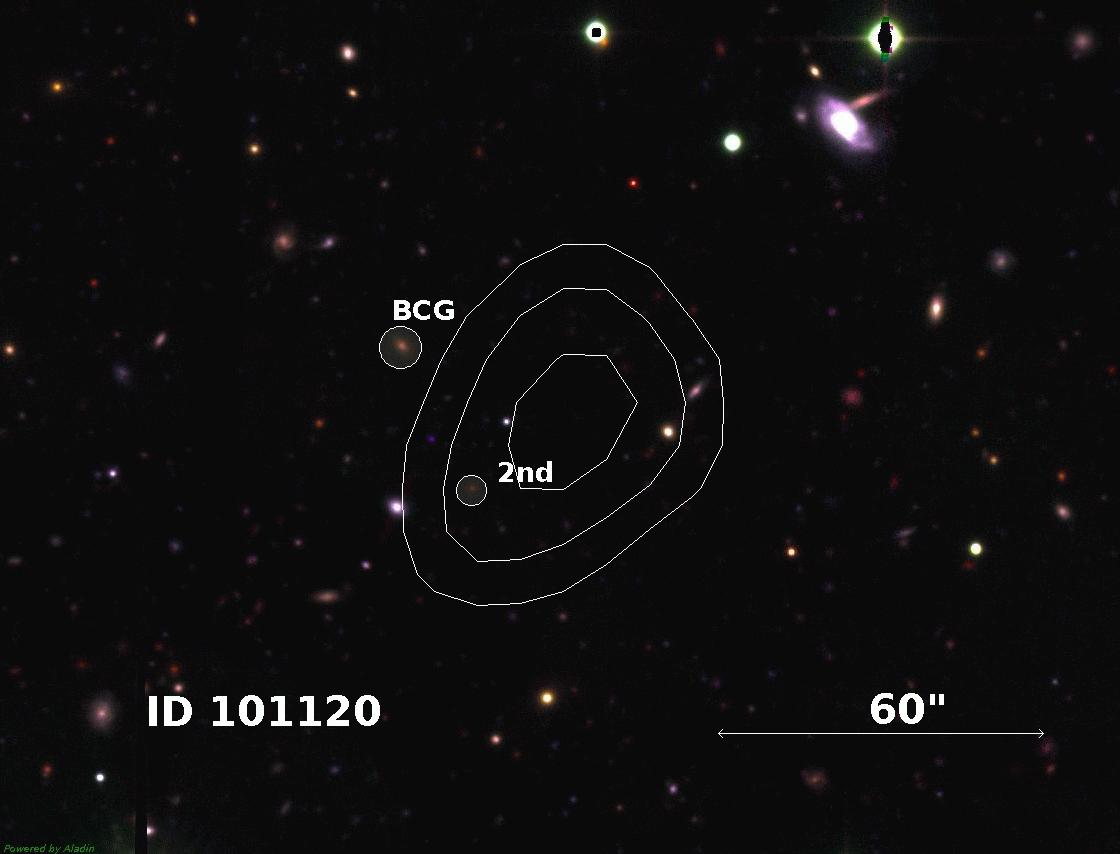}}
\resizebox{\hsize}{!}{ \includegraphics{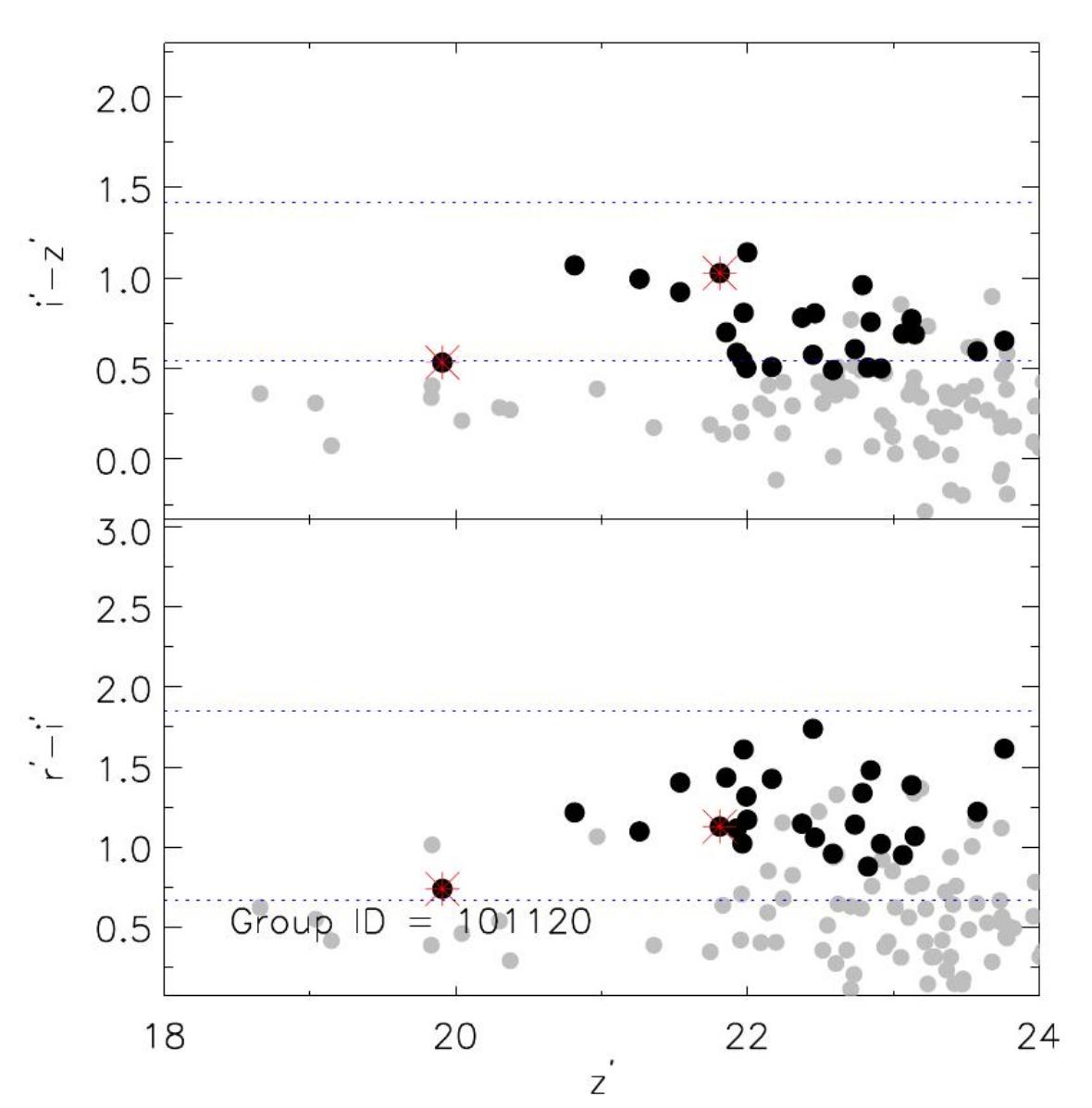}} 
 \end{center}
 \caption[flag]{Same as in Fig. \ref{f19}, but for the group 101120, at  z=1.1.} 
 \label{f25}
 \end{figure}

In Tab. A1 we present fossil group candidates catalog. Column 1
  and column 2 are group id and the photometric redshift of group.
Columns 3 and 4 present the BGG coordinate RA.(J2000) and Dec.(J2000)
in degrees. Columns 5 and 6 are BGG r$^{\prime}$-band magnitude and
$\Delta M_{1,2}$. In Figs. \ref{f1}-\ref{f25} we show the X-ray
emission as contours overlaid on the CFHTLS RGB image (top panel) and
the color magnitude diagrams of g$^{\prime}$ - r$^{\prime} $ and
r$^{\prime}$ - i$^{\prime} $ versus z$^{\prime} $ for z $< 0.6 $ and
r$^{\prime}$ - i$^{\prime} $ and i$^{\prime}$- z$^{\prime} $ versus
z$^{\prime} $ for z $> 0.6 $ for fossil group candidates listed in
Tab. \ref{fossil}.

In \S 5.2 we define a fossil group and the search radius used
  for the magnitude gap calculation. Group members selected using two
  colors according to the method described in \S 4.1. We show group
  members with filled black circles in each color magnitude
  diagrams. The first and second brightest galaxies are marked by red
  asterisks. The upper and lower limits of colors in each redshift
  have been shown with horizontal dotted blue lines.

  As it can be seen in the color magnitude diagrams, in some galaxy
  groups there are galaxies outside the 0.5$R_{200}$ that are brighter
  than the second galaxy which is used to compute the $\Delta
  M_{1,2}$. These groups are experiencing the infall of galaxies into
  search radius. This demonstrates that the magnitude gap estimate and
  the identification of fossil groups are very sensitive to the
  adopted search radius for selecting the second bright galaxy.
 
 \begin{table}
   \caption{\footnotesize Fossil group candidates.
   \label{fossil}}
  \tiny
  \begin{center}                                         
     \begin{tabular}[tc]{cccccc}
      \hline
       \hline\\
 Group ID & z$ _{phot} $& RA$ _{BGG}(J2000) $  & Dec$ _{BGG}(J2000) $ & $r^{\prime}$ & $ \Delta M_{1,2}$\\
     (1) & (2)  & (3) & (4) & (5) & (6) \\
    \hline \\
  101540   & 0.07   & 36.3736 & -4.0120  & 14.219    & 1.922    \\
 101391   & 0.07   & 36.1958 & -4.1475  & 16.929   & 3.297     \\
  102820   & 0.11   & 36.9687 & -3.8552  & 17.213   & 2.059      \\
  100130   & 0.13   & 36.8133 & -5.0756  & 16.367   & 2.645    \\
  103330   & 0.14   & 35.3946 & -4.038   & 17.142    & 6.922     \\
  101210   & 0.15   & 36.3319 & -4.317   & 16.634   & 1.747       \\
  100150   & 0.16   & 35.2821 & -4.9992  & 17.302   & 2.957     \\
   100010 & 0.22  & 36.4159 &  -5.0052 & 17.241 & 1.714\\
  102130   & 0.23   & 35.6247 & -3.8769  & 18.470   & 2.232       \\
   101290   & 0.25   & 36.6974 & -4.2531  & 18.733   & 2.514      \\
  101480   & 0.34   & 36.6486 & -4.0695  & 18.025   & 2.157      \\
  100640   & 0.34   & 35.5052& -4.6332&  18.128   & 2.022     \\
  101200   & 0.47   & 35.6688 & -4.3064  & 20.393  & 4.607       \\ 
  103320   & 0.53   & 36.5861 & -3.7339  & 20.354   & 1.791       \\
  101730   & 0.6    & 36.0241 & -3.9213  & 20.130   & 1.955       \\
  101360   & 0.6    & 36.8554 & -4.1823  & 20.489    & 2.728      \\
  100620   & 0.66  & 37.1107 &  -4.6587  & 22.320 & 2.103\\
   101000   & 0.76   & 35.8876 & -4.4227  & 21.136   & 1.727     \\
   102220    & 0.86  & 35.7790 &  -3.7906  & 22.957 & 2.187 \\ 
   100120   & 0.88   & 35.3644 & -5.067   & 22.125   & 3.080      \\
  101490    & 0.94  &  35.3728 &   -4.0977& 22.336 & 1.989 \\
   100430    &  0.98  & 36.1239 &  -4.8010 & 22.342 & 1.861 \\
  101120   & 1.1    & 35.5422 & -4.3494  & 21.181   & 2.808      \\
  \hline 
 
  \end{tabular}    
 \end{center}               
 \end{table}    
 \newpage


\begin{figure}[H]
\begin{center}  
\leavevmode
\resizebox{\hsize}{!}{\includegraphics{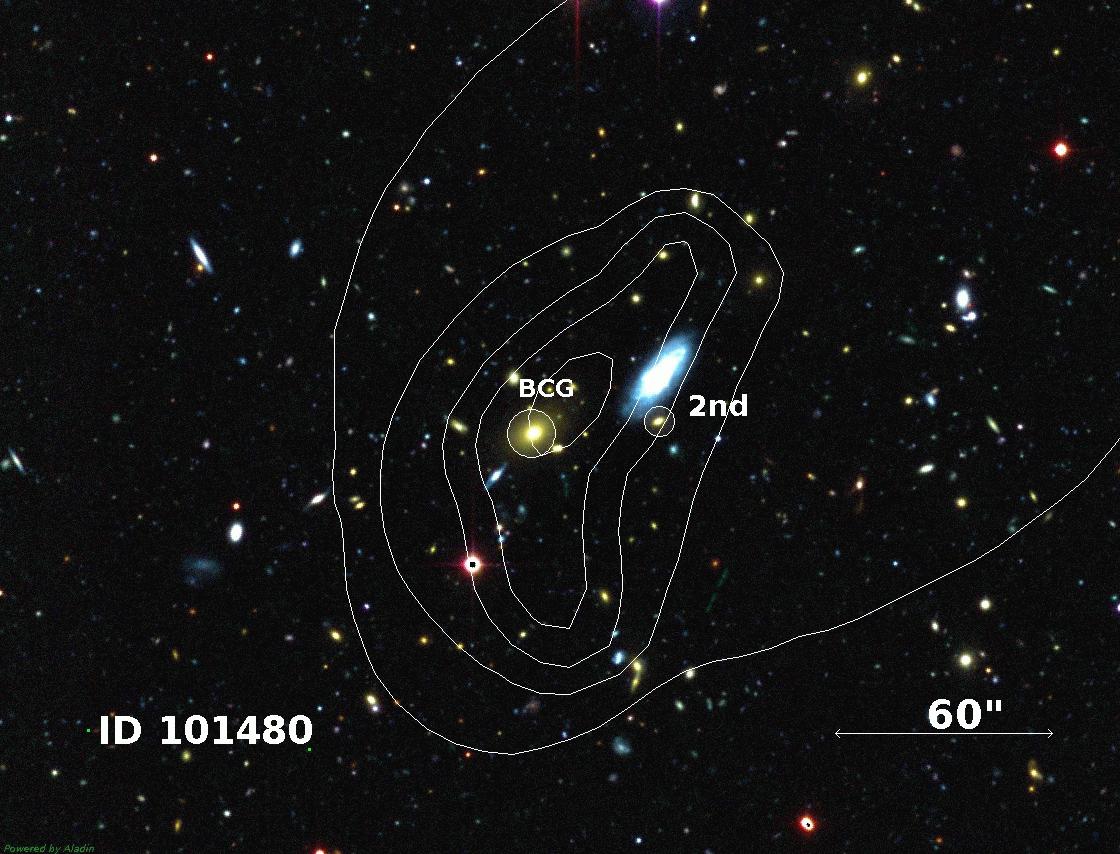}}  
\resizebox{\hsize}{!}{\includegraphics{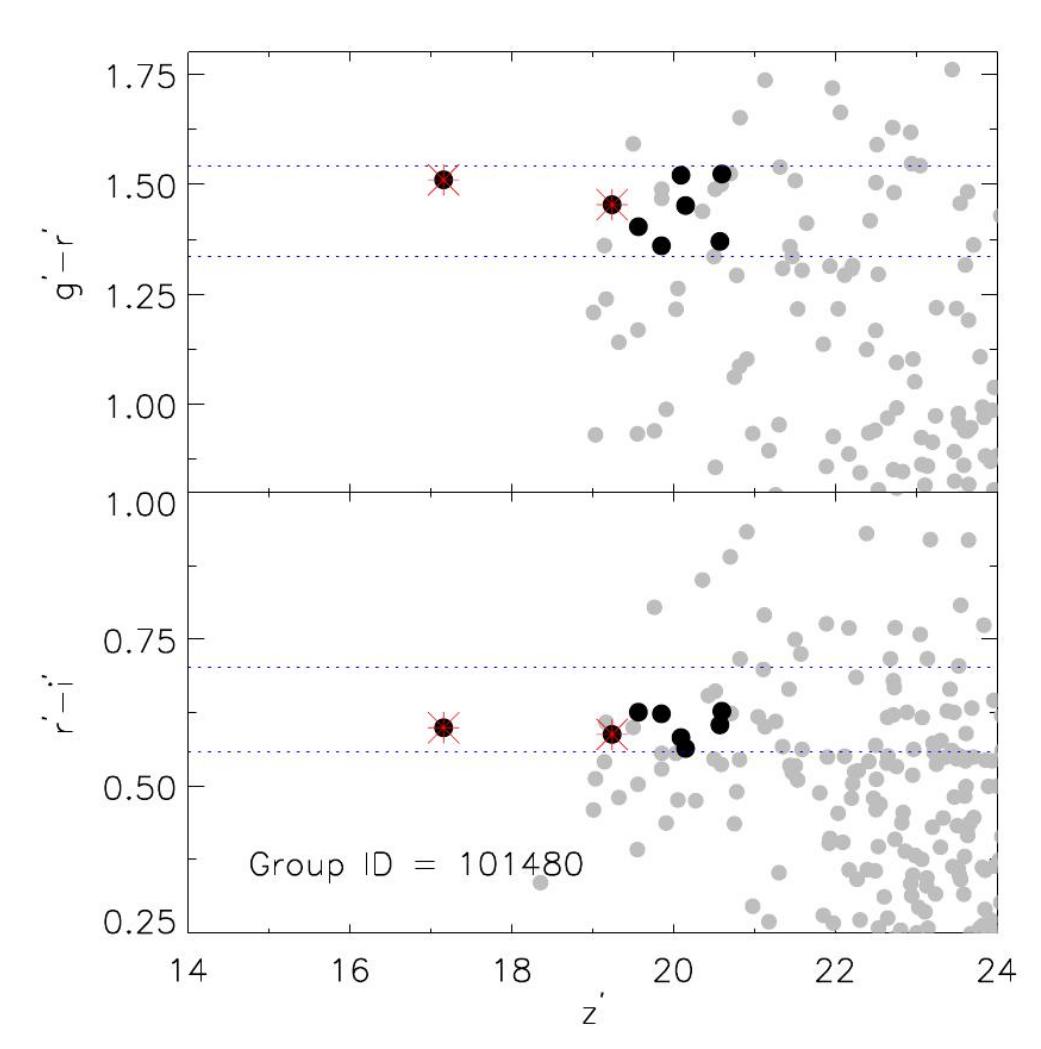}}                 
\end{center} 
\caption[flag]{Same as in Fig. \ref{f1}, but for the group 101480, at z=0.34.} 
\label{f13}
\end{figure}
\begin{figure}[H]
\begin{center}  
\leavevmode
\resizebox{\hsize}{!}{\includegraphics{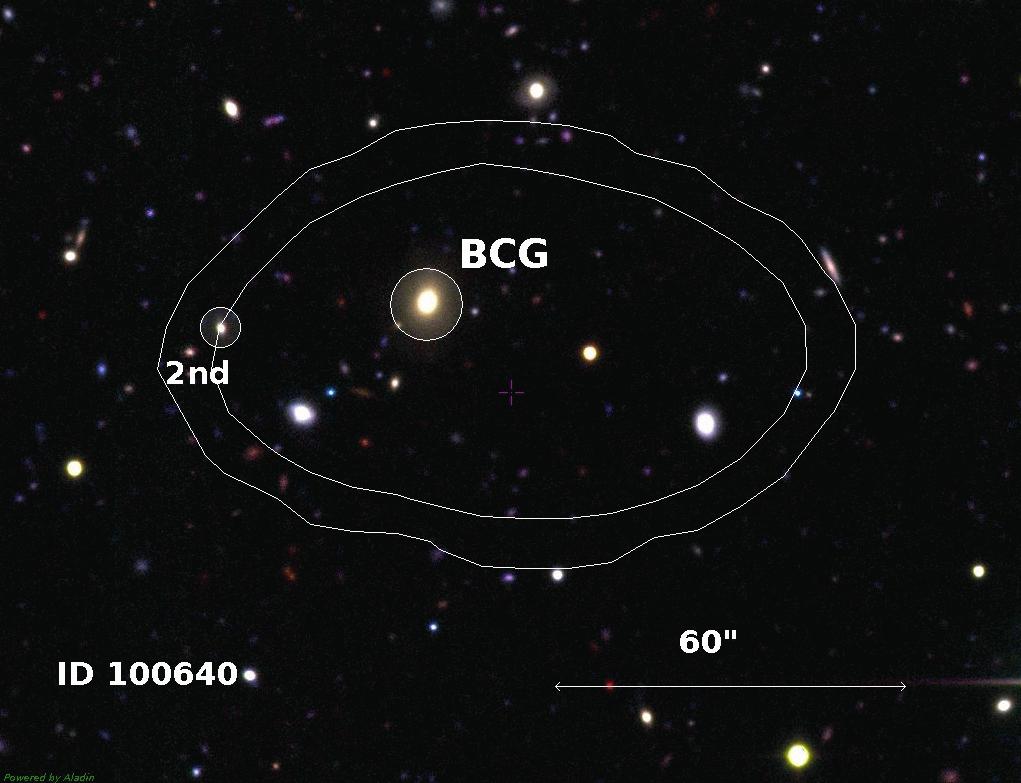}}  
\resizebox{\hsize}{!}{\includegraphics{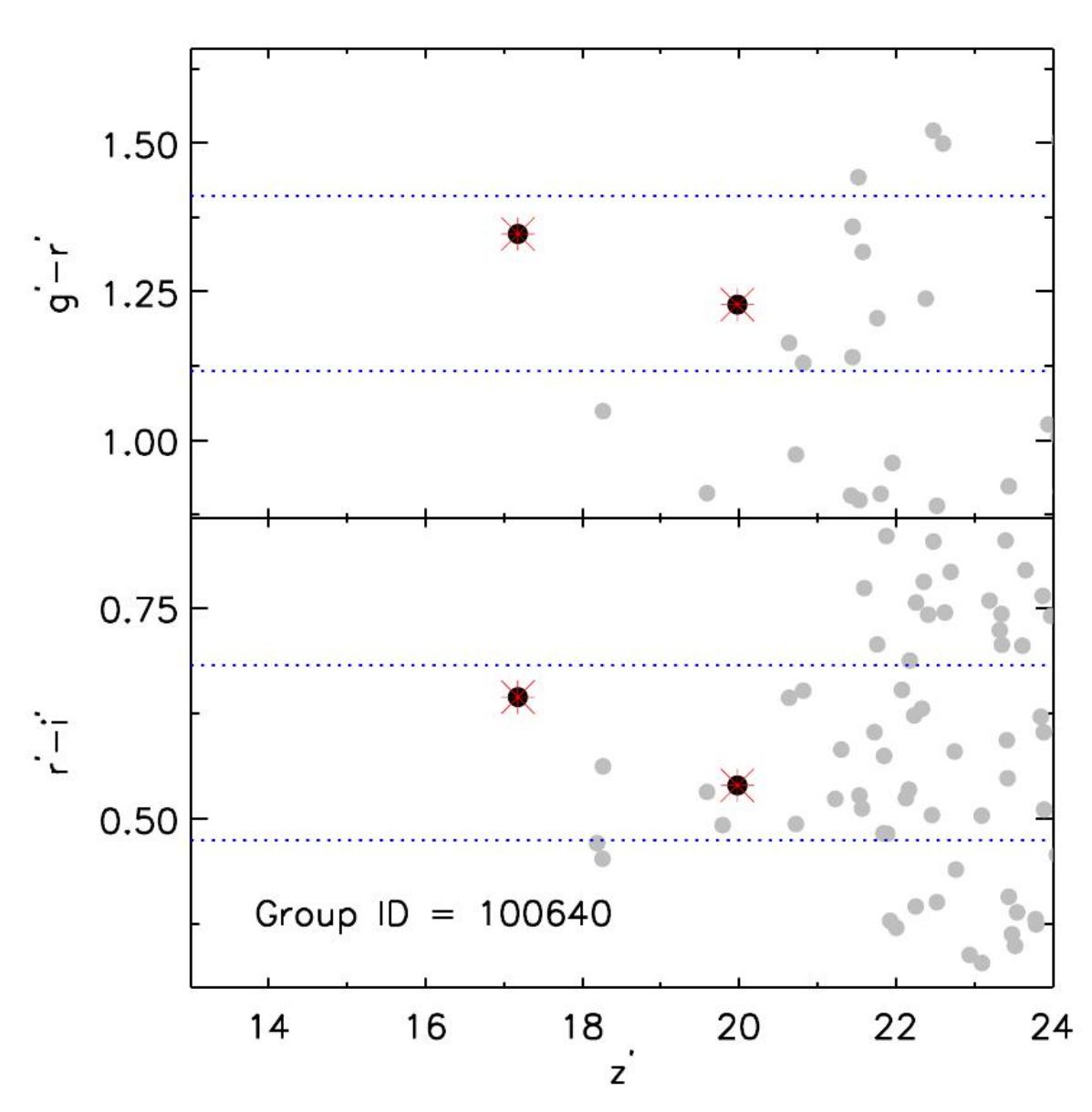}}                 
\end{center} 
\caption[flag]{Same as in Fig. \ref{f1}, but for the group 100640, at z=0.34.} 
\label{f13}
\end{figure}

                       
  
 \newpage
\begin{figure}[!ht]
\begin{center}  
\leavevmode
\resizebox{\hsize}{!}{\includegraphics{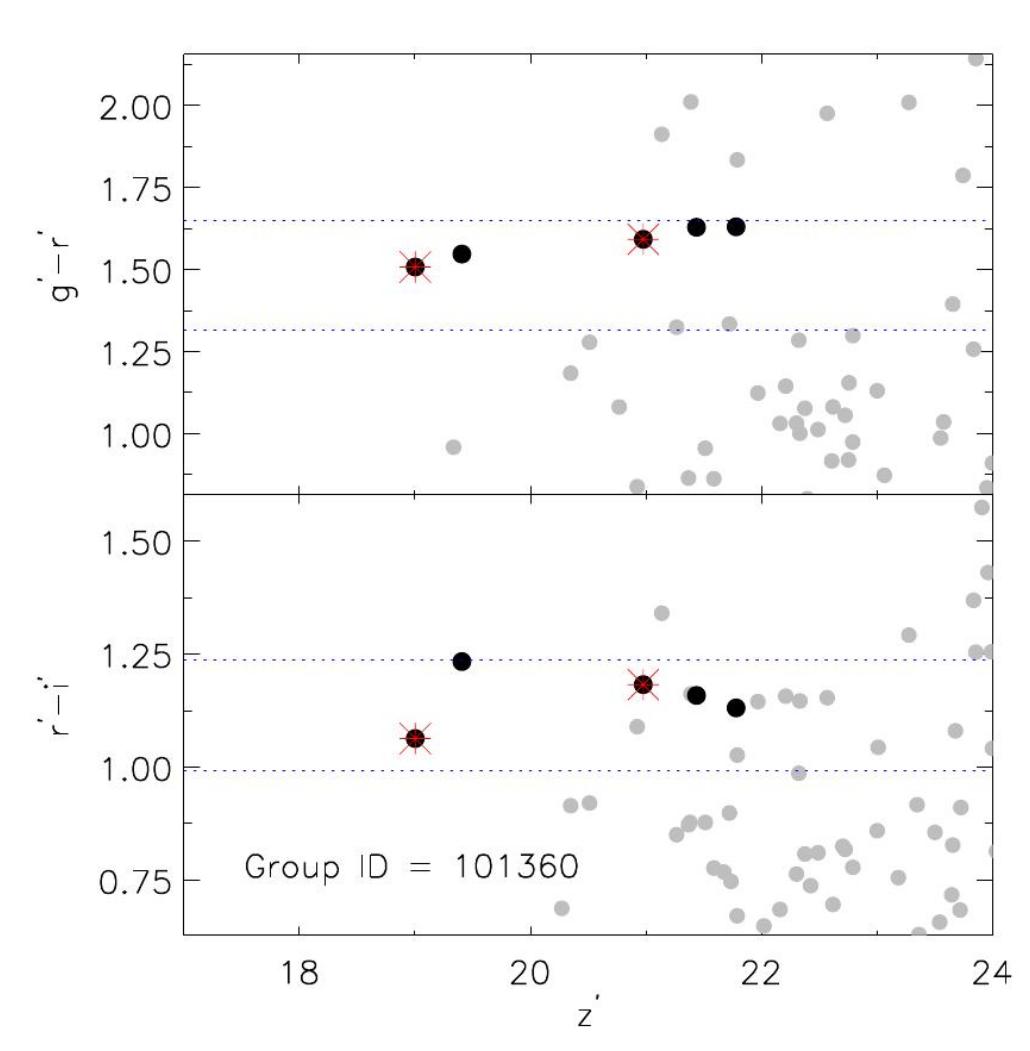}} 
\end{center}
\caption[flag]{Same as in Fig. \ref{f1}, but for the group 101360, at  z= 0.60. 
The RGB image of this group has been presented in  Fig. \ref{flag} (left bottom panel). }
\label{f18}
\end{figure}

\end{document}